\documentclass[a4paper,10pt]{article}
\usepackage[utf8]{inputenc}
\usepackage[T1]{fontenc}
\usepackage[top=2.5cm,bottom=2.5cm,left=2.5cm,right=2.5cm]{geometry}
\usepackage{textcomp}
\usepackage{fancyhdr}
\usepackage{fullpage}
\usepackage{lmodern}
\usepackage{amsmath,amssymb,bm,bbm}
\usepackage{algorithm}
\usepackage{algorithmic}
\usepackage{float}
\usepackage{extarrows}
\usepackage{caption}
\usepackage{graphicx,subfig}
\usepackage{titling}
\usepackage{yhmath}
\usepackage{cite}
\usepackage{footnote}
\usepackage{amsthm}
\usepackage{color}
\usepackage{slashed}
\usepackage{verbatim}
\usepackage[colorlinks=true,linkcolor=red,citecolor=blue]{hyperref}
\usepackage{amsfonts,euscript}
\usepackage{latexsym, multicol}
\usepackage{epstopdf}
\usepackage{psfrag}
\usepackage{mathrsfs}
\usepackage{xcolor}
\usepackage{comment}
\usepackage{authblk}
\usepackage{dsfont}
\usepackage{indentfirst}
\usepackage{lipsum}
\usepackage{tikz}
\def\pa{\parallel}
\usetikzlibrary{intersections,calc,decorations.pathmorphing}
\DeclareFontFamily{U}{mathx}{\hyphenchar\font45}
\DeclareFontShape{U}{mathx}{m}{n}{
<5> <6> <7> <8> <9> <10>
<10.95> <12> <14.4> <17.28> <20.74> <24.88>
mathx10}{}
\DeclareSymbolFont{mathx}{U}{mathx}{m}{n}
\DeclareFontSubstitution{U}{mathx}{m}{n}
\DeclareMathAccent{\widecheck}{0}{mathx}{"71}
\numberwithin{equation}{section}
\allowdisplaybreaks[3]
\def\CC{\mathcal{C}}

\def\TT{\mathcal{T}}
\def\slin{\slashed{\in}}
\def\Dk{{\mathfrak{D}}}
\def\NN{\mathcal{N}}
\def\GG{\mathcal{G}}
\def\bGa{{\mathbf{\Ga}}}
\def\bu{\bullet}
\def\vkp{\varkappa}
\def\dual{{}^*}
\def\ges{\gtrsim}
\def\omd{\om^\dagger}
\def\gb{{\overline{\g}}}
\def\Nb{\underline{N}}
\def\mub{{\underline{\mu}}}

\def\sic{\widecheck{\si}}

\def\ep{\varepsilon}
\def\Ot{\widetilde{\mo}}
\def\cuvi{H_0^{(0,\ub)}}
\def\ucuvi{\Hb_0^{(0,u)}}
\def\pb{\ov{p}}
\def\Db{\ov{\D}}
\def\Jh{\widehat{J}}

\def\VV{\mathcal{V}}
\def\HH{\mathcal{H}}

\renewcommand{\c}{\cdot}
\newcommand{\Lb}{\underline{L}}

\newcommand{\pr}{\partial}
\newcommand{\ub}{{\underline{u}}}

\newcommand{\mo}{\mathcal{O}}
\newcommand{\chib}{\underline{\chi}}
\newcommand{\omb}{\underline{\omega}}
\newcommand{\etab}{\underline{\eta}}
\newcommand{\sld}{\slashed{d}}
\def\Hb{\underline{H}}
\def\aa{{\underline{\a}}}
\def\sk{\mathfrak{s}}

\newcommand{\mr}{\mathcal{R}}
\newcommand{\ur}{\underline{\mr}}

\newcommand{\M}{\mathcal{M}}
\newcommand{\D}{\mathbf{D}}
\newcommand{\cuv}{{H_u^{(0,\ub)}}}
\newcommand{\ucuv}{{\Hb_\ub^{(0,u)}}}
\newcommand{\dd}{{\mathfrak{d}}}
\newcommand{\Ric}{{\ric}}

\newcommand{\g}{\mathbf{g}}
\newcommand{\R}{{\mathbf{R}}}
\newcommand{\W}{{\mathbf{W}}}
\newcommand{\K}{\mathbf{K}}

\def\hot{\widehat{\otimes}}

\def\La{\Lambda}

\def\la{\lambda}

\def\om{\omega}
\def\ze{\zeta}
\def\Om{\Omega}
\def\bb{{\underline{\b}}}

\def\si{\sigma}
\def\dk{\dd}
\def\dkb{\slashed{\dk}}
\def\ga{\gamma}
\def\Ga{\Gamma}
\def\xib{\underline{\xi}}
\def\a{\alpha}
\def\b{\beta}
\def\hch{\widehat{\chi}}
\def\hchb{\widehat{\chib}}
\def\trch{\tr\chi}
\def\trchb{\tr\chib}
\def\ka{\kappa}

\def\de{\delta}
\def\De{\Delta}
\def\nab{\nabla}
\def\ov{\overline}

\def\les{\lesssim}

\def\RRR{\mathbb{R}}
\def\th{\theta}
\newcommand{\T}{\mathbf{T}}

\DeclareMathOperator{\ver}{Ver}
\DeclareMathOperator{\hor}{Hor}
\DeclareMathOperator{\dvol}{dvol}
\DeclareMathOperator{\curl}{curl}

\DeclareMathOperator{\grad}{grad}
\DeclareMathOperator{\supp}{supp}

\DeclareMathOperator{\tr}{tr}
\DeclareMathOperator{\sdiv}{div}
\def\bdiv{\mathbf{Div}}
\def\ric{\mathbf{Ric}}
\def\S{\mathbf{S}}
\def\Sslh{\widehat{\Ssl}}
\def\Ssl{\slashed{\S}}
\def\slg{{\slashed{g}}}
\def\nabs{\nab}
\def\bbb{\underline{b}}
\def\Tr{\mathbf{Tr}}

\newtheorem{thm}{Theorem}[section]
\newtheorem{theorem}[thm]{Theorem}
\newtheorem{prop}[thm]{Proposition}
\newtheorem{lem}[thm]{Lemma}
\newtheorem{cor}[thm]{Corollary}
\newtheorem{rk}[thm]{Remark}
\newtheorem{remark}[thm]{Remark}
\newtheorem{df}[thm]{Definition}

\usepackage[breakable]{tcolorbox}
\newtcolorbox{warning}[1][red!75!black]{
  breakable,
  colback=white,
  colframe=#1,
  boxrule=.6pt,
  arc=0pt,
  left=5pt,right=5pt,top=4pt,bottom=4pt
}
\title{Weak null singularities for the Einstein--Vlasov system}

\author{
Dawei Shen\thanks{
Department of Mathematics, Columbia University,
New York, NY 10027, USA.
Email: \texttt{ds4350@columbia.edu}
}
\quad
Yuefeng Song\thanks{
Department of Mathematics, Stanford University,
Stanford, CA 94305, USA.
Email: \texttt{songyf@stanford.edu}
}
\quad
Jingbo Wan\thanks{
Laboratoire Jacques-Louis Lions, Sorbonne Universit\'e,
Paris, France.
Email: \texttt{jingbo.wan@sorbonne-universite.fr}
}
}
\begin{document}
\maketitle
\begin{abstract}
We study the formation and stability of weak null singularities for solutions to the massive and massless Einstein--Vlasov systems without symmetry assumptions. Extending previous results of Luk in the vacuum setting, we establish the existence of spacetimes arising from characteristic initial data that develop a singular null boundary while remaining extendible with a continuous metric.

For both massive and massless particles, the energy-momentum
tensor and Ricci curvature remain bounded and extend
continuously to the singular boundary. For genuinely singular
initial data, the Christoffel symbols in the original double-null
coordinates fail to be locally square integrable near the
boundary. We introduce new commutator vector fields that avoid both derivative loss and the most singular curvature component. Under additional curvature assumptions, we also prove Lipschitz inextendibility across the singular null boundary.
\end{abstract}
{\hypersetup{linkcolor=black}
\tableofcontents}
\section{Introduction}
\subsection{Motivation}
Recent works suggest that generic dynamical \emph{vacuum} black holes terminate at an essential weak null singularity across which the metric is continuously extendible but not Lipschitz extendible. Stable weak null singularities without symmetry assumptions were constructed locally in \cite{Luk18}; see also \cite{OriFlanagan95} for analytic examples. Indeed, the results of \cite{Daluk,Gurriaran26,LukSbierski26} show that if the black hole exterior settles down to a subextremal rotating Kerr black hole with suitable upper and lower bounds, then the black hole interior terminates at an essential weak null singularity. In the black hole interior, the required upper bound has been proven in \cite{Daluk} for initial data that are small perturbations of exact Kerr initial data, while the required lower bound is expected to hold generically; see \cite[Remark~4.8]{LukSbierski26}. Modulo the conjectured lower bounds, it follows that an essential weak null singularity forms in generic rotating black holes, at least for initial data that are sufficiently close to the corresponding data of subextremal Kerr. For related linear results, see \cite{DafermosShlapentokhRothman16,Gurriaran25,LukOh17, LukSbierski16,MaZhang23,Sbierski23}. For further references, see \cite{VandeMoortel25}.

Motivated by this picture, it is natural to investigate whether weak null singularities persist after the introduction of matter fields and how self-gravitating matter fields behave at the singularities. Several works have studied weak null singularities in the presence of matter fields. One class of results concerns null dust, Maxwell fields, and charged or uncharged, massive or massless scalar fields. Their equations of motion are luminal hyperbolic systems, propagating at the speed of light; see
\cite{Da,Dafermos05,Gautam24,Hiscock81,LukOh19a,LukOh19b,LukOhShlapentokhRothman23,PoissonIsrael89,PoissonIsrael90,VandeMoortel17,VandeMoortel21,VandeMoortel23}.

For these matter fields, the spherically symmetric picture --- both for the metric and for the matter field --- is moreover locally stable under small but nonspherically symmetric perturbations. Indeed, because the equations of motion for these matter fields have a structure similar to that of the vacuum Einstein equations, the local stability of the weak null singularities even with the presence of matter fields can be proven directly with the methods of Luk \cite{Luk18}; see \cite[Footnote 8]{Luk18}.

In recent work by the second author \cite{Song}, weak null singularities for the Einstein--Euler system without symmetry assumptions were studied. In this case, new difficulties arise because the Euler equations form a \emph{subluminal} hyperbolic system, propagating at a speed strictly lower than the speed of light. Nonetheless, for a class of characteristic initial value problems with geometric initial data exhibiting a weak null singularity and smooth, bounded initial fluid variables, it was shown that the metric continues to have a weak null singularity, while the fluid variables remain smooth in the interior and extend continuously to the singular boundary.
In particular, no shocks form. The boundedness of the fluid variables reflects additional regularity arising from the slower propagation speed of the Euler equations. In contrast to the matter fields discussed above (null dust, Maxwell fields, and scalar fields), the Ricci curvature in the Einstein--Euler system remains uniformly bounded.

In this paper, \textbf{we study essential weak null singularities for the Einstein--Vlasov system} without any symmetry assumptions. In view of the preceding discussion, one may expect the massive case and the massless case to be different, since the matter field is subluminal in the massive case and luminal in the massless case. However, \textbf{our theorem shows that \emph{independently} of the mass of the Vlasov field, a local picture similar to that in \cite{Song} holds.} As we will argue below, this is due to the fact that even in the massless case, only a set of momentum directions of zero fiber measure is tangent to the singular boundary. Moreover, one can utilize momentum-weighted commutators adapted to the weak null singularity to avoid the most singular curvature components, while the degenerate bounds of the derivatives of the distribution function remain integrable on the mass shell, thereby yielding control of the energy-momentum tensor.

We first give an informal statement of our main results. See Theorems~\ref{MainTheorem}, \ref{Luk3}, and \ref{Singularity} for the precise assumptions and conclusions.
\begin{theorem}\label{thm:first.version}
For the Einstein--Vlasov system with either massive or massless particles, consider characteristic initial data containing a metric weak null singularity, with a regular Vlasov distribution function compactly supported in momentum space; see Figure~\ref{fig:local-WNS-diamond}. Then:
\begin{enumerate}
    \item There exists a local solution arising from such data; see Theorem~\ref{MainTheorem}.
    \item The metric extends continuously across the Cauchy horizon, and the energy-momentum tensor remains bounded and extends continuously up to it; see Theorem~\ref{Luk3}.
    \item The Christoffel symbols in the double-null coordinates fail to be locally square-integrable at the Cauchy horizon, presenting an essential weak null singularity. Under additional curvature assumptions, the spacetime is also $C_{\mathrm{loc}}^{0,1}$--inextendible across it; see Theorem~\ref{Singularity}.
\end{enumerate}
\end{theorem}
 \begin{figure}[H]
\centering
\begin{minipage}[t]{0.48\textwidth}
\centering
    \begin{tikzpicture}[
        line cap=round,
        line join=round,
        every path/.style={line width=1.0pt}
        ]
        \begin{scope}[scale=1.2]
        \coordinate (L) at (3,0.18);
        \coordinate (A) at (3,3);
        \coordinate (R) at (6,0);
        \coordinate (R2) at (5.05,0.95);
        \draw[name path=rightside] (A) -- (R2);
        \draw[dashed, dash pattern=on 4pt off 3pt] (R2) -- (R);
        \path[name path=bottom] (L) .. controls (4.05,0.18) and (5.1,0.12) .. (R);
        \path[name path=rightlineB] (R2) -- (4.1,0);
        \path[name intersections={of=rightlineB and bottom, by=B4}];
        \draw (B4) -- (R2);
        \node[font=\small,right] at ($(B4)!0.5!(R2)$) {$\mathcal{H}^+$};
        \draw (L) .. controls (4.05,0.18) and (5.1,0.12) .. (R);
        \draw[decorate, decoration={zigzag,segment length=4pt,amplitude=1.5pt}]
        (L) .. controls ($(L)!0.35!(R)+(0.16,0.035)$)
             and ($(L)!0.65!(R)+(-0.16,0.035)$) .. (R);
        \node[font=\small] at ($(L)!0.5!(R)+(0,-0.32)$) {Cauchy data};
        \node[font=\small] at ($(A)!0.42!(R)+(0.34,0.44)$) {$\mathcal{CH}^+$};
            \path (R) ++(0,-0.5);
        \end{scope}
    \end{tikzpicture}
    \caption{General matter data on an asymptotically flat Cauchy hypersurface}
    \label{Figure1.0}
\end{minipage}\hfill
\begin{minipage}[t]{0.48\textwidth}
\centering
        \begin{tikzpicture}[
            line cap=round,
            line join=round,
            every path/.style={line width=1.0pt}
            ]
            \begin{scope}[scale=1.2]
            \coordinate (L) at (2,0.161682307);
            \coordinate (A) at (3,3);
            \coordinate (R) at (6,0);
            \coordinate (R2) at (5.05,0.95);
            \draw[name path=rightside] (A) -- (R2);
            \draw[dashed, dash pattern=on 4pt off 3pt] (R2) -- (R);
            \path[name path=bottom] (L) .. controls (3.364822224,0.212174847) and (4.812894415,0.158280745) .. (R);
            \coordinate (L2) at (2,0.7);
            \path[name path=leftlineB] (L2) -- (2.7,0);
            \path[name intersections={of=leftlineB and bottom, by=B2}];
            \draw[dashed, dash pattern=on 4pt off 3pt] (L2) -- (B2);
            \coordinate (R1) at (4.65,1.35);
            \path[name path=rightlineA] (R1) -- (3.3,0);
            \path[name intersections={of=rightlineA and bottom, by=B3}];
            \draw[dashed, dash pattern=on 4pt off 3pt] (B3) -- (R1);
            \path[overlay,name path=rightlineB] (R2) -- ++(-3,-3);
            \path[name intersections={of=rightlineB and bottom, by=B4}];
            \draw (B4) -- (R2);
            \node[font=\small,right] at ($(B4)!0.5!(R2)$) {$\mathcal{H}^+$};
            \draw (L) .. controls (3.364822224,0.212174847) and (4.812894415,0.158280745) .. (R);
            \draw[decorate, decoration={zigzag,segment length=4pt,amplitude=1.5pt}]
            (B2) .. controls ($(B2)!0.35!(B3)+(0.16,0.035)$)
                 and ($(B2)!0.65!(B3)+(-0.16,0.035)$) .. (B3);
            \node[font=\small] at ($(B2)!0.5!(B3)+(0,-0.32)$) {Cauchy data};
            \begin{scope}[shift={(3.92,2.08)}, rotate=-45]
                \fill[gray!30] (-0.22,0) arc[start angle=180,end angle=360,radius=0.22] -- cycle;
                \draw[dashed, dash pattern=on 3pt off 2pt]
                (-0.22,0) arc[start angle=180,end angle=360,radius=0.22];
            \end{scope}
            \node[font=\small] at ($(A)!0.42!(R)+(0.34,0.44)$) {$\mathcal{CH}^+$};
            \path (R) ++(0,-0.5);
            \end{scope}
        \end{tikzpicture}
        \caption{Matter data supported inside the black hole}
        \label{Figure1.1}
\end{minipage}
\end{figure}
\begin{remark}\label{rmk:global}
In Theorem~\ref{thm:first.version}, the initial behavior of the matter field is taken as an assumption. Nevertheless, this characteristic initial setting naturally arises from Cauchy initial data perturbing those of the generic black hole spacetime constructed in \cite{Daluk}, with Vlasov matter supported in the (would-be) black hole interior. Theorem~\ref{thm:first.version} then applies to a class of global black hole developments; see Remark~\ref{rmk:interior}.

Moreover, we emphasize that even within the restricted local setup in Figure~\ref{Figure1.1}, as compared with Figure~\ref{Figure1.0}, Theorem~\ref{thm:first.version} already shows that the behavior of Vlasov matter differs from that of a scalar field. In particular, for a characteristic initial value problem for the Einstein--scalar field system, even if the scalar field is initially smooth up to the weak null singularity, the Ricci curvature may become singular instantaneously. We prove this in Appendix~\ref{Appendix}.
\end{remark}
\begin{remark}\label{rmk:interior}
By a domain of dependence argument, Theorem~\ref{thm:first.version} allows one to construct global black hole spacetimes with nontrivial Vlasov matter supported in the black hole interior. These spacetimes have Cauchy horizons, and the metrics are continuously extendible up to the Cauchy horizons. This result is formulated in Corollary~\ref{cor1} below; see Figure~\ref{Figure1.1}. 
Notice that the interior estimates in \cite{Daluk} are consistent with the presence of a weak null singularity at the Cauchy horizon. Although these estimates alone do not establish the singularity, they suffice to apply parts (1) and (2) of Theorem~\ref{thm:first.version}.

Moreover, if a suitable lower bound holds for the curvature along the event horizon, then the result of \cite{LukSbierski26} shows that the curvature assumption required for part (3) of Theorem~\ref{thm:first.version} is satisfied, and that the Cauchy horizon is an actual weak null singularity. For this class of black hole spacetimes, the behavior of the metric and Vlasov matter at the corresponding portion of the Cauchy horizon is exactly as in Theorem~\ref{thm:first.version}.
\end{remark}
\begin{remark}
A more ambitious program will be to study the full Kerr stability problem for the Einstein--Vlasov system. The behavior of both the metric and the matter field in the black hole interior may differ from the case studied here, depending on the global dynamics in the black hole exterior, as well as the global blue shift that affects the matter field. In this context, we refer the reader to \cite{Ruiz,Weissenbacher2024} for some recent progress on the exterior problem. In the massive case, one must additionally come to terms with the dynamical implications of stably trapped geodesics on Kerr; see \cite{Eo2026} for some related discussions.
\end{remark}
\subsection{The Einstein--Vlasov system}
Let $(\mathcal{M},\g)$ be a $4$-dimensional time-oriented Lorentzian manifold, with signature $(-,+,+,+)$, and let $\D$ denote its Levi--Civita connection. We introduce the normalized null frame $\{e_1,e_2,e_3,e_4\}$ adapted to the null foliation $(u,\ub)$, see Section~\ref{doublenullfoliation}, and define the null Ricci coefficients by
\begin{align}
\begin{split}\label{defga} \chib_{AB}&:=\g(\D_Ae_3,e_B),\qquad\quad\;\;\,\chi_{AB}:=\g(\D_Ae_4,e_B),\\
\xib_A&:=\frac12\g(\D_3e_3,e_A),\qquad\qquad\xi_A:=\frac12\g(\D_4e_4,e_A),\\
\omb&:=\frac14\g(\D_3e_3,e_4),\qquad\qquad\;\;\om:=\frac14\g(\D_4e_4,e_3),\\
\etab_A&:=\frac12\g(\D_4e_3,e_A),\qquad\quad\;\;\;\eta_A:=\frac12\g(\D_3e_4,e_A),\\
\ze_A&:=\frac12\g(\D_Ae_4,e_3). \end{split}
\end{align}
For a curvature-type tensor $U$, we define its left Hodge dual as
\begin{equation*}
(\dual U)_{\alpha\beta\gamma\delta} :=\frac12\in_{\alpha\beta}{}^{\mu\nu}U_{\mu\nu\gamma\delta}.
\end{equation*}
For a horizontal $1$--form $\theta$, we set $(\dual\theta)_A:=\slin_A{}^B\theta_B$. Writing $\W$ for the Weyl tensor, its null components are defined by
\begin{align}
\begin{split}\label{defr} \a_{AB}&:=\W(e_A,e_4,e_B,e_4),\qquad\quad \aa_{AB}:=\W(e_A,e_3,e_B,e_3),\\
\b_A&:=\frac12\W(e_A,e_4,e_3,e_4),\qquad\quad\, \bb_A:=\frac12\W(e_A,e_3,e_3,e_4),\\
\rho&:=\frac14\W(e_3,e_4,e_3,e_4),\qquad\quad\;\;\;\,\si:=\frac14(\dual\W)(e_3,e_4,e_3,e_4). \end{split}
\end{align}
In particular, $\a$ and $\aa$ are symmetric and traceless horizontal $2$--tensors.

Fix $m\geq0$. Let $f=f(x,p)\geq0$ be a differentiable distribution function on the future mass shell
\begin{equation*}
P_m:=\left\{(x,p)\in T\M: \g_{\mu\nu}(x)p^\mu p^\nu=-m^2,\quad p\text{ is future-directed and nonzero} \right\}.
\end{equation*}
We write $P_{x,m}:=P_m\cap T_x\M$ for its fiber at $x$. A solution of the Einstein--Vlasov system is described by a quadruple $(\M,\g,f,m)$ satisfying
\begin{equation}\label{EVE}
\Ric_{\mu\nu}-\frac12\R\g_{\mu\nu}=\T_{\mu\nu}, \qquad Xf=0.
\end{equation}
Here $\R$ is the scalar curvature, and $X$ generates the geodesic flow on $P_m$. Writing $p=p^\mu e_\mu$, in the momentum chart $\{p^4>0\}$ with independent coordinates $(p^1,p^2,p^4)$, we have
\begin{equation*}
X=p^\mu e_\mu -\bGa_{\mu\nu}^{i'}p^\mu p^\nu\pr_{p^{i'}}, \qquad i'\in\{1,2,4\},
\end{equation*}
where $\D_{e_\mu}e_\nu=\bGa_{\mu\nu}^{\lambda}e_\lambda$. The energy-momentum tensor is
\begin{equation}\label{defT}
\T_{\mu\nu}(x):= \int_{P_{x,m}}f(x,p)p_\mu p_\nu\,\dvol_m,
\end{equation}
where $\dvol_m$ is the natural positive mass-shell measure. Taking the trace $\Tr$ of \eqref{EVE}, we obtain
\begin{equation}\label{scalarR}
\begin{aligned} \R=-\g^{\mu\nu}\T_{\mu\nu} =-\int_{P_{x,m}}f\g^{\mu\nu}p_\mu p_\nu\,\dvol_m=m^2\int_{P_{x,m}}f\,\dvol_m=:m^2N. \end{aligned}
\end{equation}
It follows that
\begin{equation*}
\Ric_{\mu\nu}=\frac12\R\g_{\mu\nu}+\T_{\mu\nu}=\frac{m^2}{2}N\g_{\mu\nu}+\T_{\mu\nu}.
\end{equation*}
With our normalization, the Schouten tensor therefore takes the form
\begin{equation}\label{Schouten=T}
\S_{\mu\nu}=\Ric_{\mu\nu}-\frac16\g_{\mu\nu}\R =\T_{\mu\nu}+\frac{m^2}{3}N\g_{\mu\nu}=\int_{P_{x,m}} \left(p_\mu p_\nu+\frac{m^2}{3}\g_{\mu\nu}\right) f\,\dvol_m.
\end{equation}
\subsection{Statement of the main theorems}
We now state our main theorem, which provides existence and upper-bound estimates and is an analogue of Theorem~2 in \cite{Luk18}.
\begin{thm}\label{MainTheorem}
For a fixed $m\geq0$, consider compatible characteristic initial data that are smooth on every shorter segment of $H_0\cap\{0\leq\ub<\de\}$ and on $\Hb_0\cap\{0\leq u\leq\de\}$, and let $D>0$ be fixed.
\begin{itemize}
\item There exists an atlas such that, in each coordinate chart with local coordinates $(\th^1,\th^2)$, the initial metric $\ga_0$ on $S_{0,0}$ obeys
\begin{align*}
D^{-1}\leq\det\ga_0\leq D
\end{align*}
and
\begin{align*}
\sum_{i+j\leq5}\left|\pr_{\th^1}^i\pr_{\th^2}^j(\ga_0)_{BC}\right|\leq D.
\end{align*}
\item The metric on $H_0$ and $\Hb_0$ satisfies the gauge conditions
\begin{align*}
\Om=1\qquad\mbox{on }\;H_0\cup\Hb_0,\qquad\quad\bbb^A=0\qquad\mbox{on }\;\Hb_0.
\end{align*}
\item Let
\begin{equation}
Z:=\{\nab_1,\nab_2,\nab_3\}.\label{dfZ}
\end{equation}
The Ricci coefficients on the initial hypersurface $H_0$ satisfy
\begin{align*}
\sum_{i=0}^4\sup_\ub\left( \left\|h^2(\ub)Z^i\chi\right\|_{L^2(S_{0,\ub})} +\left\|Z^i\ze\right\|_{L^2(S_{0,\ub})} +\left\|Z^i\trchb\right\|_{L^2(S_{0,\ub})} \right)\leq D.
\end{align*}
\item The Ricci coefficients and curvature on the initial hypersurface $\Hb_0$ satisfy
\begin{align*}
\sum_{i=0}^4\sup_u\left( \left\|Z^i\chib\right\|_{L^2(S_{u,0})} +\left\|Z^i\ze\right\|_{L^2(S_{u,0})} +\left\|Z^i\trch\right\|_{L^2(S_{u,0})} \right) +\sum_{i=0}^3\sup_u \left\|Z^i\aa\right\|_{L^2(S_{u,0})} \leq D.
\end{align*}
\item Let
\begin{equation*}
q:=\sqrt{p^3p^4},\qquad \Dk:=\left\{ e_1,e_2,e_3, q\pr_{p^1},q\pr_{p^2},p^4\pr_{p^4} \right\}.
\end{equation*}
The initial distribution function satisfies
\begin{equation*}
\sum_{i=0}^4 \left\|\Dk^i f\right\|_{L^\infty(H_0\cup\Hb_0;L^\infty(P_{x,m}))} \leq D.
\end{equation*}
Let $f_0$ denote the restriction of $f$ to $P_m|_{H_0\cup\Hb_0}$. We assume that $f_0\geq0$ has compact momentum support and that, on $\supp f_0$,
\begin{equation*}
p^3+p^4\leq D.
\end{equation*}
\item With $h$ as in Remark \ref{rkchoicef}, the initial norms in Section \ref{initialO0} satisfy
\begin{equation*}
\mo_{(0)}+\mr_{(0)}+\TT_{(0)}\leq D.
\end{equation*}
The transverse derivatives appearing above, in particular the $\nab_3$--derivatives on $H_0$, are determined by the characteristic equations, with compatibility at $S_{0,0}$.
\end{itemize}
Then, if $\ep$ is sufficiently small and
\begin{align}\label{L2ep}
\|h(\ub)^{-1}\|_{L^2_{\ub}(0,\de)}\leq\ep,
\end{align}
there exists a unique smooth Einstein--Vlasov development $(\M,\g,f)$, endowed with a double-null foliation $(u,\ub)$ in the region $0<u<\de$ and $0<\ub<\de$ that solves the Einstein--Vlasov system \eqref{EVE} with the given data.
\end{thm}
Although the weight $h$ in the spacetime norms allows the spacetime to be singular, the spacetime metric admits a continuous extension beyond the limiting hypersurfaces $H_\de$ and $\Hb_\de$. The characteristic initial value problem is illustrated in Figure~\ref{fig:local-WNS-diamond}.

\begin{figure}[ht]
\centering
\begin{tikzpicture}[x=1cm,y=1cm,>=stealth,font=\small,
  line cap=round,line join=round]
\coordinate (O) at (0,0);
\coordinate (L) at (-2.8,2.8);
\coordinate (R) at (2.8,2.8);
\coordinate (T) at (0,5.6);
\fill[blue!4] (O)--(L)--(T)--(R)--cycle;
\draw[very thick] (L)--(O)--(R);
\draw[densely dashed] (L)--(T);
\draw[red,very thick] (R)--(T);
\draw[->] (-.55,.55)--(-1.05,1.05);
\draw[->] (.55,.55)--(1.05,1.05);
\node[rotate=-45,align=center] at (-1.8,1.05) {$\Hb_0:$ regular $f_0$};
\node[rotate=45,align=center] at (1.8,1.05) {$H_0: $ regular $f_0$};
\node[rotate=45,align=center] at (-1.8,4.6)
  {$H_\de:$ regular for $\ub<\de$};
\node[rotate=-45,red,align=center] at (1.85,4.65)
  {$\Hb_\de:$ \textbf{WNS}};
\fill (O) circle (1.5pt);
\fill (L) circle (1.5pt);
\draw[fill=white] (T) circle (2pt);
\draw[red,fill=white,thick] (R) circle (2.5pt);
\node[below] at (O) {$S_{0,0}$};
\node[left] at (L) {$(\de,0)$};
\node[above] at (T) {$(\de,\de)$};
\draw[red] (3,2.8)--(3.4,2.8)
  node[right,align=left] {singular end of $H_0$\\$(u,\ub)=(0,\de)$};
\draw[->] (-4,3.25)--(-4,4.05) node[above] {future};
\node[align=center] at (0,2.8)
  {smooth Einstein--Vlasov\\development\\[5pt]
   $0<u<\de,\quad 0<\ub<\de$};
\end{tikzpicture}
{\small
\begin{gather*}
m\geq0,\quad f_0:=f|_{P_m|_{H_0\cup\Hb_0}}\geq0,\quad \supp_p f_0\text{ compact},\quad p^3+p^4\leq D\quad\text{on }\supp f_0.
\end{gather*}
}
{\footnotesize Transverse derivatives are determined by the characteristic equations, with compatibility at $S_{0,0}$.\par}
\caption{Local development of Theorem~\ref{MainTheorem}. For the singular data of Theorem~\ref{Singularity}, the red side $\Hb_\de\cap\{0<u<\de\}$ is a weak null singularity (WNS). The metric extends continuously across it; $H_\de\cap\{\ub<\de\}$ is regular (Theorem~\ref{Luk3}).}
\label{fig:local-WNS-diamond}
\end{figure}

\begin{rk}\label{rkchoicef}
The critical weight $(\de-\ub)^{1/2}$ gives a logarithmically divergent integral for $h^{-2}$. We therefore take
\begin{equation*}
h(\ub):=(\de-\ub)^{\frac12}\log^p\left(\frac1{\de-\ub}\right),\qquad p>\frac12.
\end{equation*}
Setting $x=\de-\ub$ and $y=\log(1/x)$ gives
\begin{align*}
\int_0^\de\frac{d\ub}{h(\ub)^2} =\int_0^\de\frac{dx}{x\log^{2p}(1/x)} =\int_{-\log\de}^\infty\frac{dy}{y^{2p}} =\frac{(-\log\de)^{1-2p}}{2p-1}.
\end{align*}
Thus $\de\leq\exp(-[(2p-1)\ep^2]^{1/(1-2p)})$ implies \eqref{L2ep}. Taking $\de$ smaller if necessary, we also assume $0<h(\ub)\leq1$ for $0\leq\ub<\de$.
\end{rk}
\begin{thm}\label{Luk3}
Under the assumptions of Theorem \ref{MainTheorem}, the metric $\g$ admits a continuous Lorentzian extension up to and beyond the boundaries $\Hb_\de$ and $H_\de$ in the double-null coordinates $(u,\ub,\th^1,\th^2)$. The Vlasov distribution $f$ has a continuous trace on $\Hb_\de$ in the momentum chart $\{p^4>0\}$. For $m>0$, this chart covers the whole future mass shell. In particular, the energy-momentum tensor $\T$ and the Ricci curvature remain bounded and extend continuously to these boundaries.

The induced metric and null second fundamental form are regular in the interiors of the limiting hypersurfaces. More precisely, in any transported angular coordinate chart $U(u,\ub)$, the metric components satisfy
\begin{align*}
\sum_{i+j\leq4}\sup_{0\leq u\leq\de}\left\|\pr_{\th^1}^i\pr_{\th^2}^j(\ga,\bbb,\Om)\right\|_{L^2(U(u,\de))}\leq C.
\end{align*}
Moreover, for every fixed $u_1<\de$, the tangential derivatives of the regular Ricci coefficients satisfy
\begin{align*}
\sum_{i=0}^{3}\sup_{0\leq u\leq u_1} \left\| Z^i(\chib,\eta,\etab) \right\|_{L^2(S_{u,\de})} \leq C_{u_1},
\end{align*}
with $Z$ defined in \eqref{dfZ}. The solution extends smoothly across $H_\de\cap\{\ub<\de\}$.
\end{thm}
For initially singular data, we also prove the following lower-bound estimate and $C_{\mathrm{loc}}^{0,1}$--inextendibility result.
\begin{thm}\label{Singularity}
In addition to the assumptions of Theorem \ref{MainTheorem}, assume that
\begin{align*}
\int_0^{\de}|\hch(0,\ub)|^2\,d\ub=\infty
\end{align*}
along almost every null generator of $H_0$. Then the Christoffel symbols in the coordinate system $(u,\ub,\th^1,\th^2)$ do not belong to $L^2$ in any neighborhood of any point on $\Hb_\de$.

Assume in addition that, on an open dense subset of $S_{0,0}$, the following bounds hold locally and uniformly in an orthonormal frame continuous up to $\ub=\de$, for $\ub$ sufficiently close to $\de$:
\begin{align*}
|\a_{11}|(0,\ub)&\ges(\de-\ub)^{-1}h(\ub)^{-2},& |\a_{12}|(0,\ub)&\les(\de-\ub)^{-1}h(\ub)^{-2}.
\end{align*}
The constants and the threshold in $\ub$ may depend on the angular patch. Then the spacetime is $C_{\mathrm{loc}}^{0,1}$--inextendible across $\Hb_\de\cap\{0<u<\de\}$.
\end{thm}
\begin{rk}
Theorem \ref{Singularity} shows that, if the spacetime metric is extended continuously with respect to the differentiable structure determined by the coordinate system $(u,\ub,\th^1,\th^2)$, then the Christoffel symbols are not square-integrable in the extension. However, the argument does not exclude a continuous extension with locally square-integrable Christoffel symbols in a different differentiable structure. The arguments in \cite{Sbierski26Lipschitz} do not directly apply at this regularity, since the corresponding $C^1$ extension result for the identification map is not available.
\end{rk}
For the existence argument and the black hole application below, we recall the Cauchy theory for the Einstein--Vlasov system. See \cite{Ch71} for local existence, \cite{DafermosRendall16} for continuation results under surface symmetry, and \cite{LindbladTaylor20} for global stability in the massive case. The following theorem covers both massive and massless particles; in the massless case, we follow the maximal-development formulation of \cite[Theorem~2.2]{TaylorFLRW23}.

In the following two statements, the initial distribution is assumed to be nonnegative and to have locally uniformly compact momentum support. In the massless case $m=0$, its support is additionally assumed to be locally bounded away from the zero section of the mass shell.

\begin{thm}[Smooth Cauchy data]\label{Ch71Cauchy}
Fix $m\geq0$. Any smooth Cauchy data set $(\Sigma,\g_\Sigma,k_\Sigma,f_0)$ satisfying the Einstein constraint equations admits a smooth maximal globally hyperbolic Einstein--Vlasov development $(\M,\g,f,m)$, unique up to an isometry preserving the initial data. In this development, $\Sigma$ is a Cauchy hypersurface with induced data $(\g_\Sigma,k_\Sigma,f_0)$. If the initial data are spherically symmetric, then so is the maximal development.
\end{thm}
For characteristic data, we follow the formulation in \cite[Theorem~12.2]{Taylor}. Its vacuum counterpart is due to Rendall \cite{Rendall90}. For the Einstein--Vlasov characteristic problem, see also \cite{ChoquetBruhatChrusciel12,CP12}; Dafermos and Rendall give a related result for massive particles in spherical symmetry \cite[Proposition~2]{DafermosRendall05}.
\begin{thm}[Smooth characteristic data]\label{TaylorCharacteristic}
Fix $m\geq0$. Let smooth characteristic data for the Einstein--Vlasov system be prescribed on null hypersurfaces $H_0$ and $\Hb_0$ intersecting transversely in a spacelike sphere $S_{0,0}$, and assume the characteristic constraints and compatibility conditions at $S_{0,0}$ are satisfied. Then the data admit a nonempty smooth maximal development $(\M,\g,f,m)$ whose past boundary near $S_{0,0}$ consists of portions of $H_0$ and $\Hb_0$.
\end{thm}
We also need a characteristic local existence result for the Einstein--Vlasov system that provides a slab neighborhood of the full initial null segments. The corresponding vacuum result was proved by Luk \cite{luk} using energy estimates. For the Einstein--Vlasov system, the argument additionally requires energy estimates for the Vlasov equation and control of the momentum support.
\begin{thm}[Smooth characteristic data on full null segments]
\label{LukCharacteristic}
Under the assumptions of Theorem~\ref{TaylorCharacteristic}, suppose that the data are prescribed on compact null segments
\begin{equation*}
H_0\cap\{0\leq\ub\leq L_1\}, \qquad \Hb_0\cap\{0\leq u\leq L_2\},
\end{equation*}
where $L_1,L_2>0$. Then there exists $\nu>0$ such that the data admit a smooth Einstein--Vlasov development in the region
\begin{equation*}
\{0\leq u<\nu,\ 0\leq\ub\leq L_1\} \;\cup\; \{0\leq u\leq L_2,\ 0\leq\ub<\nu\}.
\end{equation*}
The width $\nu$ depends only on $L_1,L_2$, suitable finite-order geometric and Vlasov regularity bounds of the initial data, and the momentum support bounds.
\end{thm}
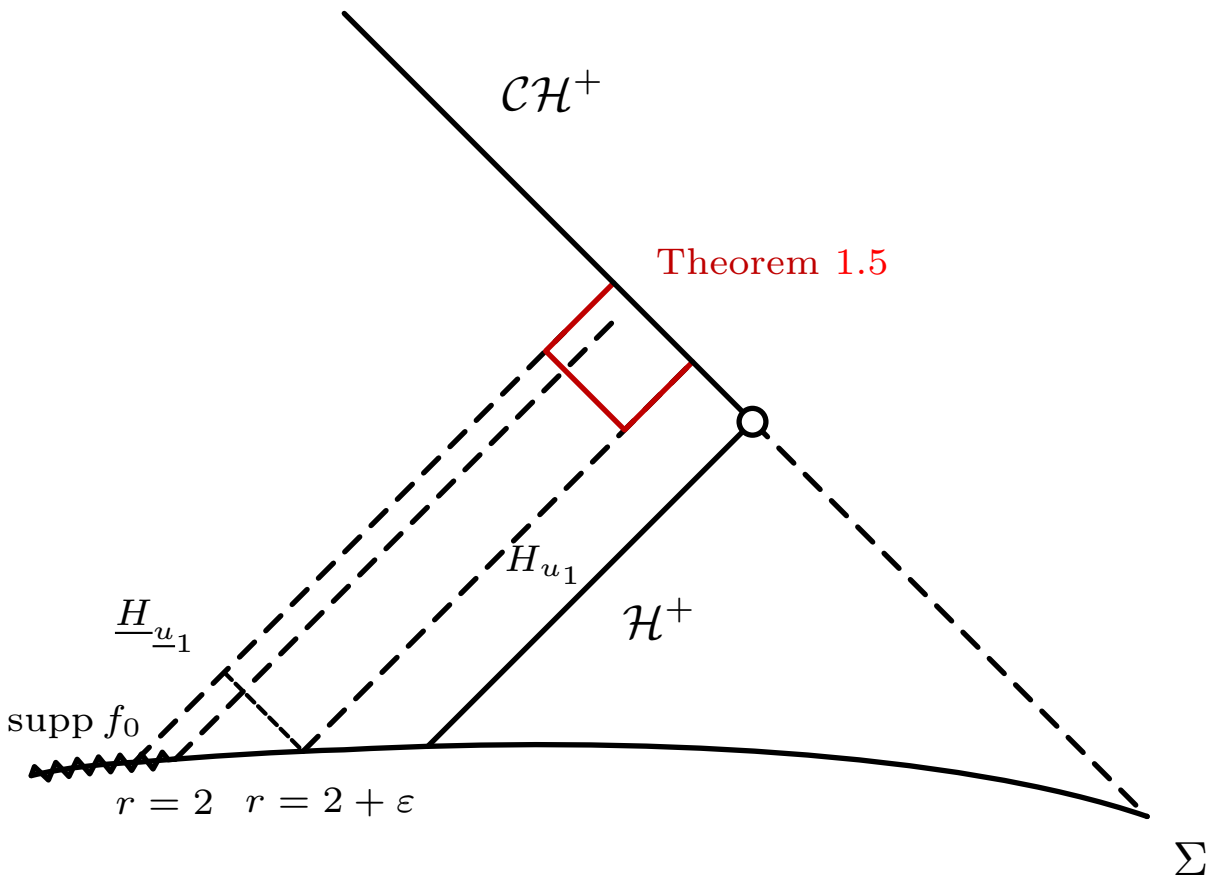
\begin{figure}[ht]
\centering
\begin{tikzpicture}[
    line cap=round,
    line join=round,
    every path/.style={line width=1.0pt}
]
\begin{scope}[scale=1.8]

    \coordinate (L) at (0,0.15);
    \coordinate (Z0) at (0.40,0.20);
    \coordinate (Z1) at (1.15,0.245);
    \coordinate (A) at (1.15,2.95);
    \coordinate (P) at (2.65,1.45);
    \coordinate (R) at (4.10,0);

    \draw[name path=bottom]
        (L) .. controls (0.12,0.18) and (0.28,0.195) .. (Z0)
            .. controls (0.62,0.22) and (0.94,0.24) .. (Z1)
            .. controls (2.20,0.30) and (3.40,0.24) .. (R);

    \coordinate (F1) at (1.96,1.78);
    \coordinate (F2) at (2.14,1.82);
    \coordinate (F3) at (2.43,1.67);
    \path[name path=fanOne]   (F1) -- (0.18,0);
    \path[name path=fanTwo]   (F2) -- (0.32,0);
    \path[name path=fanThree] (F3) -- (0.76,0);
    \path[name intersections={of=fanOne and bottom, by=B1}];
    \path[name intersections={of=fanTwo and bottom, by=B2}];
    \path[name intersections={of=fanThree and bottom, by=B3}];

    \draw[decorate, decoration={zigzag,segment length=4pt,amplitude=1.5pt}]
        (L) .. controls (0.12,0.18) and (0.28,0.195) .. (Z0) -- (B2);

    \path[name path=fanCross] (B3) -- ++(-2,2);
    \path[name intersections={of=fanCross and fanOne, by=C}];
    \draw[dashed, line width=0.8pt, dash pattern=on 2pt off 1pt] (C) -- (B3);

    \draw[dashed, dash pattern=on 4pt off 3pt] (B1) -- (F1);
    \draw[dashed, dash pattern=on 4pt off 3pt] (B2) -- (F2);
    \draw[dashed, dash pattern=on 4pt off 3pt] (B3) -- (F3);

    \coordinate (Qtop) at (2.14,1.96);
    \coordinate (Qbottom) at (2.43,1.67);
    \draw[red!75!black]
        (Qtop) -- ($(Qtop)+(-0.25,-0.25)$)
        -- ($(Qbottom)+(-0.25,-0.25)$) -- (Qbottom);
    \node[font=\scriptsize, text=red!75!black, anchor=west]
        at ($(Qtop)+(0.08,0.08)$) {Theorem \ref{MainTheorem}};
    \draw (A) -- (P);
    \draw[dashed, dash pattern=on 4pt off 3pt] (P) -- (R);
    \path[name path=rising] (P) -- (1.20,0);
    \path[name intersections={of=rising and bottom, by=B}];
    \draw (B) -- (P);
    \draw[fill=white] (P) circle[radius=1.4pt];
    \node[font=\small] at ($(A)!0.28!(P)+(0.35,0.14)$) {$\mathcal{CH}^+$};
    \node[font=\small] at ($(B)!0.48!(P)+(0.28,-0.10)$) {$\mathcal{H}^+$};
    \node[font=\scriptsize, anchor=south east] at ($(C)+(-0.02,0)$)
        {$\Hb_{\ub_1}$};
    \node[font=\scriptsize, inner sep=0pt] at ($(B3)!0.55!(F3)+(0.1,-0.1)$)
        {$H_{u_1}$};
    \node[font=\scriptsize, anchor=north] at ($(B2)+(-0.04,-0.04)$) {$r=2$};
    \node[font=\scriptsize, anchor=north] at ($(B3)+(0.1,-0.06)$) {$r=2+\ep$};
    \node[font=\scriptsize, anchor=south] at ($(L)!0.55!(B2)+(-0.13,0.03)$){$\supp f_0$};
    \node[font=\small, anchor=north west] at ($(R)+(0.02,-0.02)$) {$\Sigma$};
\end{scope}
\end{tikzpicture}
\caption{Construction of a global solution with restricted data}
\label{Figure1.3}
\end{figure}
We record an immediate consequence of
Theorem~\ref{MainTheorem} concerning the stability
of the Cauchy horizon for perturbed Kerr spacetimes
constructed in \cite{Daluk} under small, spatially
restricted Vlasov perturbations.
\begin{cor}\label{cor1}
Let $\Sigma\simeq\mathbb{S}^2\times(0,+\infty)$ be part of a two-ended admissible spacelike hypersurface in a perturbed Kerr spacetime, as constructed in \cite{Daluk}. We prescribe smooth Cauchy initial data $(g,k,f_0)$ for the Einstein--Vlasov system \eqref{EVE} on $\Sigma$ which agree exactly with the corresponding data in \cite{Daluk} on $\mathbb{S}^2\times(2,+\infty)\subset\Sigma$. Within the trapped region, we prescribe the smooth compactly supported Vlasov distribution function $f_0$ defined on $T_{\mathbb{S}^2\times [1,2)}M$, see Figure~\ref{Figure1.1}. We also assume that the initial distribution has locally uniformly compact momentum support. In the massless case, its support is locally bounded away from the zero section.

For $1\leq u_*\leq2$, denote by $H_{u_*}$ the outgoing null hypersurface emanating from $\mathbb{S}^2\times\{u_*\}\subset\Sigma$ and by $H_{u_*,C}=H_{u_*}\cap\{\ub\leq C\},\ C<+\infty$ the compact regular segment.
Then the initial data define a Cauchy initial value
problem for the Einstein--Vlasov system, which admits
a development containing a slab neighborhood of
$H_{2,C}$ for every $C<+\infty$. Moreover, the solution extends up to $\mathcal{CH}^+$, and the metric admits a continuous Lorentzian extension beyond $\mathcal{CH}^+$. In particular, the development covers the future domain of dependence of $\mathbb{S}^2\times[2-\ep,+\infty)\subset\Sigma$ for some $\ep>0$. See Figure~\ref{Figure1.3}.
\end{cor}
\begin{proof}
Solving constraint equations, we can construct a Cauchy initial value problem $(g,k,f_0)$ on $\mathbb{S}^2\times[2-\ep,+\infty)\subset\Sigma$ for the Einstein--Vlasov system. We first apply the Cauchy local existence theorem, Theorem~\ref{Ch71Cauchy}. This gives rise to a smooth development of the prescribed data in a future neighborhood of $\Sigma$. Choose a sphere $S=\mathbb{S}^2\times\{2+\ep\}\subset\Sigma$ on the side where the initial Vlasov distribution vanishes. Denote by $H_{u_1}$ and $\Hb_{\ub_1}$ the outgoing and incoming null hypersurfaces emanating from $S$, respectively. Their initial portions are contained in the Cauchy development. In particular, restricting the solution to a sufficiently short portion of $\Hb_{\ub_1}$ gives smooth characteristic data on this hypersurface.

By the support assumption, the initial Vlasov distribution vanishes outside $\mathbb{S}^2\times[1,2]\subset\Sigma$. By finite speed of propagation and uniqueness, the solution in the future domain of dependence of $\mathbb{S}^2\times(2,+\infty)\subset\Sigma$ agrees with the corresponding vacuum solution in \cite{Daluk}, in particular on $H_{u_1}$.

We now regard the continuation problem as a characteristic initial value problem with data on $H_{u_1}\cup \Hb_{\ub_1}$ and bifurcation sphere $ S=H_{u_1}\cap \Hb_{\ub_1}. $ Next, fix any $C<+\infty$, and consider a compact regular segment $H_{u_1,C}=H_{u_1}\cap\{\ub\le C\}\subset H_{u_1}$ away from the Cauchy horizon. We apply Theorem~\ref{LukCharacteristic} along $H_{u_1,C}$. Hence, $H_{u_1,C}$ has a smooth slab neighborhood contained in the development. The width may depend on the choice of $C$ and the corresponding initial norms, and thus the classical argument itself does not extend to $\mathcal{CH}^+$.

Thus, it remains to consider the extension up to the Cauchy horizon $\mathcal{CH}^+$. We consider the characteristic region represented by the red box in Figure~\ref{Figure1.3}. Recall that the data on $H_{u_1}$ are exactly the corresponding data in \cite{Daluk}, while the data on $\Hb_{\ub_1}$ are regular. After restricting the characteristic region if necessary, the induced data on its initial null boundaries satisfy the hypotheses of Theorem~\ref{MainTheorem}. Applying our main theorem gives a solution throughout the interior of this region up to $\mathcal{CH}^+$. Finally, Theorem~\ref{Luk3} gives a continuous extension of the metric to and beyond $\mathcal{CH}^+$.
We repeat this argument, obtaining at each step
a slab whose width has a positive lower bound
depending on the size of the initial data.
\end{proof}
\begin{remark}
Theorem~\ref{MainTheorem} is proved for a characteristic rectangle $\{0\leq u,\ub\leq\de\}$ with both lengths small. Our arguments extend to characteristic data prescribed on $H_0\cap\{0\leq\ub\leq\de\}$ and $\Hb_0\cap\{0\leq u\leq1\}$ with some modifications; we omit the details for simplicity. If, in addition, $\|f_0\|_{H^s}\leq\de$ for sufficiently large $s$ and sufficiently small $\de>0$, the resulting spacetime is a sufficiently small perturbation of the reference spacetime constructed in \cite{Daluk}. The adaptation of \cite{luk} to the Einstein--Vlasov system then provides slab neighborhoods of $H_{u_*,C}$ whose widths have a positive lower bound uniform in $u_*\in[1,2]$, for each fixed $C<+\infty$. Combining this continuation argument with the extended version of Theorem~\ref{MainTheorem} near the Cauchy horizon yields a development as in Corollary~\ref{cor1} covering the future domain of dependence of $\mathbb{S}^2\times[1,+\infty)\subset\Sigma$. The resulting spacetime is a small perturbation of the reference spacetime in \cite{Daluk}, and its Cauchy horizon is expected to be a weak null singularity.
\end{remark}
\subsection{Outline of the proof}
We use the notation of Section~\ref{secpreliminaries}. Dots denote terms omitted in this overview. The index ranges are $A,B,C,D\in\{1,2\}$, $i,j,k\in\{1,2,3\}$, $i',j',k'\in\{1,2,4\}$, and $\mu,\nu,\lambda\in\{1,2,3,4\}$.
\subsubsection{Geometric renormalization and closure}
We allow $\chi$ to be unbounded at the initial outgoing null boundary. On $H_0$,
\begin{equation*}
|\chi|+|\om|\les h(\ub)^{-2},\qquad h(\ub)\longrightarrow0,\qquad \int_0^\de h(\ub)^{-2}\,d\ub\leq\ep^2.
\end{equation*}
Thus $\chi,\om\in L^1_\ub$ on $H_0$, while the singular data of Theorem~\ref{Singularity} satisfy $\hch\notin L^2_\ub$ along almost every generator. By null structure equations  \eqref{basicnull},
\begin{align*}
\a&=-\nab_4\hch-(\trch+2\om)\hch,\\
|\hch(0,\ub)| &\les|\hch(0,0)|+\int_0^\ub|\a(0,\ub')|\,d\ub'.
\end{align*}
The inequality follows by an integrating factor, since $\trch+2\om\in L^1_\ub$. Hence $\hch\notin L^2_\ub$ implies $\a\notin L^1_\ub$ on these generators. We therefore seek a priori estimates for existence that do not require control of $\a$.

Following \cite{Luk18}, we use
\begin{equation*}
\K=-\rho+\frac12\hch\c\hchb+\cdots,\qquad \mu:=\K-\sdiv\eta.
\end{equation*}
The null equations give
\begin{align*}
\nab_4\K &=-\nab_4\rho+\frac12\hchb\c\nab_4\hch+\cdots =-\sdiv\b+\frac12\hchb\c\a-\frac12\hchb\c\a+\cdots,\\
\nab_4\mu &=-\sdiv(\nab_4\eta)+\nab_4\K+\cdots =\sdiv\b-\sdiv\b+\cdots.
\end{align*}
Thus $\a$ cancels from $\nab_4\K$ and $\sdiv\b$ from $\nab_4\mu$. The equations for $\sic,\vkp,\omd$ are given in Propositions~\ref{renorbianchi} and~\ref{nullrenor}. Weighted energy estimates for the renormalized Bianchi pairs, combined with transport and Hodge estimates, improve the geometry under the matter bootstrap.

We control up to three $\nab_1,\nab_2,\nab_3$ derivatives of $(\b,\K,\sic,\bb,\aa)$ in weighted null $L^2$ fluxes. The norms of Sections~\ref{secOnorms}--\ref{secRnorms} control derivatives of the Ricci coefficients and metric components up to order three, together with fourth derivatives containing at least one angular derivative. No fourth pure incoming derivative of the Ricci coefficients or metric components is required in these spacetime estimates. Proposition~\ref{estGamma} controls three derivatives of the frame connection.

The matter terms in the Bianchi equations involve first derivatives of $\T$. Commuting these equations three times therefore requires four derivatives of the relevant components of $\T$. We estimate four derivatives of $f$ in momentum-weighted null $L^2$ fluxes by applying one adapted field followed by three further derivatives. We also estimate the first adapted derivatives of $f$ pointwise. These estimates use the preceding geometric bounds without requiring a bound for $\a$.
\subsubsection{Commutators and regularity}\label{SubsecVlasovField}
We perform the commutator calculations in the $p^4$--chart $(p^1,p^2,p^4)$ on the region $p^3,p^4>0$. The mass-shell relation and Vlasov equation are
\begin{equation*}
4p^3p^4=|p^A|^2+m^2,\qquad q:=\sqrt{p^3p^4},\qquad Xf=0,\qquad X=p^\mu e_\mu-\bGa_{\mu\nu}^{i'}p^\mu p^\nu\pr_{p^{i'}}.
\end{equation*}
In \cite{Taylor}, estimates of Jacobi fields capture the geometry of spacetime. As in \cite{AMYEV}, we consider vector field commutators that need not be parallel transported along geodesics. Since $e_A$ differentiates at fixed frame momentum,
\begin{equation*}
[X,e_A]f
=p^\mu p^D e_A(\bGa_{\mu D}^C)\pr_{p^C}f+\cdots.
\end{equation*}
Three further commutations would require fourth derivatives of the frame connection; hence, fifth metric derivatives are unavailable. The horizontal lifts
$E_\mu:=e_\mu-p^\nu\bGa_{\mu\nu}^{i'}\pr_{p^{i'}}$
avoid this derivative loss by replacing connection
derivatives with curvature:
\begin{equation*}
[X,E_A]f=(p^4)^2\left(\a_A{}^B+\frac12\de_A{}^B\T_{44}\right)\pr_{p^B}f+\cdots.
\end{equation*}
In our setting, however, these commutators involve the most singular curvature component $\a$, for which no integrable bound is available. This prevents us from closing a direct Gr\"onwall estimate for the corresponding commuted derivatives of $f$.

To overcome this difficulty, we instead use $E_3$ and $H_A:=e_A-p^D\bGa_{AD}^C\pr_{p^C}$. Then
\begin{align*}
[X,H_A]f
&=p^\mu p^D
\bigl[e_A(\bGa_{\mu D}^C)-e_\mu(\bGa_{AD}^C)\bigr]
\pr_{p^C}f+\cdots\\
&=\Bigl[p^\mu p^D\R^C{}_{DA\mu}+p^Dp^3e_A(\chib_{DC})+p^Dp^4e_A(\chi_{DC})\Bigr]\pr_{p^C}f+\cdots.
\end{align*}
The curvature term $\R^C{}_{DA\mu}$ has three angular indices $C,D,A$ and thus cannot contain $\a$. Likewise, the curvature terms $\R^\nu{}_{\mu 3\lambda}$ in $[X,E_3]$ contain a $3$--index and hence exclude $\a$. In contrast, the full lift $E_A$ contains $-p^4\chi_A{}^C\pr_{p^C}$, which produces $-(p^4)^2e_4(\chi_A{}^C)\pr_{p^C}$ in $[X,E_A]$. Thus, with the same weight $q$, the singular $\a$ term in $[X,qE_A]f$ is replaced in $[X,qH_A]f$ by curvature and angular derivatives of the Ricci coefficients.

We use the families
\begin{align*}
\VV_\pa:=\{qH_1,qH_2,p^3E_3\},\qquad\VV_\perp:=\{q\pr_{p^1},q\pr_{p^2},p^4\pr_{p^4}\}, \qquad \VV:=\VV_\pa\cup\VV_\perp .
\end{align*}
We write $\VV f$ for the collection of $Vf$ with $V\in\VV$.

We now explain why we commute with $\VV$ first. 
The first commutation gives the system we estimate, written schematically as
\begin{align*}
X(\VV_\perp f)&=\VV_\pa f+\CC_1\c\VV_\perp f,\\
X(\VV_\pa f)&=\CC_1\c\VV_\pa f+\CC_2\c\VV_\perp f .
\end{align*}
In the first row, $\VV_\pa f$ denotes a sum with bounded momentum coefficients (see Corollary~\ref{corVf} for $\CC_1,\CC_2$). The fourth derivatives of $\T$ involve fourth derivatives of $f$ in the momentum integrals. We take three further derivatives of $\VV f$ using $\Dk=\{e_1,e_2,e_3\}\cup\VV_\perp$.

For the $e_1^4f$ term in $e_1^4\T_{44}$, compare
\begin{align*}
X(qe_1^4f) &=q p^\mu p^D e_1^4(\bGa_{\mu D}^C)\pr_{p^C}f+\cdots,\\
X\bigl(e_1^3(qH_1f)\bigr) &=e_1^3\bigl([X,qH_1]f\bigr) +\sum_{j=0}^2e_1^j[X,e_1]e_1^{2-j}(qH_1f).
\end{align*}
The fourth connection derivative $e_1^4(\bGa_{\mu D}^C)$ comes from $q e_1^3([X,e_1]f)$. It also appears in $(qH_1)e_1^2([X,e_1]f)$ when we expand $X((qH_1)e_1^3f)$ directly. In $e_1^3([X,qH_1]f)$, three derivatives on the coefficients give $e_1^3\R^C{}_{D1\mu}$ and $e_1^4\chi$. The Gauss equation~\eqref{gauss} expresses $\rho$ in terms of $\K$, Ricci coefficients, and Schouten terms. The Schouten terms are controlled by the matter bootstrap. In the sum, $[X,e_1]$ differentiates the connection once, and at most two outer $e_1$ derivatives hit that coefficient.

For either angular index $A$, the commutator $[X,e_1^3](qH_Af)$ involves at most three derivatives of the frame connection, while $e_1^3([X,qH_A]f)$ requires three curvature derivatives and four Ricci coefficient derivatives, with at least one angular derivative. These are controlled by the geometric estimates; see Lemmas~\ref{lem:XE_commutator}--\ref{XpHcommutator} and Corollary~\ref{corXDcomm}.
\subsubsection{Momentum weights and characteristic estimates}
The main difficulty in commuting the Vlasov equation is to avoid singular coefficients in the massless case, where $q$ need not be bounded away from zero. To overcome this difficulty, we use the weighted commutators $q\pr_{p^A}$, $p^4\pr_{p^4}$,
$qH_A$, and $p^3E_3$. To see why we use these weighted commutators, we
compare the unweighted and weighted commutators in the $p^4$--chart:
\begin{align*}
[X,\pr_{p^A}]
&=-e_A-\frac{p^A}{2p^4}e_3+\cdots,&
[X,q\pr_{p^A}]
&=-qe_A-\frac{p^A}{2q}\,p^3e_3+\cdots,\\
[X,\pr_{p^4}]
&=-e_4+\frac{p^3}{p^4}e_3+\cdots,&
[X,p^4\pr_{p^4}]
&=-X+2p^3e_3+\frac{p^A}{q}\,qe_A+\cdots .
\end{align*}
The dots denote momentum derivative terms. The last bracket
contains $p^3e_3-p^4e_4$ before substituting
$p^4e_4=X-p^Ae_A-p^3e_3+
\bGa_{\mu\nu}^{i'}p^\mu p^\nu\pr_{p^{i'}}$.
Since $Xf=0$ and $|p^A|\leq2q$, the weights $q$ and
$p^4$ remove the potentially singular coefficients,
yielding $qe_Af$ and $p^3e_3f$ with bounded coefficients,
even when $m=0$ and $q$ has no positive lower bound.
Replacing these derivatives by $qH_Af$ and $p^3E_3f$
only changes the momentum derivative terms.

To illustrate the singularity of the unweighted
commutators, take $m=0$, $p^3=a>0$, and
$(p^1,p^2)=(r,0)$ with $r>0$. Then
$p^4=r^2/(4a)$ and
\begin{equation*}
\frac{p^1}{2p^4}=\frac{2a}{r},\qquad
\frac{p^3}{p^4}=\frac{4a^2}{r^2},\qquad
\frac{p^1}{2q}=1.
\end{equation*}
To integrate the $\aa$ and $e_1\chi$ terms along
characteristics, we further normalize $qH_Af$ and
$p^3E_3f$ by $p^3+q$, using
\begin{equation*}
\frac{(p^3)^2}{p^3+q}\leq p^3,\qquad \frac{q^2}{p^3+q}\leq q,\qquad \frac{p^4q}{p^3+q}\leq p^4.
\end{equation*}
In particular, the $\aa$--term gives
\begin{equation*}
X\left(\frac{p^3E_3f}{p^3+q}\right) =-\frac{(p^3)^2}{p^3+q}\frac{p^D}{q} \aa_{CD}(q\pr_{p^C}f)+\cdots.
\end{equation*}
The $\aa_{CD}$ and $e_1(\chi_{DC})$ terms in the normalized equations multiply $q\pr_{p^C}f$ with momentum factors bounded by
\begin{equation*}
\frac{(p^3)^2}{p^3+q}\frac{|p^D|}{q}\leq 2p^3,\qquad \frac{p^4|p^D|}{p^3+q}\leq 2p^4 .
\end{equation*}
For $|p^D|\simeq q$, using $q$ alone in the first denominator gives a ratio of size $p^3/q$ to the bound $p^3$. Using $p^3$ alone in the second gives a ratio of size $q/p^3$ to the bound $p^4$. For $m=0$, these diverge as $p^4\to0$ with $p^3$ fixed and as $p^3\to0$ with $p^4$ fixed, respectively. In the angular frame, $|e_1(\chi_{DC})|\les|\nab\chi|+|\chi|$.

Along an affinely parametrized characteristic ending at $s=0$, we have $p^3ds=\Om^2du$, $p^4ds=d\ub$, and $\Om\simeq1$. Hence
\begin{align*}
\int\frac{(p^3)^2}{p^3+q}|\aa|\,ds &\leq\int p^3|\aa|\,ds \les\de^{1/2} \|\aa\|_{L^2_uL^\infty_\ub L^\infty(S)} \les\de^{1/2},\\
\int p^4(|\chi|+|\nab\chi|)\,ds &\les \|h^{-1}\|_{L^2_\ub} \|h(\chi,\nab\chi)\|_{L^2_\ub L^\infty_uL^\infty(S)} \les\ep,\\
\int q\,ds &\leq \left(\int p^3ds\int p^4ds\right)^{1/2} \les\de.
\end{align*}
Here integration runs from the initial hypersurfaces to $s=0$; the norms follow from the improved geometry and sphere Sobolev.

The same characteristic estimates control the integrals of $|X\log p^3|$ and $|X\log p^4|$. For example,
\begin{equation*}
X(\log p^3) =-\frac{\chi_{AB}p^Ap^B}{2p^3}+\cdots, \qquad\qquad \frac{|p^A|^2}{p^3}\leq4p^4.
\end{equation*}
Proposition~\ref{pmuest} gives $p^3(s)\simeq p^3(0)$ and $p^4(s)\simeq p^4(0)$, preserving bounded momentum support. These estimates also give
\begin{equation*}
\int\left( |X\log q|+|X\log(p^3+q)| \right)ds\les\ep,\qquad \int(p^3+q)\,ds\les\de.
\end{equation*}
The first bound controls derivatives of the weights. The second controls the coefficient $p^3+q$ multiplying $(p^3+q)^{-1}\VV_\pa f$ in the equation for $\VV_\perp f$. Integrating the coupled commuted equations gives
\begin{equation*}
|q\pr_{p^A}f|+|p^4\pr_{p^4}f| +\frac{|qH_Af|+|p^3E_3f|}{p^3+q}\les1;
\end{equation*}
see Proposition~\ref{Lowerorderf}. No uniform positive lower bound for $q$ is used.

\subsubsection{Top order estimates and velocity moments}

The third curvature derivatives are controlled in null
$L^2$ fluxes, so we estimate the fourth derivatives of
$f$ in flux norms. The mass-shell measure and the
momentum integral used in these estimates satisfy
\begin{equation*}
\dvol=\frac{dp^1dp^2dp^4}{p^4},\qquad
\int_{\supp f}\frac{p^3+p^4}{q}\dvol\les1.
\end{equation*}
Indeed, writing $r=|p^A|$ and using polar coordinates
in $(p^1,p^2)$, the density $(p^4/q)\dvol$ becomes
$(r/q)\,dr\,d\vartheta\,dp^4$, where
$r/q=2r/\sqrt{r^2+m^2}\leq2$.
Bounded momentum support gives the bound for the
integral of $p^4/q$. The same calculation in the
$p^3$--chart bounds the integral of $p^3/q$.
For $m=0$, momentum cutoffs near the axes are
removed in Proposition~\ref{fluxf}.

For $F:=(p^3+q)^{-1}e_1^3(qH_Af)$, the
$e_A\chi$--term gives
\begin{equation*}
XF
=\frac{p^4p^D}{p^3+q}
e_1^3e_A(\chi_{DC})(q\pr_{p^C}f)+\cdots,
\qquad
\frac{|p^D|}{p^3+q}\leq2.
\end{equation*}
We apply Lemma~\ref{VlasovL2} with weight $q^{-1}$
and use the bounds for $X\log q$ and
$X\log(p^3+q)$ when differentiating the weights.
This gives an energy inequality for the fluxes
\begin{equation*}
\int_{H_u}\int_{P_m}\frac{p^3}{q}|F|^2\dvol,
\qquad
\int_{\Hb_\ub}\int_{P_m}\frac{p^4}{q}|F|^2\dvol.
\end{equation*}
Let $V$ be the spacetime rectangle bounded by these hypersurfaces and the initial hypersurfaces, and let $B$ be the bootstrap constant. The fluxes sum to at most $B^2$. Since $|q\pr_{p^C}f|\les1$, the moment bound and Cauchy--Schwarz give
\begin{align*}
\int_V\int_{P_m}
\frac{p^4}{q}|F|\,|e_1^3e_A\chi|\dvol\les
\de^{1/2}
\|h e_1^3e_A\chi\|_{L^\infty_uL^2_\ub L^2(S)}
\left(
\int_V\int_{P_m}
\frac{p^4h^{-2}}{q}|F|^2\dvol
\right)^{1/2}
\les\ep\de^{1/2}B.
\end{align*}
Here the $h^{-2}$--weighted integral is controlled
by the incoming flux bootstrap and
$\|h^{-1}\|_{L^2_\ub}\les\ep$.
The $\aa$--terms use the $p^3/q$--flux.
For the quadratic Ricci coefficient terms we use
$\|(\Ga_{-1}\c\Ga_1)^{(3)}\|_{L^2(S)}
\les\|\Ga_1^{(3)}\|_{L^2(S)}$.
When derivatives fall on both a coefficient and
a derivative of $f$, we estimate the product by
H\"older and Sobolev inequalities on each sphere.
For the Schouten terms we use the matter bound
$\TT_\S\leq B$.
Taking $B$ large and then $\de,\ep$ sufficiently
small as in \eqref{epep0}, Proposition~\ref{fluxf}
improves the fourth-order flux bootstrap bound
from $B^2$ to $1+(\de^{1/2}+\ep)B^2\les1$. We next recover the required fourth derivatives of the stress-energy tensor from these fluxes.
Consider $e_1^4\T_{44}$. Since $p_4=-2p^3$,
and $e_1$ differentiates at fixed frame momentum
with $\dvol$ independent of $x$, we have
\begin{equation*}
e_1^4\T_{44}
=4\int_{P_m}(p^3)^2e_1^4f\,\dvol.
\end{equation*}
Taking $A=1$ in the definition of $F$, we use
\begin{equation*}
e_1^3e_Af
=\frac{p^3+q}{q}F
+e_1^3\left(
p^D\bGa_{AD}^C\pr_{p^C}f
\right).
\end{equation*}
Since $|p^D|\leq2q$, the correction is controlled
by up to three derivatives of $\bGa_{AD}^C$
and the fluxes for weighted momentum derivatives
of $f$. For the $F$--term,
\begin{equation*}
\int_{P_m}(p^3)^2\frac{p^3+q}{q}|F|\dvol\les
\left(
\int_{P_m}\frac{p^3}{q}|F|^2\dvol
\right)^{1/2}
\left(
\int_{\supp f}
\frac{(p^3)^3(p^3+q)^2}{q}\dvol
\right)^{1/2}.
\end{equation*}
The second integral is bounded by a constant times
$\int_{\supp f}(p^3/q)\dvol$, since
$(p^3)^2(p^3+q)^2$ is bounded on $\supp f$.

To explain the choice of one weighted angular
commutation and the $q^{-1}$ flux weight,
consider the massless case with $p^3$ in a
compact subinterval of $(0,\infty)$.
In the $p^3$--chart, writing $r=|p^A|$,
we have $q=r/2$ and
$\dvol=(r/p^3)\,dr\,d\vartheta\,dp^3$.
The Cauchy--Schwarz dual densities near $r=0$ are
\begin{align*}
&\frac{\bigl((p^3)^2(p^3+q)/q\bigr)^2}{p^3/q}\dvol
\simeq dr\,d\vartheta\,dp^3,
\\
&\frac{\bigl((p^3)^2(p^3+q)/q\bigr)^2}{p^3}\dvol
\simeq \frac{dr}{r}\,d\vartheta\,dp^3,
\\
&\frac{\bigl((p^3)^2(p^3+q)/q^2\bigr)^2}{p^3/q}\dvol
\simeq \frac{dr}{r^2}\,d\vartheta\,dp^3.
\end{align*}
The three cases correspond, respectively, to one
$qH_1$ commutation with the $p^3/q$ flux,
the same commutation with the $p^3$ flux,
and two $qH_1$ commutations with the $p^3/q$ flux.
Only the first dual density is integrable near $r=0$.

The $p^4/q$--flux controls the corresponding
derivatives of
$\T_{33}=4\int(p^4)^2f\,\dvol$.
To recover $e_3f$ from the normalized
$p^3E_3f$ term, use
\begin{equation*}
\frac{q(p^3+q)}{p^3}=p^4+q.
\end{equation*}
Matter conservation gives
\begin{equation*}
\D_4\T_{3A}
=-\D_3\T_{4A}+2\D^B\T_{BA}.
\end{equation*}
In $J_{3A4}$, the derivative $\D_4\S_{3A}$
equals $\D_4\T_{3A}$, so this identity replaces
it by derivatives in the $e_3$ and angular directions.
The fourth derivative estimates for $\T$ then
give weighted $L^2$ bounds for three derivatives
of the Cotton tensor components in the Bianchi
equations. In particular, the antisymmetric
structure of $\dual J_{343}$ cancels the singular
$\Ga_1$ connection terms, giving the corresponding
estimate without an $h$--weight; see
Propositions~\ref{estTtop} and~\ref{Jest}.

The resulting matter bounds improve the assumptions
used for the geometry, closing the bootstrap in
the order $\mr$, $\mo$, and $\TT$; see
Theorems~\ref{M1}--\ref{M3}.

\subsubsection{Boundary regularity and inextendibility}
The $L^1_\ub$ bound for $\Ga_1$ gives a continuous extension of $\g$. The same integrability and $|p^4e_4f|\les1+p^4|\Ga_1|$ give a locally uniform trace of $f$ on $\{p^4>0\}$, without a uniform positive lower bound for $p^4$ on the momentum support. The $\nab_3\chi$ equations propagate the initial failure of square integrability, giving non-$L^2_{\mathrm{loc}}$ Christoffel symbols; see Lemmas~\ref{hchinfty}--\ref{L2infty}.

For Lipschitz inextendibility, the $\nab_3\a$ equation and $|\Jh_4|\les1+|\Ga_1|$ give, in a $\nab_3$--parallel frame,
\begin{equation*}
\lambda\a=\a|_{H_0}+\mathcal E,\qquad \lambda\simeq1,\qquad \int_{V_{\ub'}}|\mathcal E|\,d\mathrm{Vol}_{\g}=o\bigl(h(\ub')^{-2}\bigr).
\end{equation*}
Here $\lambda>0$ is an integrating factor, and $V_{\ub'}$ is defined in Section~\ref{ssecconclusion}. The assumptions on $\a$ give a divergent signed integral of $\R(X_1,X_2,X_3,X_4)$ for continuous $(X_1,X_2,X_3,X_4)$ sufficiently close to $(e_4,e_1,e_4,e_1)$. Together with the geometric bounds, this gives inextendibility by \cite[Proposition 3.6, Theorem 3.14]{Sbierski26Lipschitz}.
\subsection{Structure of the paper}
Section \ref{secpreliminaries} fixes the geometric and mass-shell conventions. Section \ref{secmain} states the norms, intermediate estimates, and bootstrap closure, and Section \ref{secboot} records their first consequences. Sections \ref{secR}, \ref{secO}, and \ref{secVlasov} prove the curvature, connection, and Vlasov estimates, respectively. Appendix~\ref{Appendix} establishes a contrasting instantaneous blow-up result for the Einstein--real scalar field system.
\subsection*{Acknowledgements and AI statement}
\noindent\textit{Acknowledgements.} The authors thank Jonathan Luk for helpful discussions and comments, as well as his invaluable guidance and support. Y.S. is supported by NSF DMS-2304445. J.W. is supported by ERC-2023 AdG 101141855 BlaHSt.

\medskip
\noindent\textit{Disclosure of AI use.} The authors used ChatGPT to assist with checking lengthy calculations and improving the language and presentation of the manuscript. The authors reviewed the resulting suggestions and take full responsibility for the content of the manuscript.
\section{Geometric setup}\label{secpreliminaries}
\subsection{Double-null foliation}\label{doublenullfoliation} 
Let $(\M,\g)$ be a spacetime of signature $(-,+,+,+)$ with Levi--Civita connection $\D$. 
Let $u,\ub$ be smooth future-increasing optical functions with nonvanishing transversal gradients:
\begin{equation*}
\g(\grad u,\grad u)=\g(\grad\ub,\grad\ub)=0.
\end{equation*}
Their level sets are the outgoing and incoming null hypersurfaces $H_u$ and $\Hb_\ub$, respectively, and
$S_{u,\ub}:=H_u\cap\Hb_\ub$ are smooth spacelike $2$--spheres. Set
\[
L:=-2\grad u,\qquad \Lb:=-2\grad\ub,
\]
and define the positive lapse $\Om$ by
\begin{equation}\label{deflapse}
\g(L,\Lb)=-2\Om^{-2}.
\end{equation}
The normalized null pair and transport fields are
\begin{equation}\label{dfLLb}
e_3:=\Lb,\qquad e_4:=\Om^2L,\qquad \Nb:=\Om^2e_3,\qquad N:=e_4,
\end{equation}
so that
\[
\g(e_3,e_3)=\g(e_4,e_4)=0,\qquad \g(e_3,e_4)=-2.
\]
Choose a local positively oriented orthonormal frame $(e_1,e_2)$ for the induced metric $\ga$ on $S_{u,\ub}$. Then $(e_1,e_2,e_3,e_4)$ is a null frame, with
\[
\g(e_A,e_B)=\de_{AB},\qquad
\g(e_A,e_3)=\g(e_A,e_4)=0.
\]
We write $\D_\mu:=\D_{e_\mu}$, normalize the area form by $\slin_{12}=1$, and choose the spacetime orientation so that $\in_{1234}=2$. Upper-case Latin indices $A,B,C,...$ range over $\{1,2\}$, lower-case Latin indices $i,j,k,...$ over $\{1,2,3\}$, primed lower-case Latin indices $i',j',k',...$ over $\{1,2,4\}$, and Greek indices $\mu,\nu,\lambda,...$ over $\{1,2,3,4\}$. Horizontal indices are raised and lowered with $\de_{AB}$; the only nonzero inverse-metric components are
\[
\g^{AB}=\de^{AB},\qquad \g^{34}=\g^{43}=-\frac12.
\]
Transport angular coordinates $\phi^A$ along $e_4$, so that $e_4(\phi^A)=0$. Writing $\slg$ for the induced metric $\ga$, the spacetime metric takes the form
\begin{equation}\label{metricg}
\g=-2\Om^2(d\ub\otimes du+du\otimes d\ub)
+\slg_{AB}(d\phi^A-\bbb^Adu)\otimes(d\phi^B-\bbb^Bdu).
\end{equation}
With $\bbb:=\bbb^A\pr_{\phi^A}$, the frame fields satisfy
\begin{equation*}
\Nb=\pr_u+\bbb,\qquad N=e_4=\pr_{\ub},\qquad
e_3=\Om^{-2}(\pr_u+\bbb).
\end{equation*}
Since $e_3$ and $e_4$ are proportional to null gradients, the second null fundamental forms are symmetric:
\begin{equation*}
    \chi_{AB}=\chi_{BA},\qquad \chib_{AB}=\chib_{BA}.
\end{equation*}
We decompose them into their traces and traceless parts as
\begin{align*}
    \trch&:=\de^{AB}\chi_{AB},\qquad \hch_{AB}:=\chi_{AB}-\frac12\de_{AB}\trch,\\
    \trchb&:=\de^{AB}\chib_{AB},\qquad \hchb_{AB}:=\chib_{AB}-\frac12\de_{AB}\trchb.
\end{align*}
For horizontal vector fields $X,Y$, the induced connection $\nab$ and the projected null derivatives are
\begin{align*}
    \nab_XY&:=\D_XY-\frac12\chib(X,Y)e_4-\frac12\chi(X,Y)e_3,\\
    \nab_\mu X&:=\D_\mu X-\frac12\g(X,\D_\mu e_3)e_4-\frac12\g(X,\D_\mu e_4)e_3,\qquad \mu=3,4.
\end{align*}
A covariant spacetime tensor field $\psi$ is called horizontal if inserting either $e_3$ or $e_4$ into any of its slots gives zero. We extend $\nab_3$ and $\nab_4$ to horizontal tensors by applying $\D_3$ and $\D_4$, respectively, and projecting every tensorial slot onto $TS_{u,\ub}$. For a scalar function $f$, we set $\nab_3f:=e_3(f)$ and $\nab_4f:=e_4(f)$.
\begin{prop}
For the null frame defined in \eqref{dfLLb}, one has
\begin{align}
\begin{split}\label{nullidentities}
    \nab\log\Om&=\frac12(\eta+\etab),\qquad\quad \om=-\nab_4\log\Om,\qquad\quad \omb=0,\\
    \etab&=-\ze,\qquad\qquad\quad\;\;\, \xi=0,\qquad\qquad\qquad\quad \xib=0.
\end{split}
\end{align}
\end{prop}
\begin{proof}
See Section~2.3 of \cite{Taylor} and equation~(2.14) of \cite{Shen24}.
\end{proof}
We define the connection coefficients of the possibly non-coordinate null frame by
\begin{equation}\label{dfbGa}
    \bGa_{\mu\nu}^\la:=\g^{\la\tau}\g(\D_\mu e_\nu,e_\tau).
\end{equation}
Since the frame is not assumed to be coordinate, the coefficients $\bGa_{\mu\nu}^\la$ need not be symmetric in $\mu$ and $\nu$.

Metric compatibility, the normalization $\g(e_3,e_4)=-2$, and the definitions in \eqref{defga} give the Ricci formulae
\begin{equation}\label{ricciformulas}
\begin{aligned}
    \D_Ae_B
    &= \bGa_{AB}^Ce_C+\frac12\chi_{AB}e_3+\frac12\chib_{AB}e_4, \\
    \D_Ae_3
    &= \chib_A{}^Be_B+\ze_Ae_3,
    & \D_Ae_4
    &= \chi_A{}^Be_B-\ze_Ae_4, \\
    \D_3e_A
    &= \bGa_{3A}^Ce_C+\eta_Ae_3+\xib_Ae_4,
    & \D_4e_A
    &= \bGa_{4A}^Ce_C+\etab_Ae_4+\xi_Ae_3, \\
    \D_3e_3
    &= -2\omb e_3+2\xib^Ae_A,
    & \D_3e_4
    &= 2\omb e_4+2\eta^Ae_A, \\
    \D_4e_4
    &= -2\om e_4+2\xi^Ae_A,
    & \D_4e_3
    &= 2\om e_3+2\etab^Ae_A.
\end{aligned}
\end{equation}
\begin{lem}\label{Gaexpression}
The connection coefficients determined by the Ricci coefficients are
\begin{align*}
\bGa_{AB}^3&=\frac12\chi_{AB},&\bGa_{AB}^4&=\frac12\chib_{AB},&
\bGa_{B3}^A&=\chib_B{}^A,&\bGa_{B4}^A&=\chi_B{}^A,\\
\bGa_{3A}^3&=\eta_A,&\bGa_{4A}^4&=\etab_A,&
\bGa_{A3}^3&=\ze_A,&\bGa_{A4}^4&=-\ze_A,\\
\bGa_{34}^A&=2\eta^A,&\bGa_{43}^A&=2\etab^A,&
\bGa_{43}^3&=2\om,&\bGa_{44}^4&=-2\om.
\end{align*}
The coefficients
\begin{equation*}
\bGa_{3A}^4,\ \bGa_{A3}^4,\ \bGa_{A4}^3,\ \bGa_{4A}^3,\
\bGa_{33}^3,\ \bGa_{33}^4,\ \bGa_{34}^3,\ \bGa_{34}^4,\
\bGa_{43}^4,\ \bGa_{44}^3,\ \bGa_{33}^A,\ \bGa_{44}^A
\end{equation*}
vanish. The tangential coefficients $\bGa_{AB}^C$, $\bGa_{3A}^C$, and $\bGa_{4A}^C$ depend on the choice and transport of the horizontal frame and cannot be expressed solely in terms of \eqref{defga}.
\end{lem}

\begin{proof}
Since $\g^{AB}=\de^{AB}$ and $\g^{34}=\g^{43}=-\frac12$, formula \eqref{dfbGa} gives
\begin{equation*}
\bGa_{\mu\nu}^A=\de^{AB}\g(\D_\mu e_\nu,e_B),\qquad
\bGa_{\mu\nu}^3=-\frac12\g(\D_\mu e_\nu,e_4),\qquad
\bGa_{\mu\nu}^4=-\frac12\g(\D_\mu e_\nu,e_3).
\end{equation*}
Substituting \eqref{ricciformulas} and using $\xi=\xib=0$, $\omb=0$ from \eqref{nullidentities} gives every component above.
\end{proof}

We define the Schouten tensor $\S$ and the Cotton tensor $J$ by
\begin{equation}\label{dfSchouten}
    \S_{\mu\nu}:=\Ric_{\mu\nu}-\frac16\g_{\mu\nu}\R,\qquad
    J_{\la\mu\nu}:=\frac12\bigl(\D_\mu\S_{\la\nu}-\D_\nu\S_{\la\mu}\bigr),
\end{equation}
where $\R:=\g^{\mu\nu}\Ric_{\mu\nu}$ denotes the scalar curvature. In particular,
\begin{equation*}
    \S_{\mu\nu}=\S_{\nu\mu},\qquad J_{\la\mu\nu}=-J_{\la\nu\mu},\qquad J_{\la\mu\nu}+J_{\mu\nu\la}+J_{\nu\la\mu}=0,\qquad \g^{\la\mu}J_{\la\mu\nu}=0.
\end{equation*}
\begin{lem}\label{Rexpression}
The following identities hold for the curvature components:
\begin{align*}
    \R_{A3B3}&=\aa_{AB}+\frac12\de_{AB}\S_{33},\qquad\, \R_{A4B4}=\a_{AB}+\frac12\de_{AB}\S_{44},\\
    \R_{3A34}&=-2\bb_A+\S_{3A},\qquad\qquad \R_{4A34}=-2\b_A-\S_{4A},\\
    \R_{3434}&=4\rho+2\S_{34},\qquad\qquad\quad \R_{AB34}=2\si\slin_{AB},\\
    \R_{A3B4}&=-\rho\de_{AB}+\si\slin_{AB}-\S_{AB}+\frac12\de_{AB}\S_{34},\\
    \R_{ABC3}&=\slin_{AB}(\dual\bb)_C+\frac12\bigl(\de_{AC}\S_{3B}-\de_{BC}\S_{3A}\bigr),\\
    \R_{ABC4}&=-\slin_{AB}(\dual\b)_C+\frac12\bigl(\de_{AC}\S_{4B}-\de_{BC}\S_{4A}\bigr),\\
    \R_{ABCD}&=-\slin_{AB}\slin_{CD}\rho+\frac12\bigl(\de_{AC}\S_{BD}+\de_{BD}\S_{AC}-\de_{BC}\S_{AD}-\de_{AD}\S_{BC}\bigr).
\end{align*}
\end{lem}
\begin{proof}
The Weyl-Schouten decomposition associated with the normalization in \eqref{dfSchouten} is
\begin{equation}\label{WeylSchoutenDecomposition}
    \R_{\alpha\beta\gamma\delta}=\W_{\alpha\beta\gamma\delta}+\frac12\bigl(\g_{\alpha\gamma}\S_{\beta\delta}+\g_{\beta\delta}\S_{\alpha\gamma}-\g_{\beta\gamma}\S_{\alpha\delta}-\g_{\alpha\delta}\S_{\beta\gamma}\bigr).
\end{equation}
The definitions in \eqref{defr}, together with the Weyl symmetries, the trace-free identity $\g^{\alpha\gamma}\W_{\alpha\beta\gamma\delta}=0$, and the first Bianchi identity, imply
\begin{align*}
    \W_{A3B3}&=\aa_{AB},& \W_{A4B4}&=\a_{AB},\\
    \W_{3A34}&=-2\bb_A,& \W_{4A34}&=-2\b_A,\\
    \W_{3434}&=4\rho,& \W_{AB34}&=2\si\slin_{AB},\\
    \W_{A3B4}&=-\rho\de_{AB}+\si\slin_{AB},\\
    \W_{ABC3}&=\slin_{AB}(\dual\bb)_C,& \W_{ABC4}&=-\slin_{AB}(\dual\b)_C,\\
    \W_{ABCD}&=-\rho\slin_{AB}\slin_{CD}.
\end{align*}
Substituting \eqref{defr} into \eqref{WeylSchoutenDecomposition}, this concludes the proof of Lemma \ref{Rexpression}.
\end{proof}
\subsection{Hodge systems}\label{ssec7.2}
Let $S$ be an oriented spacelike $2$--sphere with induced metric $\ga$, connection $\nab$, area form $\slin$, and Gauss curvature $\K$.
\begin{df}\label{tensorfields}
We write $\sk_0:=C^\infty(S)\times C^\infty(S)$, $\sk_1:=\Gamma(T^*S)$, and $\sk_2:=\Gamma(\operatorname{Sym}_0^2T^*S)$, where $\operatorname{Sym}_0^2T^*S$ is the bundle of symmetric traceless covariant $2$--tensors.
\end{df}
\begin{df}\label{def7.2}
For $\xi,\eta\in\sk_1$ and $U,V\in\sk_2$, we set
\begin{align*}
(\dual\xi)_A&:=\slin_A{}^B\xi_B, & (\dual U)_{AB}&:=\slin_A{}^CU_{CB},\\
\xi\c\eta&:=\de^{AB}\xi_A\eta_B, & \xi\wedge\eta&:=\slin^{AB}\xi_A\eta_B,\\
(\xi\hot\eta)_{AB}&:=\xi_A\eta_B+\xi_B\eta_A-\de_{AB}(\xi\c\eta), & (\xi\c U)_A&:=\xi^BU_{AB},\\
U\wedge V&:=\slin^{AB}U_A{}^CV_{BC}. &&
\end{align*}
The tensor $\dual U$ is symmetric and traceless. We also define
\begin{equation*}
\sdiv\xi:=\nab^A\xi_A,\qquad \curl\xi:=\slin^{AB}\nab_A\xi_B,\qquad (\nab\hot\xi)_{AB}:=\nab_A\xi_B+\nab_B\xi_A-\de_{AB}\sdiv\xi.
\end{equation*}
\end{df}
\begin{df}\label{hodgeop}
We define the Hodge operators
\begin{align*}
    d_1:\sk_1&\longrightarrow\sk_0,& d_1\xi&:=(\sdiv\xi,\curl\xi),\\
    d_2:\sk_2&\longrightarrow\sk_1,& (d_2U)_A&:=\nab^BU_{AB},\\
    d_1^*:\sk_0&\longrightarrow\sk_1,& \bigl(d_1^*(f,f_*)\bigr)_A&:=-\nab_Af+\slin_A{}^B\nab_Bf_*,\\
    d_2^*:\sk_1&\longrightarrow\sk_2,& d_2^*\xi&:=-\frac12\nab\hot\xi.
\end{align*}
With respect to the $L^2(S,\ga)$ inner products, $d_1^*$ and $d_2^*$ are the formal adjoints of $d_1$ and $d_2$, respectively.
\end{df}
We define the rough Laplacian by
\begin{equation*}
    \De_0(f,f_*):=\bigl(\nab^A\nab_Af,\nab^A\nab_Af_*\bigr),\qquad (\De_1\xi)_A:=\nab^B\nab_B\xi_A,\qquad (\De_2U)_{AB}:=\nab^C\nab_CU_{AB}.
\end{equation*}
The following Weitzenb\"ock identities hold:
\begin{align}
\begin{split}\label{dddd}
    d_1^*d_1\xi&=-\De_1\xi+\K\xi,\qquad\quad d_1d_1^*(f,f_*)=-\De_0(f,f_*),\\
    d_2^*d_2U&=-\frac12\De_2U+\K U,\qquad\qquad d_2d_2^*\xi=-\frac12\bigl(\De_1\xi+\K\xi\bigr).
\end{split}
\end{align}
See equation~(2.2.2) in \cite{Ch-Kl}.
\begin{df}\label{dfdkb}
We define $\dk:=\{\dkb,\nab_3\}$, where the angular derivative $\dkb$ acts according to tensorial type:
\begin{equation*}
\dkb U:=d_2U\quad(U\in\sk_2),\qquad
\dkb\xi:=d_1\xi\quad(\xi\in\sk_1),\qquad
\dkb(f,f_*):=d_1^*(f,f_*)\quad((f,f_*)\in\sk_0).
\end{equation*}
For $\psi\in\sk_k$, $k\in\{0,1,2\}$, set $\psi^{(0)}:=\psi$ and
$\psi^{(q)}:=(\psi,\dk\psi,\ldots,\dk^q\psi)$ for $q\geq1$, where $\dk^q$
denotes all $q$--fold compositions, with $\dkb$ interpreted at each step
according to tensorial type.
\end{df}

\begin{prop}\label{ellipticLp}
Let $S$ be a smooth, compact, oriented $2$--sphere and $1<p<\infty$. Write
$\|F\|_{L^p(S)}:=(\int_S|F|^p\,d\si)^{1/p}$,
$\De:=\nab^A\nab_A$, and
$\ov\phi:=|S|^{-1}\int_S\phi\,d\si$. For a scalar $\phi$, a $1$--form
$\xi$, and a symmetric traceless $2$--tensor $U$, one has
\begin{align*}
\De\phi=f\quad&\Longrightarrow\quad
\|\nab^2\phi\|_{L^p(S)}+\|\nab\phi\|_{L^p(S)}
+\|\phi-\ov\phi\|_{L^p(S)}
\les\|f\|_{L^p(S)},\\
d_1\xi=(f,f_*)\quad&\Longrightarrow\quad
\|\nab\xi\|_{L^p(S)}+\|\xi\|_{L^p(S)}
\les\|(f,f_*)\|_{L^p(S)},\\
d_2U=F\quad&\Longrightarrow\quad
\|\nab U\|_{L^p(S)}+\|U\|_{L^p(S)}
\les\|F\|_{L^p(S)}.
\end{align*}
The implicit constants depend on $p$ and the controlled geometry of $(S,\ga)$.
\end{prop}

\begin{proof}
See Corollary~2.3.1.1 in \cite{Ch-Kl}.
\end{proof}
\subsection{Null structure equations}
We recall the null structure equations for a double null foliation, see Proposition 2.13 in \cite{ShenWan}.
\begin{prop}\label{nulles}
The null structure equations take the following form:
\begin{align}
\begin{split}\label{basicnull}
\nabs_4\eta&=-\chi\c(\eta+\ze)-\b-\frac12\Ssl_4,\\
\nabs_3\ze&=-\chib\c(\eta+\ze)-\bb+\frac12\Ssl_3,\\
\nabs_4\hch+(\trch)\hch&=-2\om\hch-\a,\\
\nabs_4\trch+\frac12(\trch)^2&=-|\hch|^2-2\om\trch-\S_{44},\\
\nabs_3\hchb+(\trchb)\hchb&=-\aa,\\
\nabs_3\trchb+\frac12(\trchb)^2&=-|\hchb|^2-\S_{33},\\
\nabs_4\hchb+\frac12(\trch)\hchb&=-\nabs\hot\ze+2\om\hchb-\frac12\trchb\,\hch+\ze\hot\ze+\Sslh,\\
\nabs_3\hch+\frac12(\trchb)\hch&=\nabs\hot\eta-\frac12\trch\,\hchb+\eta\hot\eta+\Sslh,\\
\nabs_4\trchb+\frac12(\trch)\trchb&=2\om\trchb+2\rho-\hch\c\hchb-2\sdiv\ze+2|\ze|^2+\Tr\S,\\
\nabs_3\trch+\frac12(\trchb)\trch&=2\rho-\hch\c\hchb+2\sdiv\eta+2|\eta|^2+\Tr\S.
\end{split}
\end{align}
Here,
\begin{equation*}
    (\Ssl_3)_A:=\S_{3A},\qquad (\Ssl_4)_A:=\S_{4A},\qquad \Ssl_{AB}:=\S_{AB},\qquad \tr\Ssl:=\de^{AB}\Ssl_{AB},\qquad \Tr\S:=\g^{\mu\nu}\S_{\mu\nu},
\end{equation*}
and $\Sslh$ denotes the traceless part of $\Ssl$. We also have the Codazzi equations
\begin{align}
\begin{split}\label{codazzi}
\sdiv\hch&=\frac12\nabs\trch-\ze\c\left(\hch-\frac12\trch\right)-\b+\frac12\Ssl_4,\\
\sdiv\hchb&=\frac12\nabs\trchb+\ze\c\left(\hchb-\frac12\trchb\right)+\bb+\frac12\Ssl_3,
\end{split}
\end{align}
the torsion equation
\begin{equation}\label{torsion}
    \curl\eta=\curl\ze=\si-\frac12\hch\wedge\hchb,
\end{equation}
and the Gauss equation
\begin{equation}\label{gauss}
    \K=-\frac14\trch\trchb+\frac12\hch\c\hchb-\rho+\frac12\tr\Ssl.
\end{equation}
Moreover, the lapse equation in this gauge is
\begin{align}
\nabs_3\om=2\eta\c\ze+|\ze|^2+\rho+\frac12\S_{34}.
\end{align}
\end{prop}
\subsection{Bianchi equations}\label{ssec7.4}
We recall the Bianchi equations in the double-null foliation under the gauge conditions
$\omb=0$ and $\etab=-\ze$. We first note that, with our conventions, for
$\xi,\eta\in\sk_1$ and $U,V\in\sk_2$,
\begin{equation*}
\xi\c\dual\eta=\xi\wedge\eta,\qquad
U\c\dual V=U\wedge V,\qquad
U\wedge V=-V\wedge U.
\end{equation*}
\begin{prop}\label{bianchiequations}
The Bianchi equations take the following form:
\begin{align*}
\nabs_3\a+\frac12\trchb\,\a
&=\nabs\hot\b-3(\hch\rho+\dual\hch\,\si)+(\ze+4\eta)\hot\b+\Jh_4,\\
\nabs_4\b+2\trch\,\b
&=\sdiv\a-2\om\b+\ze\c\a-J_{4\bu4},\\
\nabs_3\b+\trchb\,\b
&=\nabs\rho+\dual\nabs\si+2\hch\c\bb+3(\eta\rho+\dual\eta\,\si)+J_{3\bu4},\\
\nabs_4\rho+\frac32\trch\,\rho
&=\sdiv\b-\frac12\hchb\c\a-\ze\c\b-\frac12J_{434},\\
\nabs_4\si+\frac32\trch\,\si
&=-\curl\b+\frac12\hchb\wedge\a+\ze\wedge\b+\frac12\dual J_{434},\\
\nabs_3\rho+\frac32\trchb\,\rho
&=-\sdiv\bb-\frac12\hch\c\aa+(\ze-2\eta)\c\bb-\frac12J_{343},\\
\nabs_3\si+\frac32\trchb\,\si
&=-\curl\bb-\frac12\hch\wedge\aa+(\ze-2\eta)\wedge\bb-\frac12\dual J_{343},\\
\nabs_4\bb+\trch\,\bb
&=-\nabs\rho+\dual\nabs\si+2\om\bb+2\hchb\c\b
+3(\ze\rho-\dual\ze\,\si)-J_{4\bu3},\\
\nabs_3\bb+2\trchb\,\bb
&=-\sdiv\aa+(2\ze-\eta)\c\aa+J_{3\bu3},\\
\nabs_4\aa+\frac12\trch\,\aa
&=-\nabs\hot\bb+4\om\aa-3(\hchb\rho-\dual\hchb\,\si)
+5\ze\hot\bb+\Jh_3.
\end{align*}
Here, we denote
\begin{align*}
(\Jh_4)_{AB}&:=J_{AB4}+J_{BA4}-\frac12J_{434}\de_{AB},&
(\Jh_3)_{AB}&:=J_{AB3}+J_{BA3}-\frac12J_{343}\de_{AB},\\
\dual J_{434}&:=(J_{AB4}-J_{BA4})\slin^{AB},&
\dual J_{343}&:=(J_{AB3}-J_{BA3})\slin^{AB},&
(J_{\la\bu\nu})_A&:=J_{\la A\nu}.
\end{align*}
The tensor $J_{\la\mu\nu}$ is the Cotton tensor defined in
\eqref{dfSchouten}.
\end{prop}
\begin{proof}
See Proposition~2.15 of \cite{ShenWan} and the corresponding null Bianchi equations in
\cite{Luk18}. We have specialized them to the gauge $\omb=0$, $\etab=-\ze$, and to the conventions used in this paper.
\end{proof}
We next record the Bianchi equations for the renormalized curvature quantities.
\begin{prop}\label{renorbianchi}
Recall from \eqref{gauss} that
$\K=-\frac14\trch\trchb+\frac12\hch\c\hchb-\rho+\frac12\tr\Ssl$,
and set
\begin{equation*}
\sic:=\si-\frac12\hch\wedge\hchb.
\end{equation*}
Then the following renormalized Bianchi equations hold:
\begin{align*}
\nabs_3\b+\trchb\,\b
&=d_1^*(\K,\sic)+2\hch\c\bb-3(\eta\K-\dual\eta\,\sic)\\
&+\frac12\left(\nabs(\hch\c\hchb)+\dual\nabs(\hch\wedge\hchb)\right)
+\frac32\left(\eta\,\hch\c\hchb+\dual\eta\,\hch\wedge\hchb\right)\\
&-\frac14\left((\nabs\trch)\trchb+\trch\,\nabs\trchb\right)
-\frac34\eta\,\trch\trchb
+\frac12\nabs\tr\Ssl+\frac32\eta\,\tr\Ssl+J_{3\bu4},\\
\nabs_4\K+\trch\,\K
&=-\sdiv\b+\ze\c\b-\frac12\hch\c(\nabs\hot\ze)
+\frac12\hch\c(\ze\hot\ze)\\
&+\frac12\trch\,\sdiv\ze-\frac12\trch|\ze|^2
+\frac12\sdiv\Ssl_4-\frac12\ze\c\Ssl_4,\\
\nabs_4\sic+\frac32\trch\,\sic
&=-\curl\b+\ze\wedge\b
+\frac12\hch\wedge(\nabs\hot\ze)
-\frac12\hch\wedge(\ze\hot\ze)
-\frac12\hch\wedge\Sslh+\frac12\dual J_{434},\\
\nabs_3\K+\trchb\,\K
&=\sdiv\bb-(\ze-2\eta)\c\bb
+\frac12\hchb\c(\nabs\hot\eta)
+\frac12\hchb\c(\eta\hot\eta)\\
&-\frac12\trchb\,\sdiv\eta-\frac12\trchb|\eta|^2
+\frac12\sdiv\Ssl_3+\frac12(2\eta-\ze)\c\Ssl_3,\\
\nabs_3\sic+\frac32\trchb\,\sic
&=-\curl\bb+(\ze-2\eta)\wedge\bb
+\frac12\hchb\wedge(\nabs\hot\eta)
+\frac12\hchb\wedge(\eta\hot\eta)+\frac12\hchb\wedge\Sslh-\frac12\dual J_{343},\\
\nabs_4\bb+\trch\,\bb
&=-d_1^*(\K,-\sic)+2\om\bb+2\hchb\c\b
-3(\ze\K+\dual\ze\,\sic)\\
&-\frac12\left(\nabs(\hch\c\hchb)
-\dual\nabs(\hch\wedge\hchb)\right)
+\frac32\left(\ze\,\hch\c\hchb
-\dual\ze\,\hch\wedge\hchb\right)\\
&+\frac14\left((\nabs\trch)\trchb+\trch\,\nabs\trchb\right)
-\frac34\ze\,\trch\trchb
-\frac12\nabs\tr\Ssl+\frac32\ze\,\tr\Ssl-J_{4\bu3}.
\end{align*}
\end{prop}
\begin{proof} 
They follow from Proposition~\ref{bianchiequations}, the Gauss equation, and the null structure equations, specialized to the gauge $\omb=0$, $\etab=-\ze$; see \cite{Luk18,ShenWan}. 
\end{proof}
\begin{prop}\label{metricequations}
The metric and shift components satisfy
\begin{align*}
    e_4(\slg_{AB})&=2\chi_{AB},&
    e_4(\bbb^A)&=-2\Om^2(\eta^A-\etab^A).
\end{align*}
Here, the horizontal indices denote coordinate components with respect to
$(\pr_{\phi^1},\pr_{\phi^2})$.
\end{prop}
\begin{proof}
See Proposition~2.20 of \cite{ShenKerr}, after translating to the present
normalization.
\end{proof}
\subsection{Commutation identities}
We recall the following commutation formulae; compare Lemma~7.3.3 in \cite{Ch-Kl}.
\begin{lem}\label{comm}
Let $U_{A_1\ldots A_k}$ be a horizontal $k$--covariant tensor field on $(\M,\g)$. Then,
\begin{align*}
[\nabs_4,\nabs_B]U_{A_1\ldots A_k}
&=-\chi_{BC}\nabs_CU_{A_1\ldots A_k}+\sum_{i=1}^k
\left(-\chi_{A_iB}\ze_C+\chi_{BC}\ze_{A_i}+\R_{CA_iB4}
\right)U_{A_1\ldots C\ldots A_k},\\
[\nabs_3,\nabs_B]U_{A_1\ldots A_k}
&=-\chib_{BC}\nabs_CU_{A_1\ldots A_k}+(\eta_B-\ze_B)\nabs_3U_{A_1\ldots A_k}\\
&+\sum_{i=1}^k\left(\chib_{A_iB}\eta_C-\chib_{BC}\eta_{A_i}+\R_{CA_iB3}\right)U_{A_1\ldots C\ldots A_k},\\
[\nabs_3,\nabs_4]U_{A_1\ldots A_k}
&=-2\om\,\nabs_3U_{A_1\ldots A_k}+2(\eta_C+\ze_C)\nabs_CU_{A_1\ldots A_k}\\
&+\sum_{i=1}^k\left(-2\ze_{A_i}\eta_C+2\eta_{A_i}\ze_C-\R_{CA_i34}
\right)U_{A_1\ldots C\ldots A_k},\\
[\nabs_A,\nabs_B]U_{A_1\ldots A_k}
&=\K\sum_{i=1}^k\left(\slg_{AA_i}U_{A_1\ldots B\ldots A_k}-\slg_{BA_i}U_{A_1\ldots A\ldots A_k}\right).
\end{align*}
\end{lem}
\subsection{Sasaki metric on the mass shell}
We write $\pb^\mu$ for the fiber coordinates on $T\M$ and $p^\mu$ for their restrictions to the mass shell $P_m$. On the region where $p^4>0$, the Vlasov estimates are written in the $p^4$--chart
\[
(u,\ub,\phi^1,\phi^2,p^1,p^2,p^4),
\]
with $p^3$ determined by the mass-shell relation. The $p^3$--chart is also used below, in particular for the initial norms and momentum integrals. The ambient Sasaki metric $\gb$ \cite{Sasaki} restricts nondegenerately to $P_m$ for $m>0$ and degenerately for $m=0$.

\subsubsection{Sasaki metric}\label{ssec:Sasaki}
\begin{df}\label{dfhorver}
Given $(x,\pb)\in T\M$ and $v\in T_x\M$, the vertical lift is
\begin{equation*}
\ver(v):=\ver_{(x,\pb)}(v):=\left.\frac{d}{ds}\right|_{s=0}(x,\pb+sv).
\end{equation*}
Let $c(0)=x$, $c'(0)=v$, and let $\pb(s)$ be the parallel translate of $\pb$ along $c$, so that $\D_{c'}\pb=0$. The horizontal lift is
\begin{equation*}
\hor(v):=\hor_{(x,\pb)}(v):=\left.\frac{d}{ds}\right|_{s=0}(c(s),\pb(s)).
\end{equation*}
Thus vertical motion changes the momentum at fixed $x$, whereas horizontal motion changes $x$ while parallel-transporting the momentum.
\end{df}
Given $v=v^\mu e_\mu\in T_x\M$, we have
\begin{equation}\label{horver}
\ver(v)=v^\mu\pr_{\pb^\mu},\qquad\quad \hor(v)=v^\mu e_\mu-v^\mu\pb^\nu\bGa_{\mu\nu}^{\la}\pr_{\pb^\la}.
\end{equation}
The first term in $\hor(v)$ moves the base point, while the second term compensates for the variation of the frame so that $\pb$ is parallel-transported rather than kept with fixed frame components.
\begin{rk}
The generator of the geodesic flow is
\begin{equation}\label{X=horp}
X=\pb^\mu e_\mu-\pb^\mu\pb^\nu\bGa_{\mu\nu}^{\la}\pr_{\pb^\la}=\hor(\pb).
\end{equation}
Its integral curves satisfy $x'(s)=\pb(s)$ and $\D_{x'}\pb=0$. Thus the base curve is a geodesic and $\pb$ is its velocity.
\end{rk}
The vertical and horizontal subspaces are
\begin{equation*}
\VV_{(x,\pb)}:=\ver_{(x,\pb)}(T_x\M),\qquad\quad \HH_{(x,\pb)}:=\hor_{(x,\pb)}(T_x\M).
\end{equation*}
The connection gives the direct-sum decomposition $T_{(x,\pb)}T\M=\HH_{(x,\pb)}\oplus\VV_{(x,\pb)}$. Thus, every infinitesimal variation of a position-momentum pair decomposes uniquely into the motion of the base point and genuine variation of the momentum after subtracting parallel transport.
\begin{df}\label{dfSasaki}
The Sasaki metric $\gb$ on $T\M$ is defined by
\begin{equation*}
\gb(\hor(X),\hor(Y))=\g(X,Y),\qquad \gb(\hor(X),\ver(Y))=0,\qquad \gb(\ver(X),\ver(Y))=\g(X,Y).
\end{equation*}
\end{df}
In other words, the connection splitting identifies $T_{(x,\pb)}T\M\simeq T_x\M\oplus T_x\M$, and $\gb$ corresponds to $\g\oplus\g$. The first copy of $T_x\M$ measures the motion of the base point, while the second measures the vertical variation of momentum.
\begin{prop}\label{nabb}
Let $\Db$ be the Levi-Civita connection of the Sasaki metric $\gb$ and let $X,Y\in\Ga(T\M)$. Then, we have
\begin{align*}
\Db_{\hor(X)}\hor(Y)&=\hor(\D_XY)-\frac12\ver(\R(X,Y)\pb),&
\Db_{\hor(X)}\ver(Y)&=\ver(\D_XY)+\frac12\hor(\R(\pb,Y)X),\\
\Db_{\ver(X)}\hor(Y)&=\frac12\hor(\R(\pb,X)Y),& \Db_{\ver(X)}\ver(Y)&=0.
\end{align*}
\end{prop}
\begin{proof}
See Proposition 4.4 in \cite{Taylor}.
\end{proof}
These formulae show that the geometry of $(T\M,\gb)$ is controlled by the geometry of $(\M,\g)$. The new terms are curvature terms, which arise because the horizontal distribution is defined by parallel transport.
\begin{cor}\label{cornabb}
Under the assumptions of Proposition \ref{nabb}, we have
\begin{align*}
[\hor(X),\hor(Y)]&=\hor([X,Y])-\ver(\R(X,Y)\pb),&
[\hor(X),\ver(Y)]&=\ver(\D_XY).
\end{align*}
\end{cor}
\begin{proof}
By Proposition \ref{nabb},
\begin{align*}
[\hor(X),\hor(Y)]&=\Db_{\hor(X)}\hor(Y)-\Db_{\hor(Y)}\hor(X)\\ 
&=\hor(\D_XY-\D_YX)-\ver(\R(X,Y)\pb)\\ 
&=\hor([X,Y])-\ver(\R(X,Y)\pb),
\end{align*}
and
\begin{align*}
[\hor(X),\ver(Y)]=\Db_{\hor(X)}\ver(Y)-\Db_{\ver(Y)}\hor(X)=\ver(\D_XY).
\end{align*}
\end{proof}
\begin{rk}
The fiber-independence assumption in Corollary \ref{cornabb} is important. Since the tautological momentum $\pb$ depends on the fiber, the geodesic spray $X=\hor(\pb)$ satisfies
\begin{equation*}
[X,\ver(Y)]=-\hor(Y)+\ver(\D_{\pb}Y)
\end{equation*}
for a fiber-independent spacetime vectorfield $Y$. Thus one cannot substitute the tautological field $\pb$ directly into the second identity of Corollary \ref{cornabb} without the additional term $-\hor(Y)$.
\end{rk}
\begin{rk}
By Corollary \ref{cornabb}, the obstruction to integrability of the horizontal distribution is $-\ver(\R(X,Y)\pb)$: infinitesimal parallel transport around a parallelogram changes momentum by the curvature term $\R(X,Y)\pb$.
\end{rk}
\subsubsection{Mass shell}\label{ssec:massshell}
Fix $m\geq0$. The future mass shell is
\begin{equation*}
P_m:=\{(x,\pb)\in T\M:\g(\pb,\pb)=-m^2,\ \pb\text{ future directed}\}.
\end{equation*}
Using the null frame, write $\pb=\pb^\mu e_\mu$. The mass-shell relation is
\begin{equation}\label{massshelldf}
-4\pb^3\pb^4+(\pb^1)^2+(\pb^2)^2=-m^2.
\end{equation}
We denote the restrictions of $\pb^\mu$ to $P_m$ by $p^\mu$. Only three of the four momentum components are independent.

On $P_m\cap\{p^4>0\}$, we use the $p^4$-chart $(u,\ub,\phi^1,\phi^2,p^1,p^2,p^4)$, where
\begin{equation}\label{eq:p3-from-p4}
\pb^1=p^1,\qquad \pb^2=p^2,\qquad \pb^4=p^4,\qquad \pb^3=p^3:=\frac{(p^1)^2+(p^2)^2+m^2}{4p^4}.
\end{equation}
Thus the four spacetime coordinates and three independent momentum coordinates give $\dim P_m=7$. For fixed $x$, the fiber $P_{x,m}$ is the future mass shell in $T_x\M$.

For comparison, the analogous $p^3$-chart on $P_m\cap\{p^3>0\}$ is determined by
\begin{equation}\label{eq:p4-from-p3}
\pb^1=p^1,\qquad \pb^2=p^2,\qquad \pb^3=p^3,\qquad \pb^4=p^4:=\frac{(p^1)^2+(p^2)^2+m^2}{4p^3}.
\end{equation}
\subsubsection{Mass-shell volume form}
For fixed $x$, the metric and volume form on the ambient fiber $T_x\M$ are
\begin{equation}\label{eq:ambient-fiber-volume}
\g|_{T_x\M}=-4d\pb^3d\pb^4+(d\pb^1)^2+(d\pb^2)^2,\qquad \dvol_{T_x\M}=2d\pb^3\wedge d\pb^4\wedge d\pb^1\wedge d\pb^2.
\end{equation}
Set $\La(\pb):=\g(\pb,\pb)$. Since $P_{x,m}=\{\La=-m^2\}$, the one-form $d\La$ is normal to $P_{x,m}$, with
\begin{equation}\label{eq:dLambda-half}
\frac12d\La=-2\pb^4d\pb^3-2\pb^3d\pb^4+\pb^1d\pb^1+\pb^2d\pb^2.
\end{equation}
In the $p^4$-chart and, analogously, in the $p^3$-chart, the oriented mass-shell volume forms are
\begin{equation}\label{eq:dvolm-p4,p3}
\dvol_m^{(4)}:=-\frac{dp^1\wedge dp^2\wedge dp^4}{p^4},\qquad \dvol_m^{(3)}:=\frac{dp^1\wedge dp^2\wedge dp^3}{p^3}.
\end{equation}
Indeed, both satisfy $\frac12d\La\wedge\dvol_m^{(4)}=\dvol_{T_x\M}$ and $\frac12d\La\wedge\dvol_m^{(3)}=\dvol_{T_x\M}$. Differentiating \eqref{eq:p3-from-p4} gives $dp^1\wedge dp^2\wedge dp^3=-(p^3/p^4)dp^1\wedge dp^2\wedge dp^4$. Hence, as positive densities and oriented forms, respectively,
\begin{equation*}
\frac{dp^1dp^2dp^3}{p^3}=\frac{dp^1dp^2dp^4}{p^4},\qquad  \frac{dp^1\wedge dp^2\wedge dp^3}{p^3}=-\frac{dp^1\wedge dp^2\wedge dp^4}{p^4}.
\end{equation*}
The factor $(p^4)^{-1}$ is the Jacobian arising from solving the mass-shell relation for $p^3$, and is the null-frame analogue of the usual relativistic density $d^3p/p^0$. Therefore,
\begin{align}
\T_{\mu\nu}&=\int_0^\infty\int_{\RRR^2}\frac{fp_\mu p_\nu}{p^4}\,dp^1dp^2dp^4,\label{eq:expressionTm-p4}\\ 
\T_{\mu\nu}&=\int_0^\infty\int_{\RRR^2}\frac{fp_\mu p_\nu}{p^3}\,dp^1dp^2dp^3,\label{eq:expressionTm-p3}
\end{align}
where the second expression is the equivalent formula in the $p^3$-chart.
\subsubsection{Vertical coordinate derivatives}
\begin{prop}\label{prop:T19}
In the $p^4$-chart, the coordinate vectorfields tangent to $P_m$ are
\begin{equation}\label{T19eq-p4}
\pr_{p^A}^{(4)}=\pr_{\pb^A}+\frac{p^A}{2p^4}\pr_{\pb^3},\qquad A=1,2,\qquad\quad \pr_{p^4}^{(4)}=\pr_{\pb^4}-\frac{p^3}{p^4}\pr_{\pb^3}.
\end{equation}
Analogously, in the $p^3$-chart,
\begin{equation}\label{T19eq}
\pr_{p^A}^{(3)}=\pr_{\pb^A}+\frac{p^A}{2p^3}\pr_{\pb^4},\qquad A=1,2,\qquad\quad \pr_{p^3}^{(3)}=\pr_{\pb^3}-\frac{p^4}{p^3}\pr_{\pb^4}.
\end{equation}
\end{prop}
\begin{proof}
Differentiating \eqref{eq:p3-from-p4} with respect to the independent variables gives, regardless of the value of $m$,
\begin{equation*}
\pr_{p^A}p^3=\frac{p^A}{2p^4},\qquad\quad \pr_{p^4}p^3=-\frac{p^3}{p^4}.
\end{equation*}
This proves \eqref{T19eq-p4}. Formula \eqref{T19eq} follows in the same way from \eqref{eq:p4-from-p3}.
\end{proof}
Geometrically, these are the coordinate vectorfields tangent to the embedded hypersurface $P_m\subset T\M$. Varying an independent momentum component also changes the dependent component in order to preserve the mass-shell relation. In particular, $\pr_{p^A}^{(3)}$ and $\pr_{p^A}^{(4)}$ are different tangent vectorfields because they keep different null momentum components fixed. On the common domain of the two charts,
\begin{equation}\label{eq:chart-transition}
p^3\pr_{p^3}^{(3)}=-p^4\pr_{p^4}^{(4)}, \qquad \pr_{p^A}^{(3)}=\pr_{p^A}^{(4)}+\frac{p^A}{2p^3}\pr_{p^4}^{(4)}.
\end{equation}
\subsubsection{A frame on the mass shell}
In the $p^4$-chart, define
\begin{equation}\label{dfEI-p4}
E_\mu^{(4)}=\hor(e_\mu),\qquad \mu=1,2,3,4,\qquad E_{4+A}^{(4)}=\pr_{p^A}^{(4)},\quad A=1,2,\qquad E_8^{(4)}=\pr_{p^4}^{(4)}.
\end{equation}
Using \eqref{T19eq-p4}, we obtain
\begin{equation}\label{eq:vertical-frame-p4}
E_{4+A}^{(4)}=\ver\left(e_A+\frac{p^A}{2p^4}e_3\right),\qquad A=1,2,\qquad\quad E_8^{(4)}=\ver\left(e_4-\frac{p^3}{p^4}e_3\right).
\end{equation}
For comparison, the analogous frame in the $p^3$-chart is
\begin{equation}\label{dfEI-p3}
E_\mu^{(3)}=\hor(e_\mu),\qquad \mu=1,2,3,4,\qquad E_{4+A}^{(3)}=\pr_{p^A}^{(3)},\quad A=1,2,\qquad E_7^{(3)}=\pr_{p^3}^{(3)},
\end{equation}
with
\begin{equation}\label{eq:vertical-frame-p3}
E_{4+A}^{(3)}=\ver\left(e_A+\frac{p^A}{2p^3}e_4\right),\qquad A=1,2,\qquad\quad E_7^{(3)}=\ver\left(e_3-\frac{p^4}{p^3}e_4\right).
\end{equation}
\section{Main theorem}\label{secmain}
\subsection{Fundamental norms}
We use the notation
\begin{align*}
&V:=V(u,\ub):=\{(u',\ub')\in[0,u]\times[0,\ub]\},\qquad
V_*:=V(u_*,\ub_*),\qquad V_f:=V(\de,\de),\\
&L^p_u:=L^p_u(0,u_*),\qquad
L^p_\ub:=L^p_\ub(0,\ub_*),\qquad S:=S_{u,\ub}.
\end{align*}
\subsubsection{Schematic notation \texorpdfstring{$\Ga_k$}{} and \texorpdfstring{$\GG$}{}}
We introduce the following schematic notation, which records the expected singular weight of Ricci coefficients, curvature components, and matter fields.
\begin{df}\label{gammag}
We set
\begin{align*}
&\Ga_1:=\{\trch,\hch,\om,\omd,\bGa_{4A}^B\},\qquad
\Ga_0:=\{\eta,\etab\},\qquad
\Ga_{-1}:=\{\trchb,\hchb\},\\
&Z:=\{\nab_1,\nab_2,\nab_3\},\qquad
\nab:=\{\nab_1,\nab_2\}.
\end{align*}
For $1\leq i\leq3$, we define
\begin{align*}
\Ga_1^{(i)}&:=Z^{\leq i}\Ga_1\cup Z^{\leq i-1}\{\b\},&
\Ga_0^{(i)}&:=Z^{\leq i}\Ga_0\cup Z^{\leq i-1}\{\K,\sic,\bGa_{AB}^C\},\\
\Ga_{-1}^{(i)}&:=Z^{\leq i}\Ga_{-1}\cup Z^{\leq i-1}\{\bb,\bGa_{3A}^B\},&
\Ga_{-2}^{(i)}&:=Z^{\leq i-1}\{\aa\}.
\end{align*}
At the fourth order, we define
\begin{align*}
\Ga_1^{(4)} &:=Z^{\leq3}\Ga_1 \cup Z^{\leq3}\nab\Ga_1 \cup Z^{\leq3}\{\b\},&
\Ga_0^{(4)} &:=Z^{\leq3}\Ga_0 \cup Z^{\leq3}\nab\Ga_0 \cup Z^{\leq3}\{\K,\sic,\bGa_{AB}^C\},\\
\Ga_{-1}^{(4)} &:=Z^{\leq3}\Ga_{-1} \cup Z^{\leq3}\nab\Ga_{-1} \cup Z^{\leq3}\{\bb,\bGa_{3A}^B\},&
\Ga_{-2}^{(4)} &:=Z^{\leq3}\{\aa\}.
\end{align*}
Thus, $\Ga_k^{(3)}$ contains all third order $Z$--derivatives of the
fields in $\Ga_k$ for $k=-1,0,1$, including the pure derivatives $\nab_3^3\Ga_k$,
whereas $\Ga_k^{(4)}$ contains fourth order derivatives with at least
one angular derivative, but not the pure derivatives
$\nab_3^4\Ga_k$.

We also introduce the metric classes
\begin{align*}
\GG_0:=\{\log\Om,\slg_{AB}\},\qquad
\GG_{-1}:=\{\bbb^A\},\qquad \GG:=\GG_0\cup\GG_{-1}.
\end{align*}
For $1\leq i\leq3$, we define
\begin{align*}
\GG_0^{(i)}
&:=Z^{\leq i}\GG_0
\cup Z^{\leq i-1}\{\Ga_0,\Ga_{-1},\bGa_{AB}^C\},\\
\GG_{-1}^{(i)}
&:=Z^{\leq i}\GG_{-1}
\cup Z^{\leq i-1}\{\Ga_0,\Ga_{-1},\bGa_{3A}^B\},
\end{align*}
and at the fourth order,
\begin{align*}
\GG_0^{(4)}
&:=Z^{\leq3}\GG_0
\cup Z^{\leq3}\nab\GG_0
\cup Z^{\leq3}\{\Ga_0,\Ga_{-1},\bGa_{AB}^C\},\\
\GG_{-1}^{(4)}
&:=Z^{\leq3}\GG_{-1}
\cup Z^{\leq3}\nab\GG_{-1}
\cup Z^{\leq3}\{\Ga_0,\Ga_{-1},\bGa_{3A}^B\}.
\end{align*}
Finally, set $\GG^{(i)}:=\GG_0^{(i)}\cup\GG_{-1}^{(i)}$, $1\leq i\leq4$. As for $\Ga_k^{(4)}$, no pure $\nab_3^4$ derivative is included in the fourth order metric classes.
\end{df}
\begin{rk}\label{rk:metric-components}
Here and below, $\slg_{AB}$ and $\bbb^A$ denote the coordinate components with respect to the transported angular coordinates $(\phi^1,\phi^2)$. Whenever $Z$ or $\nab$ is applied to these
coordinate components, it is understood componentwise, namely as
ordinary frame differentiation of the corresponding scalar functions,
rather than as the covariant derivative of the horizontal metric tensor.
\end{rk}
\subsubsection{Sphere norms}\label{secOnorms}
We define
\begin{align*}
\mo_3:=&
\left\|\Ga_0^{(3)}\right\|_{L^\infty_u L^\infty_\ub L^2(S)}
+\left\|\Ga_{-1}^{(3)}\right\|_{L^\infty_u L^\infty_\ub L^2(S)}
+\left\|h(\ub)\Ga_1^{(3)}\right\|_{L^2_\ub L^\infty_u L^2(S)}
+\left\|\Ga_{-2}^{(3)}\right\|_{L^2_u L^\infty_\ub L^2(S)},\\
\Ot_4:=&
\left\|h(\ub)\Ga_1^{(4)}\right\|_{L^\infty_u L^2_\ub L^2(S)}
+\left\|\Ga_0^{(4)}\right\|_{L^\infty_u L^2_\ub L^2(S)}
+\left\|\GG^{(4)}\right\|_{L^\infty_u L^\infty_\ub L^2(S)}\\
&+\left\|h(\ub)\Ga_0^{(4)}\right\|_{L^\infty_\ub L^2_u L^2(S)}
+\left\|\Ga_{-1}^{(4)}\right\|_{L^\infty_\ub L^2_u L^2(S)},\\
\mo:=&\mo_3+\Ot_4.
\end{align*}

\begin{rk}
The mixed norms in $\mo_3$ of type $L^2L^\infty$ with respect to the two null directions are stronger than the corresponding $L^\infty L^2$ norms.
\end{rk}

\subsubsection{Flux norms}\label{secRnorms}
For $0\leq i\leq3$, we define
\begin{align*}
\mr_i:=&
\left\|h(\ub)Z^i\b\right\|_{L^\infty_u L^2_\ub L^2(S)}
+\left\|Z^i(\K,\sic,\bb)\right\|_{L^\infty_u L^2_\ub L^2(S)},\\
\ur_i:=&
\left\|h(\ub)Z^i(\K,\sic)\right\|_{L^\infty_\ub L^2_u L^2(S)}
+\left\|Z^i(\bb,\aa)\right\|_{L^\infty_\ub L^2_u L^2(S)},\\
\mr:=&\sum_{i=0}^3\big(\mr_i+\ur_i\big).
\end{align*}

\subsubsection{Matter norms}\label{initialTT}
We first introduce the following norms for the distribution function. On the region where $p^4>0$, in the $(p^1,p^2,p^4)$--chart, set
\begin{equation*}
q:=\sqrt{p^3p^4},\qquad
E_3:=\hor(e_3),\qquad
H_A:=e_A-\bGa_{AD}^Cp^D\pr_{p^C},\qquad A=1,2,
\end{equation*}
and define
\begin{equation*}
\VV_\pa:=\left\{qH_1,qH_2,p^3E_3\right\},\qquad
\VV_\perp:=\left\{q\pr_{p^1},q\pr_{p^2},p^4\pr_{p^4}\right\},
\end{equation*}
as well as
\begin{equation*}
\Dk:=
\left\{
e_1,e_2,e_3,
q\pr_{p^1},q\pr_{p^2},p^4\pr_{p^4}
\right\}.
\end{equation*}
We define
\begin{align*}
\TT_{f,\mathrm{low}}
:=&
\left\|
\left(
\VV_\perp f,\frac{\VV_\pa f}{p^3+q}
\right)
\right\|_{L^\infty_uL^\infty_\ub L^\infty(S;L^\infty(P_{x,m}))},\\
\TT_{f,4}:={}&
\left\|
\left(
\Dk^{\leq3}(\VV_\perp f),
(p^3+q)^{-1}\Dk^{\leq3}(\VV_\pa f)
\right)
\right\|_{L^\infty_uL^2_\ub L^2(S;L^2_p(p^3q^{-1}\dvol))}\\
&+\left\|
\left(
\Dk^{\leq3}(\VV_\perp f),
(p^3+q)^{-1}\Dk^{\leq3}(\VV_\pa f)
\right)
\right\|_{L^\infty_\ub L^2_uL^2(S;L^2_p(p^4q^{-1}\dvol))}.
\end{align*}
Moreover,
\begin{align*}
\TT_\S:=&
\sum_{\nu=1}^4
\left\|Z^{\leq4}(\S_{4\nu},\S_{AB})\right\|_{L^\infty_uL^2_\ub L^2(S)}
+\sum_{\nu=1}^4
\left\|Z^{\leq4}(\S_{3\nu},\S_{AB})\right\|_{L^\infty_\ub L^2_uL^2(S)},\\
\TT_J:=&
\left\|h(\ub)J_{3\bu4}^{(3)}\right\|_{L^2_\ub L^2_uL^2(S)}
+\left\|h(\ub)J_{4\bu3}^{(3)}\right\|_{L^2_\ub L^2_uL^2(S)}
+\left\|h(\ub)\dual J_{434}^{(3)}\right\|_{L^2_\ub L^2_uL^2(S)}\\
&+\left\|\dual J_{343}^{(3)}\right\|_{L^2_\ub L^2_uL^2(S)}
+\left\|h(\ub)J_{3\bu3}^{(3)}\right\|_{L^2_\ub L^2_uL^2(S)}
+\left\|h(\ub)\Jh_3^{(3)}\right\|_{L^2_\ub L^2_uL^2(S)}.
\end{align*}
Finally, set
\begin{equation*}
\TT_f:=\TT_{f,\mathrm{low}}+\TT_{f,4},
\qquad
\TT:=\TT_f+\TT_\S+\TT_J.
\end{equation*}

\subsubsection{Initial data}\label{initialO0}
We introduce the following norms on the initial hypersurfaces
$H_0^{(0,\de)}\cup \Hb_0^{(0,\de)}$:
\begin{align*}
\mo_{(0)}
:=&\sum_{i=0}^4\bigg(
\left\|h(\ub)^2Z^i(\chi,\om)\right\|_{L^\infty_\ub(0,\de)L^2(S_{0,\ub})}
+\left\|Z^i\ze\right\|_{L^\infty_\ub(0,\de)L^2(S_{0,\ub})}
+\left\|Z^i\trchb\right\|_{L^\infty_\ub(0,\de)L^2(S_{0,\ub})}\\
&+\left\|Z^i\chib\right\|_{L^\infty_u(0,\de)L^2(S_{u,0})}
+\left\|Z^i(\ze,\slg,\bbb)\right\|_{L^\infty_u(0,\de)L^2(S_{u,0})}
+\left\|Z^i\trch\right\|_{L^\infty_u(0,\de)L^2(S_{u,0})}
\bigg),
\end{align*}
and
\begin{align*}
\mr_{(0)}
:=\sum_{i=0}^3\bigg(
\left\|h(\ub)Z^i\b\right\|_{L^2(H_0^{(0,\de)})}
+\left\|Z^i(\K,\sic,\bb)\right\|_{L^2(H_0^{(0,\de)})}
+\left\|Z^i(\b,\K,\sic,\bb,\aa)\right\|_{L^2(\Hb_0^{(0,\de)})}
\bigg).
\end{align*}

For the initial distribution function, define
\begin{equation*}
\TT_{(0)}
:=\sum_{i=0}^4
\left\|\Dk^if\right\|_{L^\infty(H_0\cup\Hb_0;L^\infty(P_{x,m}))}.
\end{equation*}
We also assume compact momentum support and
\begin{equation*}
|p^\mu|\leq D
\qquad\text{on }\supp f\cap(H_0\cup\Hb_0).
\end{equation*}
\subsection{Auxiliary constants}
Let $D>0$ measure the initial data norms, $B>0$ the bootstrap norms, $\de>0$ the size of $V_f$, and $\ep>0$ the size of $\|h(\ub)^{-1}\|_{L^2_\ub(0,\ub_*)}$. We write $A_1\les A_2$ if $A_1\leq C(D)A_2$, where $C(D)>0$ depends only on $D$; $A_1=O(A_2)$ if $|A_1|\les|A_2|$; and $A_1\simeq A_2$ if $A_1\les A_2$ and $A_2\les A_1$. The notation $A_1\ll A_2$ means that $A_1$ is sufficiently small compared with $A_2$, with smallness depending only on the constants implicit in $\les$ throughout the proof.
We take $B$ sufficiently large and $\de$ and $\ep$ sufficiently small so that
\begin{equation}\label{epep0}
\de\leq e^{-[(2p-1)\ep^2]^{\frac{1}{1-2p}}}\ll\ep^2,\qquad\quad
\ep\ll B^{-20},\qquad\quad
1\leq D\ll B,
\end{equation}
where $p>\frac12$ is a fixed constant.
\subsection{Main intermediate results}\label{intermediate}
\begin{thm}\label{M1}
Assume that
\begin{equation}
    \mo_{(0)}\les 1,\qquad \mr_{(0)}\les 1,\qquad \TT_{(0)}\les 1,\qquad
    \mo\leq B,\qquad \mr\leq B,\qquad \TT\leq B.
\end{equation}
Then,
\begin{equation}
    \mr\les 1.
\end{equation}
\end{thm}
Theorem \ref{M1} is proved in Section \ref{secR}. The proof is based on energy estimates for paired Bianchi equations, following an argument similar to that of \cite{Luk18}.
\begin{thm}\label{M2}
Assume that
\begin{equation}
    \mo_{(0)}\les 1,\qquad \mr_{(0)}\les 1,\qquad \TT_{(0)}\les 1,\qquad
    \mo\leq B,\qquad \mr\les 1,\qquad \TT\leq B.
\end{equation}
Then,
\begin{equation}
    \mo\les 1.
\end{equation}
\end{thm}
Theorem \ref{M2} is proved in Section \ref{secO}. The proof is based on integrating the null structure equations along the outgoing and incoming null hypersurfaces.
\begin{thm}\label{M3}
Assume that
\begin{equation}
    \mo_{(0)}\les 1,\qquad \mr_{(0)}\les 1,\qquad \TT_{(0)}\les 1,\qquad
    \mo\les 1,\qquad \mr\les 1,\qquad \TT\leq B.
\end{equation}
Then,
\begin{equation}
    \TT\les 1.
\end{equation}
 
 \end{thm}
Theorem \ref{M3} is proved in Section \ref{secVlasov}. We first estimate the distribution function $f$ by integrating the Vlasov equation along its characteristics. The estimates for the Schouten and Cotton tensors then follow from the representation of the energy-momentum tensor in \eqref{eq:expressionTm-p4} and the definition \eqref{dfSchouten}.
\subsection{Proof of the main theorem}\label{proofmain}
\begin{proof}[Proof of Theorem~\ref{MainTheorem}]
Fix $0<\ub_1<\de$. The initial data are smooth on
$H_0\cap\{0\leq\ub\leq\ub_1\}$ and on $\Hb_0$. Theorems
\ref{TaylorCharacteristic} and \ref{LukCharacteristic} give a smooth
development containing $V(u_0,\ub_1)$ for some $u_0>0$.

Choose $B$ sufficiently large depending on $D$, and then choose
$\ep,\de$ as in \eqref{epep0}. Let $u_*$ be the supremum of
$u\leq\de$ for which the development is smooth on $V(u,\ub_1)$ and
satisfies
\begin{equation}\label{B2}
\mo\leq B,\qquad\quad\mr\leq B,\qquad\quad\TT\leq B.
\end{equation}
The initial bounds and continuity give $u_*>0$. Applying Theorems
\ref{M1}--\ref{M3} successively yields
\begin{equation*}
\mr\les1,\qquad\quad\mo\les1,\qquad\quad\TT\les1.
\end{equation*}
For $B$ sufficiently large, these estimates give
\begin{equation*}
\max\{\mo,\mr,\TT\}\leq\frac B2.
\end{equation*}
This bound is uniform in $u<u_*$ and $\ub_1<\de$. Following the continuation argument in \cite[Theorem~4]{Song} and using the classical local existence results in Theorems~\ref{Ch71Cauchy} and~\ref{TaylorCharacteristic}, we obtain a smooth development throughout $0<u<\de$, $0<\ub<\de$.
\end{proof}
\subsection{Nature of the singular boundary}\label{ssecconclusion}
\begin{proof}[Proof of Theorem \ref{Luk3}]
Commuting \eqref{basicnull} and the metric equations,
and using \eqref{nullidentities}, \eqref{gauss},
Propositions~\ref{commutation},
\ref{estmetricangular}, \ref{estGamma},
and Remark~\ref{Gawrk}, we obtain
\begin{align*}
\pr_\ub\pr_\th^{\leq4}(\GG_0,\GG_{-1})&\in\Ga_1^{(4)}+\Ga_0^{(4)},&
\pr_\ub Z^{\leq3}(\Ga_0,\Ga_{-1})&\in\Ga_1^{(4)}+\Ga_0^{(4)}.
\end{align*}
We compare components on different spheres using the transported angular coordinates; the corresponding $L^2$ norms are uniformly equivalent.
For $F\in\pr_\th^{\leq4}(\GG_0,\GG_{-1})\cup Z^{\leq3}(\Ga_0,\Ga_{-1})$, Theorems \ref{M1}--\ref{M3} and Cauchy--Schwarz imply
\begin{align*}
\sup_u\|F(u,\ub_2)-F(u,\ub_1)\|_{L^2(S)}
&\les\sup_u\int_{\ub_1}^{\ub_2}\left(\|\Ga_1^{(4)}\|_{L^2(S)}+\|\Ga_0^{(4)}\|_{L^2(S)}\right)d\ub\\
&\les\|h^{-1}\|_{L^2(\ub_1,\ub_2)}+(\ub_2-\ub_1)^{\frac12}\longrightarrow0
\end{align*}
as $\ub_1,\ub_2\to\de$, since $h^{-1}\in L^2(0,\de)$.
These estimates give the stated traces and tangential bounds. Applying Proposition \ref{sobolev} to differences of the metric components gives uniform convergence to continuous boundary values. Uniform nondegeneracy then gives a continuous Lorentzian extension of the metric. Passing to the limit in the integrated metric equation yields
\begin{equation*}
(\pr_u+\mathcal L_{\bbb})\slg=2\Om^2\chib\qquad\mbox{on }\Hb_\de.
\end{equation*}
For $f$, the equation $Xf=0$, Lemma \ref{Gaexpression}, and Proposition \ref{Lowerorderf} give
\begin{align}
p^4e_4f&=-p^3e_3f-p^Ae_Af+p^\mu p^\nu\bGa_{\mu\nu}^{k'}\pr_{p^{k'}}f,\qquad k'=1,2,4,\notag\\
|p^4e_4f|&\les(p^3+q)(1+|\GG^{(1)}|)+p^4|\Ga_1|\les1+p^4|\Ga_1|.\label{boundary:weighted-e4f}
\end{align}
Indeed,
\begin{align*}
qe_Af&=qH_Af+q\GG^{(1)}\c\VV_\perp f,\\
p^3e_3f&=p^3E_3f+(p^3+q)\GG^{(1)}\c\VV_\perp f.
\end{align*}
The remaining vertical terms have the weights $p^3,q,p^4$ of $\CC_1$. At fixed frame momentum, on any compact set with $p^4\geq c>0$, sphere Sobolev and Cauchy--Schwarz give
\begin{equation*}
\sup_{u,S,p}|f(u,\ub_2,p)-f(u,\ub_1,p)|
\les_c(\ub_2-\ub_1)+\|h^{-1}\|_{L^2(\ub_1,\ub_2)}\longrightarrow0.
\end{equation*}
Thus $f$ has a continuous trace on these sets. Every point with $p^4>0$ has a neighborhood on which $p^4$ is bounded away from zero, so these local traces agree on overlaps and define a continuous trace throughout the $p^4$--chart, including the massless axis $p^3=0$, $p^4>0$. No uniform positive lower bound for $p^4$ on the support is required for this local conclusion.
For $m>0$, Proposition \ref{pmuest} and $4p^3p^4\geq m^2$ bound $p^4$ away from zero on the support, giving a trace on the whole shell. The bound $|f|\leq\|f_0\|_\infty$, bounded momentum support, and \eqref{Schouten=T} also give
\begin{equation*}
|\T|+|\Ric|\les\|f_0\|_\infty\int_{\{0<p^3,p^4\leq C(D)\}}(p^3+p^4)^2\,\dvol\les1.
\end{equation*}
The massless axes have zero fiber measure. The continuous trace of $f$ for almost every frame momentum, its uniform boundedness, and its uniformly bounded momentum support then imply, by dominated convergence, that $\T$ and $\Ric$ extend continuously to these boundaries.

The continuation argument in Section~\ref{proofmain} also gives a smooth extension across $H_\de\cap\{\ub<\de\}$.
\end{proof}
Let $\Phi_u:H_0\to H_u$ be the flow of $\Om^2e_3=\pr_u+\bbb$. It preserves $\ub$, and
\begin{equation*}
\frac{d}{du}(F\circ\Phi_u)=(\Om^2e_3F)\circ\Phi_u
\end{equation*}
for scalar functions $F$.
\begin{lem}\label{hchinfty}
Suppose that the assumptions of Theorem \ref{MainTheorem} hold. If
\begin{equation*}
\int_0^\de|\hch(\ub)|^2\big|_\gamma\,d\ub=\infty
\end{equation*}
along an outgoing null generator $\gamma$ of $H_0$, then, for every $0\leq u<\de$,
\begin{equation*}
\int_0^\de\left(|\trch(\ub)|^2+|\hch(\ub)|^2\right)\big|_{\Phi_u(\gamma)}\,d\ub=\infty.
\end{equation*}
\end{lem}
\begin{proof}
Fix $u<\de$. If $\trch|_{\Phi_u(\gamma)}\notin L^2_\ub(0,\de)$, then the conclusion follows immediately. We may therefore assume that
$\trch|_{\Phi_u(\gamma)}\in L^2_\ub(0,\de)$. By Theorems \ref{M1}--\ref{M3} and sphere Sobolev,
\begin{equation*}
\Om+\Om^{-1}+|\K|+|\eta|+|\nab\eta|+|\chib|
+|\Ssl|+|\S_{34}|\les1.
\end{equation*}
The Gauss equation gives
\begin{equation*}
2\rho-\hch\c\hchb
=-2\K-\frac12\trch\trchb+\tr\Ssl.
\end{equation*}
Substitution in \eqref{basicnull} yields
\begin{equation*}
\nab_3\trch+\trchb\,\trch
=-2\K+2\sdiv\eta+2|\eta|^2+\tr\Ssl+\Tr\S.
\end{equation*}
Along the flow, set
\begin{equation*}
\lambda(u',\ub):=\exp\left(\frac12\int_0^{u'}\Om^2\trchb\big|_{\Phi_s(\gamma)}\,ds\right)\simeq1.
\end{equation*}
Multiplying the trace equation by $\lambda^2$, using
$\frac{d}{du}=\Om^2\nab_3$ along the flow, and integrating between
$u'$ and $u$ gives
\begin{equation*}
|\trch(\ub)|\big|_{\Phi_{u'}(\gamma)}
\les
|\trch(\ub)|\big|_{\Phi_u(\gamma)}+1,
\qquad 0\leq u'\leq u.
\end{equation*}
Choose a $\nab_3$--parallel orthonormal frame along each integral curve of the flow. The shear equation in \eqref{basicnull} then gives
\begin{align*}
\lambda(u,\ub)\hch\big|_{\Phi_u(\gamma)}
&=\hch\big|_\gamma
+\int_0^u\lambda\Om^2
\left(
\nab\hot\eta-\frac12\trch\,\hchb
+\eta\hot\eta+\Sslh
\right)\big|_{\Phi_s(\gamma)}\,ds.
\end{align*}
Using the preceding pointwise bounds and the estimate for $\trch$ along the flow, we obtain
\begin{equation*}
|\hch|\big|_{\Phi_u(\gamma)}
\geq
c|\hch|\big|_\gamma
-C\left(1+|\trch|\big|_{\Phi_u(\gamma)}\right).
\end{equation*}
Consequently, for every $\ub_0<\de$,
\begin{align*}
\|\hch\|_{L^2_\ub(0,\ub_0;\Phi_u(\gamma))}
\geq{}&
c\|\hch\|_{L^2_\ub(0,\ub_0;\gamma)}
-C\left(
\sqrt{\ub_0}
+\|\trch\|_{L^2_\ub(0,\ub_0;\Phi_u(\gamma))}
\right).
\end{align*}
The first term on the right-hand side tends to infinity as
$\ub_0\to\de$, whereas the remaining terms stay bounded. This proves the claim.
\end{proof}

\begin{lem}\label{L2infty}
Under the assumptions of Theorem \ref{Singularity}, for every $p\in\Hb_\de$ and every neighborhood $\NN$ of $p$,
\begin{equation*}
\int_{\NN\cap V_f}|\chi|^2\,d\mathrm{Vol}_{\g}=\infty.
\end{equation*}
\end{lem}
\begin{proof}
The bounds for $\bbb$, $\pr_\th\bbb$, and
\begin{equation*}
e_4\bbb=-2\Om^2(\eta-\etab)
\end{equation*}
imply that $\Phi_u$ extends continuously to $\ub=\de$. Moreover, the angular Jacobian of $\Phi_u$ is bounded above and away from zero. Indeed, its angular differential satisfies an ordinary differential equation whose coefficients are controlled by $\pr_\th\bbb$.

Consequently, every neighborhood $\NN$ of $p$ contains a flow box of the form
\begin{equation*}
\left\{\Phi_{u'}(0,\ub',\phi):(u',\ub',\phi)\in I\times(\de-\nu,\de)\times\mathcal O\right\},
\end{equation*}
where $\nu>0$, $I\subset[0,\de)$ is an interval of positive length, and $\mathcal O$ is a nonempty open angular set.

By the assumption of Theorem \ref{Singularity}, the initial shear fails to be square integrable along almost every generator corresponding to $\phi\in\mathcal O$. Lemma \ref{hchinfty} therefore gives, for every $u'\in I$ and almost every $\phi\in\mathcal O$,
\begin{equation*}
\int_0^\de
\left(|\trch|^2+|\hch|^2\right)
\bigl(\Phi_{u'}(0,\ub',\phi)\bigr)\,d\ub'
=\infty.
\end{equation*}
Since the solution is smooth for $\ub'<\de$, the integral over every compact subinterval of $[0,\de)$ is finite. Hence,
\begin{equation*}
\int_{\de-\nu}^\de
\left(|\trch|^2+|\hch|^2\right)
\bigl(\Phi_{u'}(0,\ub',\phi)\bigr)\,d\ub'
=\infty
\end{equation*}
for every $u'\in I$ and almost every $\phi\in\mathcal O$.

The uniform nondegeneracy of the metric and the bounds for the angular Jacobian imply that the pullback of the spacetime volume form under the flow coordinates is comparable to
\begin{equation*}
du'\,d\ub'\,d\phi.
\end{equation*}
Since
\begin{equation*}
|\chi|^2=\frac12|\trch|^2+|\hch|^2,
\end{equation*}
Tonelli's theorem gives
\begin{align*}
\int_{\NN\cap V_f}|\chi|^2\,d\mathrm{Vol}_{\g}
&\ges
\int_I\int_{\mathcal O}\int_{\de-\nu}^\de
\left(|\trch|^2+|\hch|^2\right)
\bigl(\Phi_{u'}(0,\ub',\phi)\bigr)
\,d\ub'\,d\phi\,du'=\infty.
\end{align*}
\end{proof}
\begin{proof}[Proof of Theorem \ref{Singularity}]
For horizontal vector fields, \eqref{ricciformulas} gives
\begin{equation*}
\D_XY=\nab_XY+\frac12\chib(X,Y)e_4+\frac12\chi(X,Y)e_3.
\end{equation*}
Thus, in the original double null coordinates,
\begin{equation*}
\Ga_{AB}^u=\frac12\Om^{-2}\chi_{AB},\qquad
\sum_{A,B=1}^2|\Ga_{AB}^u|^2\simeq|\chi|^2,
\end{equation*}
where $A,B$ are coordinate indices and the equivalence follows from
the uniform nondegeneracy of the metric. Lemma \ref{L2infty} gives
the claimed failure of local square integrability in these coordinates.
Together with the continuous metric extension, this proves the first
assertion.

For the last assertion, Theorems \ref{M1}--\ref{M3},
Proposition \ref{estGamma}, sphere Sobolev, and the matter bounds
in the proof of Theorem \ref{Luk3} give
\begin{equation*}
|\Ga_0|+|\Ga_{-1}|+|\nab\eta|+|\K|+|\sic|+|\S|\les1.
\end{equation*}
The Gauss equation and the definition of $\sic$ imply
\begin{equation*}
|\rho|+|\si|\les1+|\Ga_1|.
\end{equation*}
Consequently,
\begin{equation*}
\|h\Ga_1\|_{L^2_\ub L^\infty_uL^\infty(S)}
+\|h(\rho,\si)\|_{L^2_\ub L^\infty_uL^\infty(S)}
\les1.
\end{equation*}
To estimate $\Jh_4$, use \eqref{boundary:weighted-e4f} and
$|qe_Af|\les p^3+q$ from its proof. Since
$|p^Ap^B|\leq4p^3p^4$, $|p^A|\leq2q$, and $p_4=-2p^3$,
\eqref{vlasov:moment-bound} gives
\begin{align*}
|e_4\T_{AB}|
&\les\int_{\supp f}p^3\left(1+p^4|\Ga_1|\right)\dvol
\les1+|\Ga_1|,\\
|e_A\T_{B4}|
&\les\int_{\supp f}p^3(p^3+q)\dvol\les1.
\end{align*}
The derivatives are taken at fixed frame momentum; the moment bounds
justify removal of the momentum cutoffs at the massless axes.
By \eqref{Schouten=T}, \eqref{dfSchouten}, and metric compatibility,
\begin{equation}\label{boundary-Jh4}
(\Jh_4)_{AB}
=\left[
\frac12(\D_A\T_{B4}+\D_B\T_{A4})-\D_4\T_{AB}
\right]^{\mathrm{tf}},
\qquad
|\Jh_4|\les1+|\Ga_1|.
\end{equation}
Here $\mathrm{tf}$ denotes the angular trace-free part, which removes
the trace term in $\S-\T$. The connection terms are bounded by
$(1+|\Ga_1|)|\T|$. The estimate holds for both $m>0$ and $m=0$.

To integrate the equation for $\a$, transport the initial angular
frame in the curvature assumptions by $\nab_3e_A=0$ along $\Phi_u$.
In the original metric-dependent frame, the proof of
Proposition \ref{estGamma} expresses $\bGa_{3A}^B$ as a smooth
combination of $\log\Om$, $\slg$, $\chib$, and $\pr_\th\bbb$.
The traces established in the proof of Theorem \ref{Luk3}, together
with sphere Sobolev, therefore give continuous limits for these
coefficients at $\ub=\de$. Since the initial frame extends
continuously to $\ub=\de$, continuous dependence for the frame
rotation equation along $\Phi_u$ shows that the transported frame
also extends continuously to this boundary.

Set
\begin{equation*}
\lambda(u,\ub,\th)
:=\exp\left(
\frac12\int_0^u
\Om^2\trchb(s,\ub,\Phi_s\th)\,ds
\right)\simeq1.
\end{equation*}
In the transported frame, Proposition \ref{bianchiequations} yields
\begin{align*}
\lambda\a_{AB}(u,\ub,\Phi_u\th)
&=\a_{AB}(0,\ub,\th)+\mathcal E_{AB},\\
\mathcal E_{AB}
&:=\int_0^u\lambda\Om^2
\left[
\nab\hot\b-3(\hch\rho+\dual\hch\si)
+(\ze+4\eta)\hot\b+\Jh_4
\right]_{AB}(s,\ub,\Phi_s\th)\,ds.
\end{align*}
In a connected patch satisfying the initial curvature assumptions,
choose a compact rectangle $U$ and a compact interval
$I\Subset(0,\de)$, both with nonempty interiors.
The constants below may also depend on this patch.
For $\ub_0<\ub'<\de$, set
\begin{align*}
V_{\ub'}
&:=\{(u,\ub,\Phi_u\th):
(u,\ub,\th)\in I\times[\ub_0,\ub']\times U\},& w(\ub)&:=(\de-\ub)^{-1}h(\ub)^{-2}.
\end{align*}
Let $V_\de$ denote the closure of
$\bigcup_{\ub'<\de}V_{\ub'}$ in the continuous extension.

By Proposition \ref{sobolev} and Theorem \ref{M1},
\begin{align*}
\int_0^\de\!\int_0^\de
\|\nab\b\|_{L^\infty(S)}\,du\,d\ub
&\les
\|h^{-1}\|_{L^2_\ub}
\int_0^\de
\|h\nab^{\leq3}\b\|_{L^2_\ub L^2(S)}\,du
\les\de\ep\mr\les 1.
\end{align*}
The terms $(\ze+4\eta)\hot\b$ and $\Jh_4$ are also integrable,
by the curvature flux bounds and \eqref{boundary-Jh4}.
Moreover,
\begin{equation*}
\hch\rho+\dual\hch\si\in\Ga_1\c\Ga_1.
\end{equation*}
Using the uniform bounds for $\lambda$, $\Om$, and the angular
Jacobian of $\Phi_u$, we obtain
\begin{equation*}
\int_{V_{\ub'}}|\mathcal E|\,d\mathrm{Vol}_{\g}
\les 1+\int_{\ub_0}^{\ub'}
\|\Ga_1\|_{L^\infty_u L^\infty(S)}^2\,d\ub
=o\left(h(\ub')^{-2}\right).
\end{equation*}
Indeed, the weighted estimates and sphere Sobolev give
$h\|\Ga_1\|_{L^\infty_u L^\infty(S)}\in L^2_\ub$.
Since $h'\leq 0$, we have for
$\ub_0<r<\ub'<\de$ sufficiently close to $\de$,
\begin{align*}
h(\ub')^2\int_{\ub_0}^{\ub'}
\|\Ga_1\|_{L^\infty_u L^\infty(S)}^2\,d\ub
&\leq h(\ub')^2\int_{\ub_0}^{r}
\|\Ga_1\|_{L^\infty_u L^\infty(S)}^2\,d\ub+\int_r^\de h(\ub)^2
\|\Ga_1\|_{L^\infty_u L^\infty(S)}^2\,d\ub.
\end{align*}
Letting first $\ub'\to\de$ and then $r\to\de$ proves the claim.

For $\ub_0$ sufficiently close to $\de$,
$\a_{11}(0,\ub,\th)$ is nonzero and has constant sign on the
connected patch. Choose $\varsigma\in\{-1,1\}$ with
$\varsigma\a_{11}>0$. The hypotheses and $\a_{22}=-\a_{11}$ give
\begin{equation*}
\varsigma\a_{11}(0,\ub,\th)\ges w(\ub),\qquad
|\a(0,\ub,\th)|\les\varsigma\a_{11}(0,\ub,\th).
\end{equation*}
Define
\begin{equation*}
A_{\ub'}
:=\int_{V_{\ub'}}
\varsigma\lambda^{-1}\a_{11}(0,\ub,\th)
\,d\mathrm{Vol}_{\g},
\end{equation*}
where the initial curvature component is evaluated using
the flow labels $(\ub,\th)$.
The initial curvature lower bound and the choice of weight
in Remark~\ref{rkchoicef} give
\begin{equation*}
A_{\ub'}
\ges\int_{\ub_0}^{\ub'}w(\ub)\,d\ub
\simeq h(\ub')^{-2}\longrightarrow\infty
\qquad\text{as }\ub'\to\de.
\end{equation*}
The preceding error estimate is therefore $o(A_{\ub'})$.
Using the transport identity and the initial component bounds,
we conclude that
\begin{align*}
\int_{V_{\ub'}}\varsigma\a_{11}\,d\mathrm{Vol}_{\g}
&=A_{\ub'}+o(A_{\ub'}),\\
\int_{V_{\ub'}}|\a|\,d\mathrm{Vol}_{\g}
&\les A_{\ub'}+o(A_{\ub'}).
\end{align*}
Theorems \ref{M1}--\ref{M3}, the Gauss equation, and
$h^{-1}\in L^2_\ub$ give
$\b,\rho,\si,\bb,\aa,\S\in L^1(V_\de)$.
Since
\begin{equation*}
\R_{4A4B}=\a_{AB}+\frac12\de_{AB}\S_{44},
\end{equation*}
multilinearity and Lemma \ref{Rexpression} give, for all continuous
vector fields $X_i$ on $V_\de$ satisfying
\begin{equation*}
\sup_{V_\de}
|(X_1,X_2,X_3,X_4)-(e_4,e_1,e_4,e_1)|<\ep_0
\end{equation*}
in a fixed coordinate norm,
\begin{align*}
|\R(X_1,X_2,X_3,X_4)-\a_{11}|
\les
\ep_0|\a|
+|\b|+|\rho|+|\si|+|\bb|+|\aa|+|\S|.
\end{align*}
Thus, we obtain for $C\ep_0<\frac{1}{2}$
\begin{align*}
\varsigma\int_{V_{\ub'}}
\R(X_1,X_2,X_3,X_4)\,d\mathrm{Vol}_{\g}
&\geq
(1-C\ep_0)A_{\ub'}-o(A_{\ub'})-C
\longrightarrow\infty.
\end{align*}
To apply \cite{Sbierski26Lipschitz}, set
$\th=\Phi_u(\ub,\phi)$ and $\Phi_0(\ub,\phi)=\phi$.
Proposition \ref{metricequations} gives, along $\Phi_u$,
\begin{align*}
\pr_u\pr_\ub\Phi_u
&=(\pr_\th\bbb)\pr_\ub\Phi_u
-2\Om^2(\eta-\etab),&
\bbb'
&=-(\pr_\phi\Phi_u)^{-1}\pr_\ub\Phi_u.
\end{align*}
Here the right-hand side of the first equation is evaluated along
$\Phi_u$, and the components of $\eta-\etab$ are raised using
the coordinate sphere metric.
The metric and Ricci coefficient estimates, together with the
differentiated flow equations, give
\begin{equation*}
|\pr_\phi\Phi_u|
+|(\pr_\phi\Phi_u)^{-1}|
+|\pr_\ub\Phi_u|
+|\pr_\phi\bbb'|\les1.
\end{equation*}
The boundary traces also give a $C^1$ extension of this coordinate
change with invertible differential at $\ub=\de$. Rescale $(e'_3,e'_4)=(\Om^2e_3,\Om^{-2}e_4)$.
In the new coordinates,
\begin{equation*}
e'_3=\pr_u,\qquad
e'_4=\Om^{-2}\left(\pr_\ub+\bbb'{}^A\pr_{\phi^A}\right).
\end{equation*}
The Ricci coefficients satisfy
\begin{align*}
(\chib',\chi')
&=(\Om^2\chib,\Om^{-2}\chi),&
(\eta',\etab')&=(\eta,\etab),&
\omb'&=-\Om^2e_3\log\Om.
\end{align*}
Propositions \ref{estmetricangular} and \ref{estGamma}, together
with the preceding bounds and $\int_0^\de |\Ga_1|\,d\ub<\infty$,
imply
\begin{align*}
|\omb'|+|\eta'|+|\etab'|+|\chib'|&\les1,&
\int_0^\de\sup_{u,S}
\left(|\chi'|+|\pr_\phi\bbb'|\right)d\ub&<\infty.
\end{align*}
These bounds and the continuous metric extension verify the
geometric hypotheses of
\cite[Proposition 3.6, Theorem 3.14]{Sbierski26Lipschitz}.
The curvature-integral divergence established above is preserved
under this coordinate change and gives the required blow-up
condition. Finally, the density assumption and the boundary flow
map supply such a patch in every boundary neighborhood.
This proves the last assertion.
\end{proof}
\section{Bootstrap assumptions and first consequences}\label{secboot}
In the rest of the paper, we always make the following bootstrap assumptions:
\begin{equation}\label{B1}
    \mo\leq B,\qquad\quad\mr\leq B,\qquad\quad\TT\leq B.
\end{equation}
\begin{lem}\label{decayGagGabGaa}
Under the bootstrap assumptions \eqref{B1}, we have
\begin{align*}
\left\|h(\ub)\Ga_1^{(3)}\right\|_{L^2_\ub L^\infty_u L^2(S)}
&\les B,\qquad
\left\|\Ga_0^{(3)}\right\|_{L^\infty_u L^\infty_\ub L^2(S)}
\les B,\qquad
\left\|\Ga_{-1}^{(3)}\right\|_{L^\infty_u L^\infty_\ub L^2(S)}
\les B,\\
\left\|\Ga_{-2}^{(3)}\right\|_{L^2_u L^\infty_\ub L^2(S)}
&\les B,\qquad\;
\left\|\Ga_{-2}^{(4)}\right\|_{L^\infty_\ub L^2_u L^2(S)}
\les B,\\
\left\|h(\ub)\Ga_0^{(4)}\right\|_{L^\infty_\ub L^2_uL^2(S)}
&\les B,\qquad\;
\left\|\Ga_0^{(4)}\right\|_{L^\infty_u L^2_\ub L^2(S)}
\les B,\\
\left\|h(\ub)\Ga_1^{(4)}\right\|_{L^\infty_u L^2_\ub L^2(S)}
&\les B,\qquad\;
\left\|\Ga_{-1}^{(4)}\right\|_{L^\infty_\ub L^2_uL^2(S)}
\les B.
\end{align*}
\end{lem}
\begin{proof}
This follows directly from Definition \ref{gammag}, the definitions of
$\mo$, $\mr$, and $\TT$, and the bootstrap assumptions \eqref{B1}.
\end{proof}
\begin{rk}\label{Gawrk}
In the sequel, we use the following conventions:
\begin{itemize}
\item For a quantity $F$ satisfying the same or better decay and regularity as
$\Ga_i^{(q)}$, we write
\begin{equation*}
    F\in\Ga_i^{(q)}.
\end{equation*}
Here $i=-1,0,1$ and $q\geq0$, or $i=-2$ and $q\geq1$.
\item In a sum of schematic terms, we omit terms with the same or better decay
and regularity. For example, we write
\begin{equation}\label{simplerGamma}
\Ga_1\c\Ga_{-1}+\Ga_0\c\Ga_0=\Ga_1\c\Ga_{-1},\qquad \Ga_{-1}+\Ga_0=\Ga_0,\qquad \GG\c\Ga_k=B\Ga_k.
\end{equation}
\item The Schouten components are not included in the definition of the
classes $\Ga_k$. Nevertheless, whenever only their schematic weight
and regularity are relevant, we use the conventions
\begin{equation*}
\Ssl_4\in\Ga_1,\qquad
(\S_{34},\Ssl)\in\Ga_0,\qquad
\Ssl_3\in\Ga_{-1},\qquad \S_{33}\in\Ga_{-2}
\end{equation*}
and similarly for their commuted versions. Their norms are always
controlled by $\TT_\S$ rather than by $\mo$.
\end{itemize}
\end{rk}
The following consequences of \eqref{B1} will be used frequently throughout the paper.
\begin{lem}\label{evolution}
For any $S$--tangent tensor field $\phi$ and $p\in[2,\infty]$, we have
\begin{align*}
\|\phi\|_{L^p(S_{u,\ub})}
&\les\|\phi\|_{L^p(S_{u,0})}
+\int_0^\ub\|\nab_4\phi\|_{L^p(S_{u,\ub'})}\,d\ub',\\
\|\phi\|_{L^p(S_{u,\ub})}
&\les\|\phi\|_{L^p(S_{0,\ub})}
+\int_0^u\|\Om^2\nab_3\phi\|_{L^p(S_{u',\ub})}\,du'.
\end{align*}
\end{lem}
\begin{proof}
The estimates follow from the standard transport inequalities along the null
hypersurfaces; see Propositions 5 and 6 in \cite{Luk18}. Notice that in the
present gauge $\pr_u+\bbb=\Om^2e_3$.
\end{proof}
\begin{prop}\label{sobolev}
We have
\begin{equation*}
    \|\phi\|_{L^\infty(S_{u,\ub})}
    \les\sum_{i=0}^2\|\nab^i\phi\|_{L^2(S_{u,\ub})}.
\end{equation*}
\end{prop}
\begin{proof}
See Proposition 7 in \cite{Luk18}.
\end{proof}
\begin{prop}\label{commutation}
We have the following schematic commutator formulae:
\begin{align*}
[\nab_4,\nab]
&=\Ga_1\c\nab+B\Ga_1^{(1)},\\
[\nab_3,\nab]
&=\Ga_{-1}\c\nab+\Ga_0\c\nab_3+B\Ga_{-1}^{(1)},\\
[\nab_3,\nab_4]
&=\Ga_1\c\nab_3+\Ga_0\c\nab
+\Ga_1\c\Ga_{-1}+B\Ga_0^{(1)},\\
[\nab_A,\nab_B]
&=B\Ga_0^{(1)}.
\end{align*}
\end{prop}
\begin{proof}
By Lemma \ref{comm}, Definition \ref{gammag}, and \eqref{simplerGamma}, we have schematically
\begin{align*}
[\nab_4,\nab]
&=\Ga_1\c\nab+\Ga_0\c\Ga_1+\Ga_1^{(1)}
=\Ga_1\c\nab+B\Ga_1^{(1)},\\
[\nab_3,\nab]
&=\Ga_{-1}\c\nab+\Ga_0\c\nab_3
+\Ga_0\c\Ga_{-1}+\Ga_{-1}^{(1)}
=\Ga_{-1}\c\nab+\Ga_0\c\nab_3+B\Ga_{-1}^{(1)}.
\end{align*}
For the third commutator, we use
\begin{equation*}
\R_{AB34}=2\si\slin_{AB},
\qquad
\si=\sic+\frac12\hch\wedge\hchb,
\end{equation*}
and hence schematically
\begin{equation*}
\R_{AB34}
=\Ga_0^{(1)}+\Ga_1\c\Ga_{-1}.
\end{equation*}
Therefore,
\begin{align*}
[\nab_3,\nab_4]
&=\Ga_1\c\nab_3+\Ga_0\c\nab
+\Ga_0\c\Ga_0
+\Ga_1\c\Ga_{-1}
+\Ga_0^{(1)}\\
&=\Ga_1\c\nab_3+\Ga_0\c\nab
+\Ga_1\c\Ga_{-1}+B\Ga_0^{(1)}.
\end{align*}
Finally, the last identity follows from the Gauss equation on $S_{u,\ub}$ and the fact that $\K\in\Ga_0^{(1)}$.
\end{proof}
\begin{prop}\label{nullrenor}
We define the renormalized lapse quantity, its auxiliary potential, and the mass aspect functions by
\begin{align}
\vkp&:=d_1^*(-\om,\omd)-\b,\label{renorvkp}\\
\nabs_3\omd&=\si,\qquad \omd|_{H_0}=0,\label{dfomd}\\
\mu&:=-\sdiv\eta+\K,\qquad \mub:=-\sdiv\etab+\K.\label{murenor}
\end{align}
Then we have
\begin{align*}
\nab_4\mu
&=(\Ga_1\c\Ga_0)^{(1)}
+\Ga_1\c\Ga_0\c\Ga_0
+\sdiv\Ssl_4+\Ga_0\c\Ssl_4,\\
\nab_3\mub
&=(\Ga_{-1}\c\Ga_0)^{(1)}
+\Ga_{-1}\c\Ga_0\c\Ga_0
+\sdiv\Ssl_3+\Ga_0\c\Ssl_3,\\
\nab_3\vkp
&=(\Ga_1\c\Ga_{-1})^{(1)}
+\Ga_0\c\Ga_0^{(1)}
+\nab^{\leq 1}(\Ssl,\S_{34})-J_{3\bu4},
\end{align*}
where $\omd\in\Ga_1$ is included in the bootstrap norms. Its lower order bounds follow from \eqref{dfomd}, and its top order bounds follow from \eqref{renorvkp} and Proposition \ref{ellipticLp}, together with those of $\om$.
\end{prop}
\begin{proof}
The equations for $\mu$ and $\mub$ follow from Proposition \ref{nulles},
the Gauss equation, and Definition \ref{gammag}. The equation for $\vkp$
follows from the lapse equation for $\om$ in Proposition \ref{nulles},
the Bianchi equation for $\b$ in Proposition \ref{bianchiequations},
and the definition \eqref{dfomd}.
\end{proof}
\begin{lem}\label{initialtracesrepair}
Under the initial data assumptions of Theorem~\ref{MainTheorem}, we have
\begin{align*}
\left\|\GG^{(4)}\right\|_{L^\infty(H_0\cup\Hb_0;L^2(S))}
+\left\|Z^{\leq3}(\eta,\chib)\right\|_{L^\infty_\ub L^2(S_{0,\ub})}
+\left\|Z^{\leq3}\nab\eta\right\|_{L^2(H_0)}
&\les1,\\
\left\|Z^{\leq3}\mu\right\|_{L^2(H_0)}
+\left\|h(\ub)Z^{\leq3}\mu\right\|_{L^\infty_\ub L^2(S_{0,\ub})}
+\left\|Z^{\leq3}\mub\right\|_{L^\infty_\ub L^2(S_{0,\ub})}
+\left\|h(\ub)Z^{\leq3}\vkp\right\|_{L^2(H_0)}
&\les1,\\
\left\|Z^{\leq3}(\mu,\mub)\right\|_{L^\infty_uL^2(S_{u,0})}
&\les1.
\end{align*}
\end{lem}
\begin{proof}
These estimates follow from the initial data assumptions, the null
constraint and transport equations restricted to $H_0\cup\Hb_0$,
and the definitions of $\mu$, $\mub$, and $\vkp$. In particular,
the gauge identities on the initial hypersurfaces control $\eta$,
while the null structure and metric equations give the required initial
bounds for $\chib$ and $\GG^{(4)}$. The remaining estimates follow
from \eqref{murenor}, \eqref{renorvkp}, and Proposition~\ref{ellipticLp}.
\end{proof}
\section{Curvature estimates}\label{secR}
In this section, we prove Theorem \ref{M1} using weighted energy estimates for Bianchi pairs. Similar arguments have been used in \cite{An,AnLuk,holzegel,Luk18,Shen22,Shen23,Shen24,ShenWan}.
\subsection{Estimates for general Bianchi pairs}
The following proposition provides a general weighted estimate for Bianchi pairs and will be used repeatedly throughout this section. See Lemma 4.2 and Proposition 4.4 in \cite{Shen22} for an analogous version.
\begin{prop}\label{keyintegral}
Let $w\in C^1([0,\de))$ be positive with $w'\leq0$, and let $k\in\{1,2\}$. Assume that $(\psi_{(1)},\psi_{(2)})$ satisfies
\begin{align}
\begin{split}\label{bianchi1}
\nab_3\psi_{(1)}&=k d_k^*\psi_{(2)}+N_{(1)},\\
\nab_4\psi_{(2)}&=-d_k\psi_{(1)}+N_{(2)},
\end{split}
\end{align}
or
\begin{align}
\begin{split}\label{bianchi2}
\nab_3\psi_{(1)}&=d_k\psi_{(2)}+N_{(1)},\\
\nab_4\psi_{(2)}&=-k d_k^*\psi_{(1)}+N_{(2)}.
\end{split}
\end{align}
Then
\begin{align}
\begin{split}\label{rpest}
&\int_\cuv w(\ub)|\psi_{(1)}|^2
+\int_\ucuv w(\ub)|\psi_{(2)}|^2\\
\les{}&
\int_{\cuvi}w(\ub)|\psi_{(1)}|^2
+\int_{\ucuvi}w(0)|\psi_{(2)}|^2\\
&+\int_Vw(\ub)\left(
N_{(1)}\c\psi_{(1)}+N_{(2)}\c\psi_{(2)}
+|\psi_{(1)}|^2\c\Ga_{-1}
+|\psi_{(2)}|^2\c\Ga_1
+\Ga_0\c\psi_{(1)}\c\psi_{(2)}
\right).
\end{split}
\end{align}
\end{prop}
\begin{proof}
Recall that, in the present gauge $\omb=0$,
\begin{equation*}
\bdiv e_4=\trch-2\om\in\Ga_1,\qquad
\bdiv e_3=\trchb\in\Ga_{-1},\qquad
e_3(\ub)=0,\qquad e_4(\ub)=1.
\end{equation*}
Suppose first that $(\psi_{(1)},\psi_{(2)})$ satisfies \eqref{bianchi1}. Substituting the equations gives
\begin{align*}
&\bdiv\left(w(\ub)|\psi_{(1)}|^2e_3\right)
+k\bdiv\left(w(\ub)|\psi_{(2)}|^2e_4\right)\\
={}&2kw(\ub)\left(
\psi_{(1)}\c d_k^*\psi_{(2)}
-\psi_{(2)}\c d_k\psi_{(1)}
\right)
+kw'(\ub)|\psi_{(2)}|^2\\
&+2w(\ub)\psi_{(1)}\c N_{(1)}
+2kw(\ub)\psi_{(2)}\c N_{(2)}
+w(\ub)|\psi_{(1)}|^2\c\Ga_{-1}
+w(\ub)|\psi_{(2)}|^2\c\Ga_1.
\end{align*}
The weight derivative term has a favorable sign. Moreover,
\begin{equation*}
\psi_{(1)}\c d_k^*\psi_{(2)}
-\psi_{(2)}\c d_k\psi_{(1)}
=-\sld_k(\psi_{(1)}\c\psi_{(2)}),
\end{equation*}
where
\begin{equation*}
\sld_1\bigl(\xi\c(f,f_*)\bigr)
:=\sdiv(f\xi)+\curl(f_*\xi),\qquad\sld_2\xi:=\sdiv\xi,
\end{equation*}
for $\xi\in\sk_1$ and $(f,f_*)\in\sk_0$. Since the spacetime volume form contains the factor $\Om^2$, integration by parts on the spheres gives
\begin{align*}
\left|\int_Vw(\ub)\sld_k(\psi_{(1)}\c\psi_{(2)})\right|
\les\int_Vw(\ub)|\nab\log\Om||\psi_{(1)}||\psi_{(2)}|\les\int_Vw(\ub)\Ga_0\c\psi_{(1)}\c\psi_{(2)}.
\end{align*}
Integrating the divergence identity over $V=V(u,\ub)$ and using $k\in\{1,2\}$ proves \eqref{rpest}.

For \eqref{bianchi2}, the same argument applies to
\begin{equation*}
k\,w(\ub)|\psi_{(1)}|^2e_3+w(\ub)|\psi_{(2)}|^2e_4.
\end{equation*}
The principal angular terms cancel in the same way, while
$w'(\ub)|\psi_{(2)}|^2\leq0$. This completes the proof.
\end{proof}
\subsection{Estimate for \texorpdfstring{$(\b,(\K,\sic))$}{}}
\begin{prop}\label{estbr}
We have the following estimate for $0\leq i\leq 3$:
\begin{align*}
\left\|h(\ub)Z^i\b\right\|_{L^\infty_uL^2_\ub L^2(S)}
+\left\|h(\ub)Z^i(\K,\sic)\right\|_{L^\infty_\ub L^2_uL^2(S)}
\les 1.
\end{align*}
\end{prop}
\begin{proof}
By Propositions \ref{renorbianchi} and \ref{commutation}, the commuted equations have the following schematic form for $0\leq i\leq3$:
\begin{equation}\label{curvature-commuted-pair}
\begin{aligned}
\nab_3(\dk^i\b)
&=\dkb(\dk^i(\K,\sic))+B(\Ga_{-1}\c\Ga_1)^{(4)}+B\Ssl^{(4)}+(J_{3\bu4})^{(3)},\\
\nab_4\bigl(\dk^i(\K,\sic)\bigr)
&=-\dkb(\dk^i\b)+B(\Ga_0\c\Ga_1)^{(4)}+B\Ssl_4^{(4)}+(\Ga_1\c\Sslh)^{(3)}+(\dual J_{434})^{(3)}.
\end{aligned}
\end{equation}
Write $L^2(V):=L^2_uL^2_\ub L^2(S)$. Using $L^\infty(S)\cdot L^2(S)$ when one factor has at most one derivative, and $L^4(S)\cdot L^4(S)$ for the $2+2$ terms, sphere Sobolev gives
\begin{equation}\label{curvature-product-split}
\|(\Ga_a\c\Ga_b)^{(4)}\|_{L^2(S)}
\les\|\Ga_a^{(4)}\|_{L^2(S)}\|\Ga_b^{(3)}\|_{L^2(S)}
+\|\Ga_a^{(3)}\|_{L^2(S)}\|\Ga_b^{(4)}\|_{L^2(S)}.
\end{equation}
Here $a,b\in\{-1,0,1\}$, and the fourth order products contain an angular derivative, as in Definition \ref{gammag}. Applying Proposition \ref{keyintegral} with $w(x)=h(x)^2$ gives
\begin{align*}
&\int_\cuv h(\ub)^2|\dk^i\b|^2+\int_\ucuv h(\ub)^2|\dk^i(\K,\sic)|^2\\
\les{}&1+B\int_Vh(\ub)^2\left(
\Ga_1^{(4)}\c(\Ga_{-1}\c\Ga_1)^{(4)}
+\Ga_0^{(4)}\c(\Ga_0\c\Ga_1)^{(4)}\right)+B\int_Vh(\ub)^2\left(\Ga_1^{(4)}\c\Ssl^{(4)}+\Ga_0^{(4)}\c\Ssl_4^{(4)}\right)\\
&+\int_Vh(\ub)^2\left(\Ga_1^{(4)}\c J_{3\bu4}^{(3)}+\Ga_0^{(4)}\c(\dual J_{434})^{(3)}
+\Ga_0^{(4)}\c(\Ga_1\c\Sslh)^{(3)}\right).
\end{align*}
By H\"older and \eqref{curvature-product-split},
\begin{equation}\label{curvature-weighted-products}
\begin{aligned}
\int_Vh(\ub)^2\Ga_1^{(4)}\c\Ga_{-1}^{(4)}\c\Ga_1
&\les\de^{\frac12}
\left\|h(\ub)\Ga_1^{(4)}\right\|_{L^\infty_uL^2_\ub L^2(S)}
\left\|\Ga_{-1}^{(4)}\right\|_{L^\infty_\ub L^2_uL^2(S)}
\left\|h(\ub)\Ga_1^{(3)}\right\|_{L^2_\ub L^\infty_uL^2(S)}
\les\de^{\frac12}B^3,\\
\int_Vh(\ub)^2\Ga_1^{(4)}\c\Ga_{-1}\c\Ga_1^{(4)}
&\les\de^{\frac12}
\left\|h(\ub)\Ga_1^{(4)}\right\|_{L^\infty_uL^2_\ub L^2(S)}^2
\left\|\Ga_{-1}^{(3)}\right\|_{L^2_uL^\infty_\ub L^2(S)}
\les\de^{\frac12}B^3.
\end{aligned}
\end{equation}
For the remaining geometric product,
\begin{equation}\label{curvature-central-product}
\begin{aligned}
&\int_Vh(\ub)^2\Ga_0^{(4)}\c(\Ga_0\c\Ga_1)^{(4)}\\
\les{}&B\|h(\ub)\Ga_0^{(4)}\|_{L^2(V)}
\|h(\ub)\Ga_1^{(4)}\|_{L^2(V)} +\|h(\ub)\Ga_0^{(4)}\|_{L^\infty_\ub L^2_uL^2(S)}
\|\Ga_0^{(4)}\|_{L^2(V)}
\|h(\ub)\Ga_1^{(3)}\|_{L^2_\ub L^\infty_uL^2(S)}\\
\les{}&\de^{\frac12}B^3.
\end{aligned}
\end{equation}
By the bootstrap bounds for the Schouten and Cotton tensors,
\begin{equation}\label{curvature-matter-pairing}
\begin{aligned}
&\int_Vh(\ub)^2\left(\Ga_1^{(4)}\c\Ssl^{(4)}+\Ga_0^{(4)}\c\Ssl_4^{(4)}\right)\\
\les{}&\|h(\ub)\Ga_1^{(4)}\|_{L^2(V)}\|h(\ub)\Ssl^{(4)}\|_{L^2(V)}
+\|h(\ub)\Ga_0^{(4)}\|_{L^2(V)}\|h(\ub)\Ssl_4^{(4)}\|_{L^2(V)}
\les\de B^2,
\end{aligned}
\end{equation}
and
\begin{equation*}
\begin{aligned}
&\int_Vh(\ub)^2\left(\Ga_1^{(4)}\c J_{3\bu4}^{(3)}+\Ga_0^{(4)}\c(\dual J_{434})^{(3)}\right)\\
\les{}&\|h(\ub)\Ga_1^{(4)}\|_{L^2(V)}\|h(\ub)J_{3\bu4}^{(3)}\|_{L^2(V)}
+\|h(\ub)\Ga_0^{(4)}\|_{L^2(V)}\|h(\ub)(\dual J_{434})^{(3)}\|_{L^2(V)}
\les\de^{\frac12}B^2.
\end{aligned}
\end{equation*}
For the lower Schouten factors, Lemma \ref{evolution}, the initial bounds, and $\TT_\S\leq B$ give
\begin{equation}\label{curvature-schouten-trace}
\begin{aligned}
\|Z^{\leq3}(\S_{33},\S_{34},\Ssl_3,\Ssl)\|_{L^\infty_uL^\infty_\ub L^2(S)}
\les1+\de^{1/2}\|Z^{\leq4}(\S_{33},\S_{34},\Ssl_3,\Ssl)\|_{L^\infty_\ub L^2_uL^2(S)} 
\les1+\de^{1/2}B\les1.
\end{aligned}
\end{equation}
The same sphere splitting and \eqref{curvature-schouten-trace} give
\begin{equation}\label{curvature-schouten-product}
\int_Vh(\ub)^2\Ga_0^{(4)}\c(\Ga_1\c\Sslh)^{(3)}
\les{}\|h(\ub)\Ga_0^{(4)}\|_{L^2(V)}
\|h(\ub)\Ga_1^{(3)}\|_{L^2(V)}
\|\Sslh^{(3)}\|_{L^\infty_uL^\infty_\ub L^2(S)}
\les{}\de B^3.
\end{equation}
Consequently,
\begin{align*}
\int_\cuv h(\ub)^2|\dk^i\b|^2+\int_\ucuv h(\ub)^2|\dk^i(\K,\sic)|^2
\les1+\de^{1/2}B^4\les1.
\end{align*}
Finally, Proposition \ref{ellipticLp}, the definition of $\dk$, and Proposition \ref{commutation} recover the $Z^i$ estimates.
\end{proof}
\subsection{Estimate for \texorpdfstring{$((\K,-\sic),\bb)$}{}}
\begin{prop}\label{estrb}
We have the following estimate for $0\leq i\leq 3$:
\begin{align*}
\left\|Z^i(\K,\sic)\right\|_{L^\infty_uL^2_\ub L^2(S)}
+\left\|Z^i\bb\right\|_{L^\infty_\ub L^2_uL^2(S)}
\les 1.
\end{align*}
\end{prop}
\begin{proof}
Proceeding as in Proposition \ref{estbr}, we have, for $0\leq i\leq3$,
\begin{align*}
\nab_3\bigl(\dk^i(\K,-\sic)\bigr)
&=\dkb(\dk^i\bb)+B(\Ga_0\c\Ga_{-1})^{(4)}+(\Ssl_3)^{(4)}
+(\Ga_{-1}\c\Ssl)^{(3)}+(\dual J_{343})^{(3)},\\
\nab_4(\dk^i\bb)
&=-\dkb\bigl(\dk^i(\K,-\sic)\bigr)+B(\Ga_{-1}\c\Ga_1)^{(4)}
+\Ssl^{(4)}+J_{4\bu3}^{(3)}.
\end{align*}
Applying Proposition \ref{keyintegral} with $w(x)=1$, we obtain
\begin{align*}
\int_\cuv|\dk^i(\K,-\sic)|^2+\int_\ucuv|\dk^i\bb|^2
\les{}&1+B\int_V\left(\Ga_0^{(4)}\c(\Ga_0\c\Ga_{-1})^{(4)}
+\Ga_{-1}^{(4)}\c(\Ga_{-1}\c\Ga_1)^{(4)}\right)\\
&+\int_V\left(\Ga_0^{(4)}\c\Ssl_3^{(4)}+\Ga_{-1}^{(4)}\c\Ssl^{(4)}
+\Ga_0^{(4)}\c(\Ga_{-1}\c\Ssl)^{(3)}\right)\\
&+\int_V\left(\Ga_{-1}^{(4)}\c J_{4\bu3}^{(3)}+\Ga_0^{(4)}\c(\dual J_{343})^{(3)}\right).
\end{align*}
As in \eqref{curvature-central-product}, the first geometric term satisfies
\begin{equation}\label{curvature-unweighted-central}
\int_V\Ga_0^{(4)}\c(\Ga_0\c\Ga_{-1})^{(4)}
\les B\|\Ga_0^{(4)}\|_{L^2(V)}
\left(\|\Ga_{-1}^{(4)}\|_{L^2(V)}+\|\Ga_0^{(4)}\|_{L^2(V)}\right) \les\de B^3.
\end{equation}
The counterparts of \eqref{curvature-weighted-products} now use $\|h^{-1}\|_{L^2_\ub}$:
\begin{equation}\label{curvature-unweighted-products}
\begin{aligned}
\int_V\Ga_{-1}^{(4)}\c\Ga_{-1}^{(4)}\c\Ga_1
&\les
\left\|\Ga_{-1}^{(4)}\right\|_{L^\infty_\ub L^2_uL^2(S)}^2
\left\|h(\ub)\Ga_1^{(3)}\right\|_{L^2_\ub L^\infty_uL^2(S)}
\left\|h(\ub)^{-1}\right\|_{L^2_\ub}
\les\ep B^3,\\
\int_V\Ga_{-1}^{(4)}\c\Ga_{-1}\c\Ga_1^{(4)}
&\les\left\|h(\ub)^{-1}\right\|_{L^2_\ub}
\left\|\Ga_{-1}^{(4)}\right\|_{L^\infty_\ub L^2_uL^2(S)}
\left\|\Ga_{-1}^{(3)}\right\|_{L^2_uL^\infty_\ub L^2(S)}
\left\|h(\ub)\Ga_1^{(4)}\right\|_{L^\infty_uL^2_\ub L^2(S)}
\les\ep B^3.
\end{aligned}
\end{equation}
As in \eqref{curvature-matter-pairing}, the unweighted Schouten fluxes give $\int_V\left(\Ga_0^{(4)}\c\Ssl_3^{(4)}+\Ga_{-1}^{(4)}\c\Ssl^{(4)}\right)\les\de B^2$.
For the Cotton terms, we use $\|h^{-1}\|_{L^2_\ub}\leq\ep$:
\begin{equation}\label{curvature-unweighted-cotton}
\begin{aligned}
&\int_V\left(\Ga_{-1}^{(4)}\c J_{4\bu3}^{(3)}+\Ga_0^{(4)}\c(\dual J_{343})^{(3)}\right)\\
\les{}&\ep\left\|\Ga_{-1}^{(4)}\right\|_{L^\infty_\ub L^2_uL^2(S)}
\left\|h(\ub)J_{4\bu3}^{(3)}\right\|_{L^2(V)}
+\de^{\frac12}\left\|\Ga_0^{(4)}\right\|_{L^\infty_uL^2_\ub L^2(S)}
\left\|(\dual J_{343})^{(3)}\right\|_{L^2(V)}\\
\les{}&(\ep+\de^{\frac12})B^2.
\end{aligned}
\end{equation}
The Schouten product is estimated as in \eqref{curvature-schouten-product}, using \eqref{curvature-schouten-trace}: $\int_V\Ga_0^{(4)}\c(\Ga_{-1}\c\Ssl)^{(3)}\les\de^{\frac12}B^3$.
Consequently,
\begin{align*}
\int_\cuv|\dk^i(\K,-\sic)|^2+\int_\ucuv|\dk^i\bb|^2
\les1+(\ep+\de^{\frac12})B^4\les1.
\end{align*}
The $Z^i$ estimates follow as at the end of the proof of Proposition \ref{estbr}.
\end{proof}
\subsection{Estimate for \texorpdfstring{$(\bb,\aa)$}{}}
\begin{prop}\label{estba}
We have the following estimate for $0\leq i\leq 3$:
\begin{align*}
\left\|Z^i\bb\right\|_{L^\infty_uL^2_\ub L^2(S)}+\left\|Z^i\aa\right\|_{L^\infty_\ub L^2_uL^2(S)}\les 1.
\end{align*}
\end{prop}
\begin{proof}
Proceeding as in Propositions \ref{estbr} and \ref{estrb}, we have, for
$0\leq i\leq 3$,
\begin{align*}
\nab_3(\dk^i\bb)&=-\dkb(\dk^i\aa)+B(\Ga_{-1}\c\Ga_{-1})^{(4)}+B(\Ga_0\c\Ga_{-2})^{(4)}+J_{3\bu3}^{(3)},\\
\nab_4(\dk^i\aa)&=\dkb^*(\dk^i\bb)+B(\Ga_1\c\Ga_{-2})^{(4)}+B(\Ga_0\c\Ga_{-1})^{(4)}+\Jh_3^{(3)}.
\end{align*}
Here, the principal angular terms are written schematically,
with the appropriate adjoint Hodge operators and fixed
numerical coefficients understood.
Applying Proposition \ref{keyintegral} with $w(x)=1$, we obtain
\begin{align*}
&\int_\cuv |\dk^i\bb|^2+\int_\ucuv |\dk^i\aa|^2\\
\les{}&1
+B\int_V\bigg[
\Ga_{-1}^{(4)}\c
\left((\Ga_{-1}\c\Ga_{-1})^{(4)}+(\Ga_0\c\Ga_{-2})^{(4)}\right)+\Ga_{-2}^{(4)}\c
\left((\Ga_1\c\Ga_{-2})^{(4)}+(\Ga_0\c\Ga_{-1})^{(4)}\right)
\bigg]\\
&+\int_V\left(\Ga_{-1}^{(4)}\c J_{3\bu3}^{(3)}+\Ga_{-2}^{(4)}\c\Jh_3^{(3)}\right).
\end{align*}
The terms containing $(\Ga_{-1}\c\Ga_{-1})^{(4)}$ are estimated
exactly as in Proposition \ref{estrb}. By the bootstrap, Sobolev, and
matter estimates, the remaining nonlinear terms satisfy
\begin{align*}
&B\int_V
\Ga_{-1}^{(4)}\c(\Ga_0\c\Ga_{-2})^{(4)}
+B\int_V
\Ga_{-2}^{(4)}\c(\Ga_1\c\Ga_{-2})^{(4)}
\les
(\ep+\de^{\frac12})B^4,
\end{align*}
while
\begin{align*}
\int_V
\left(
\Ga_{-1}^{(4)}\c J_{3\bu3}^{(3)}
+\Ga_{-2}^{(4)}\c\Jh_3^{(3)}
\right)
\les
(\ep+\de^{\frac12})B^2.
\end{align*}
Consequently,
\begin{align*}
\int_\cuv |\dk^i\bb|^2+\int_\ucuv |\dk^i\aa|^2
\les
1+(\ep+\de^{\frac12})B^4
\les 1.
\end{align*}
Finally, Proposition \ref{ellipticLp}, the definition of $\dk$, and
Proposition \ref{commutation} recover the corresponding $Z^i$
estimates. This concludes the proof of Proposition \ref{estba}.
\end{proof}
Combining Propositions \ref{estbr}--\ref{estba}, we conclude the proof
of Theorem \ref{M1}.
\section{Ricci coefficient and metric estimates}\label{secO}
The goal of this section is to prove Theorem \ref{M2} by using null structure equations as transport equations.
\subsection{Lower order estimates}
\begin{prop}\label{estetaetab}
We have the following estimate:
\begin{align*}
\left\|Z^{\leq3}(\eta,\etab)\right\|_{L^\infty_uL^\infty_\ub L^2(S)}
\les1.
\end{align*}
\end{prop}
\begin{proof}
By Proposition \ref{nulles}, $\nab_4\eta=-\b+\Ga_1\c\Ga_0-\frac12\Ssl_4$.
Commuting with $Z^i$ and using Proposition \ref{commutation}, for $0\leq i\leq3$, gives
\begin{align*}
\nab_4(Z^i\eta)=Z^{\leq i}\b+(\Ga_1\c\Ga_0)^{(i)}+\Ssl_4^{(3)}.
\end{align*}
Proposition \ref{sobolev}, with the lower derivative factor in $L^\infty(S)$, gives
$\|(\Ga_1\c\Ga_0)^{(3)}\|_{L^2(S)}\les B\|\Ga_1^{(3)}\|_{L^2(S)}$.
Lemma \ref{evolution} and Cauchy--Schwarz therefore give
\begin{align}
\|Z^i\eta\|_{L^2(S_{u,\ub})}
\les{}&\|Z^i\eta\|_{L^2(S_{u,0})}\notag
+\int_0^\ub\left(\|Z^{\leq3}\b\|_{L^2(S)}+\|(\Ga_1\c\Ga_0)^{(3)}\|_{L^2(S)}+\|\Ssl_4^{(3)}\|_{L^2(S)}\right)d\ub'\\
\les{}&1+\left\|h(\ub)^{-1}\right\|_{L^2_\ub}
\left(\left\|h(\ub)Z^{\leq3}\b\right\|_{L^2_\ub L^2(S)}
+B\left\|h(\ub)\Ga_1^{(3)}\right\|_{L^2_\ub L^2(S)}\right)
+\de^{\frac12}\left\|\Ssl_4^{(3)}\right\|_{L^2_\ub L^2(S)}\notag\\
\les{}&1+\ep B+\ep B^2+\de^{\frac12}B\les1.\label{ricci-outgoing-transport}
\end{align}
For $\etab$, the corresponding commuted equation is
\begin{align*}
\nab_3(Z^i\etab)=Z^{\leq i}\bb+(\Ga_{-1}\c\Ga_0)^{(i)}+\Ssl_3^{(3)}.
\end{align*}
Lemma \ref{evolution} similarly gives
\begin{align}
\|Z^i\etab\|_{L^2(S_{u,\ub})}
&\les\|Z^i\etab\|_{L^2(S_{0,\ub})}
+u^{\frac12}\left(\|Z^{\leq3}\bb\|_{L^2_uL^2(S)}
+B\|\Ga_{-1}^{(3)}\|_{L^2_uL^2(S)}+\|\Ssl_3^{(3)}\|_{L^2_uL^2(S)}\right)\notag\\
&\les1+\de^{\frac12}(B+B^2)\les1.\label{ricci-incoming-transport}
\end{align}
The bounded factor $\Om^2$ is absorbed in the implicit constant.
Taking the supremum in $u$ and $\ub$ concludes the proof.
\end{proof}
\begin{prop}\label{estomomb}
We have the following estimates:
\begin{align*}
\left\|h(\ub)Z^{\leq3}(\trch,\om)\right\|_{L^2_\ub L^\infty_uL^2(S)}
+\left\|h(\ub)Z^{\leq3}\hch\right\|_{L^2_\ub L^\infty_uL^2(S)}
\les1.
\end{align*}
\end{prop}

\begin{proof}
By Proposition \ref{nulles}, $\nab_4\trch=\Ga_1\c\Ga_1-\S_{44}$.
Commuting with $Z^i$, $0\leq i\leq3$, and using Proposition \ref{commutation}, we obtain
\begin{align*}
\nab_4(Z^i\trch)=B(\Ga_1\c\Ga_1)^{(3)}+B\S_{44}^{(3)}.
\end{align*}
Since $h$ is decreasing, $h(\ub')\geq h(\ub)$ for $\ub'\leq\ub$. Lemma \ref{evolution} and Proposition \ref{sobolev} therefore give
\begin{align}
\|Z^i\trch\|_{L^2(S_{u,\ub})}
&\les1+B\int_0^\ub\|\Ga_1^{(3)}\|_{L^2(S)}^2\,d\ub'
+B\int_0^\ub\|\S_{44}^{(3)}\|_{L^2(S)}\,d\ub'\notag\\
&\les1+Bh(\ub)^{-2}\int_0^\ub\|h(\ub')\Ga_1^{(3)}\|_{L^2(S)}^2\,d\ub'
+\de^{\frac12}B^2
\les1+B^3h(\ub)^{-2}+\de^{\frac12}B^2.\label{ricci-quadratic-primitive}
\end{align}
Multiplying by $h(\ub)$ and using $\|h^{-1}\|_{L^2_\ub}\leq\ep$, we obtain the asserted bound for $\trch$ with right hand side $1+\ep B^3+\de^{1/2}B^2\les1$.
For $\om$, Proposition \ref{nulles}, the Gauss equation, and \eqref{simplerGamma} give
\begin{align*}
\nab_3\om&=-\K+\Ga_1\c\Ga_{-1}+\Ssl,\\
\nab_3(Z^i\om)&=BZ^{\leq i}\K+B(\Ga_1\c\Ga_{-1})^{(3)}+B\Ssl^{(3)}.
\end{align*}
Applying \eqref{ricci-incoming-transport} to $Z^i\om$ and taking $h(\ub)L^2_\ub$, we obtain
\begin{align}
\left\|h(\ub)Z^{\leq3}\om\right\|_{L^2_\ub L^\infty_uL^2(S)}
&\les1+\de^{\frac12}B\left\|h(\ub)Z^{\leq3}\K\right\|_{L^2_\ub L^2_uL^2(S)}
+\de^{\frac12}B^2\left\|h(\ub)\Ga_1^{(3)}\right\|_{L^2_\ub L^\infty_uL^2(S)}
+\de^{\frac12}B^2\notag\\
&\les1+\de^{\frac12}B^3\les1,\label{ricci-incoming-weighted-flux}
\end{align}
where we used Proposition \ref{estbr} and placed $\Ga_{-1}^{(3)}$ in $L^\infty_\ub L^2_uL^2(S)$.
Finally, we use the $e_3$ equation for $\hch$ to avoid $\a$:
\begin{align*}
\nab_3\hch+\frac12\trchb\,\hch
&=\nab\hot\eta-\frac12\trch\,\hchb+\eta\hot\eta+\Sslh,\\
\nab_3\hch&=B\Ga_0^{(1)}+B\Ga_{-1}\c\Ga_1,\\
\nab_3(Z^i\hch)&=B\Ga_0^{(4)}+B(\Ga_{-1}\c\Ga_1)^{(3)}.
\end{align*}
The product is the same as in \eqref{ricci-incoming-weighted-flux}. Replacing $Z^{\leq3}\K$ there by $\Ga_0^{(4)}$, whose $h(\ub)L^2_\ub L^2_uL^2(S)$ norm is $\les B$, gives $1+\de^{1/2}(B^2+B^3)\les1$.
\end{proof}

\begin{prop}\label{estbbbrho}
We have the following estimates:
\begin{align*}
&\left\|h(\ub)Z^{\leq2}\b\right\|_{L^2_\ub L^\infty_uL^2(S)}
+\left\|Z^{\leq2}(\K,\sic,\bb)\right\|_{L^\infty_uL^\infty_\ub L^2(S)}
+\left\|Z^{\leq2}\aa\right\|_{L^2_uL^\infty_\ub L^2(S)}
\les1.
\end{align*}
\end{prop}

\begin{proof}
For $\b$, Proposition \ref{renorbianchi}, \eqref{simplerGamma}, and Proposition \ref{commutation} give, for $0\leq i\leq2$,
\begin{align*}
\nab_3(Z^i\b)=BZ^{\leq3}(\K,\sic)+B(\Ga_{-1}\c\Ga_1)^{(3)}+B\Ssl^{(3)}+B(J_{3\bu4})^{(2)}.
\end{align*}
Apply \eqref{ricci-incoming-weighted-flux} to $Z^i\b$, with $Z^{\leq3}(\K,\sic)$ in place of $Z^{\leq3}\K$. The product is unchanged, and the Schouten and Cotton sources have bounded $h(\ub)L^2_\ub L^2_uL^2(S)$ norms. The first bound follows with right hand side $1+\de^{1/2}B^3\les1$.
For $\bb$, Proposition \ref{bianchiequations} gives
\begin{align*}
\nab_3\bb+2\trchb\,\bb&=-\sdiv\aa+(2\ze-\eta)\c\aa+J_{3\bu3},\\
\nab_3(Z^i\bb)&=BZ^{\leq3}\aa+B(\Ga_{-1}\c\bb)^{(2)}
+B(\Ga_0\c\aa)^{(2)}+B(J_{3\bu3})^{(2)}.
\end{align*}
By the definition of the Cotton tensor,
$(J_{3\bu3})_A=J_{3A3}=\frac12\left(\D_A\S_{33}-\D_3\S_{3A}\right)$.
Only angular and $e_3$ derivatives occur. Expanding the covariant derivatives therefore gives
\begin{align*}
\left|(J_{3\bu3})^{(2)}\right|\les B\left|(\S_{3\nu})^{(3)}\right|.
\end{align*}
By Proposition \ref{estba}, Sobolev, and the displayed Cotton reduction, the source has $L^\infty_\ub L^2_uL^2(S)$ norm $\les B^3$. Equation \eqref{ricci-incoming-transport} gives $1+\de^{1/2}B^3\les1$.
We next estimate $\aa$. By Proposition \ref{bianchiequations},
\begin{align*}
\nab_4\aa+\frac12\trch\,\aa
=-\nab\hot\bb+4\om\aa-3(\hchb\rho-\dual\hchb\,\si)+5\ze\hot\bb+\Jh_3.
\end{align*}
Using the definitions of $\K$ and $\sic$, we obtain schematically
\begin{align*}
\nab_4\aa
=2d_2^*\bb+\Ga_1\c\Ga_{-2}
+\Ga_{-1}\c(\K,\sic)+\Ga_0\c\Ga_{-1}
+\Ga_{-1}\c\Ssl+\Jh_3.
\end{align*}
After commuting,
\begin{align*}
\nab_4(Z^i\aa)
={}&BZ^{\leq3}\bb+B(\Ga_1\c\Ga_{-2})^{(3)}
+B(\Ga_{-1}\c(\K,\sic))^{(3)}\\
&+B(\Ga_0\c\Ga_{-1})^{(3)}
+B(\Ga_{-1}\c\Ssl)^{(3)}
+B(\Jh_3)^{(2)}.
\end{align*}
Apply the first line of \eqref{ricci-outgoing-transport} to $Z^i\aa$ and take $L^2_u$. For the singular product,
\begin{align*}
\left\|(\Ga_1\c\Ga_{-2})^{(3)}\right\|_{L^2_uL^1_\ub L^2(S)}
&\les\|h^{-1}\|_{L^2_\ub}
\|h(\ub)\Ga_1^{(3)}\|_{L^2_\ub L^\infty_uL^2(S)}
\|\Ga_{-2}^{(3)}\|_{L^2_uL^\infty_\ub L^2(S)}
\les\ep B^2.
\end{align*}
Here $(\Ga_1\c\Ga_{-2})^{(3)}$ contains at most two $Z$--derivatives of $\Ga_1\c\aa$. The Cotton term satisfies
\begin{align*}
\left\|(\Jh_3)^{(2)}\right\|_{L^2_uL^1_\ub L^2(S)}
\les\left\|h^{-1}\right\|_{L^2_\ub}
\left\|h(\ub)(\Jh_3)^{(2)}\right\|_{L^2_\ub L^2_uL^2(S)}
\les\ep B.
\end{align*}
The other sources have $L^2_uL^1_\ub L^2(S)$ norm $\les\de^{1/2}B^2$, by Proposition \ref{estba}, the preceding estimate for $\bb$, and \eqref{curvature-schouten-trace}. Thus the asserted bound is $1+(\ep+\de^{1/2})B^3\les1$.
Finally, Proposition \ref{renorbianchi} gives
\begin{align*}
\nab_3(\K,-\sic)
=d_1\bb+B(\Ga_0\c\Ga_{-1})^{(1)}+\Ssl_3^{(1)}
+(\Ga_0+\Ga_{-1})\c(\Ssl+\Ssl_3)^{(0)}
+(\dual J_{343})^{(0)}.
\end{align*}
By the definition of the Cotton tensor and the symmetry of $\S$,
\begin{align*}
\dual J_{343}
=(J_{AB3}-J_{BA3})\slin^{AB}
=\frac12\left(\D_B\S_{A3}-\D_A\S_{B3}\right)\slin^{AB}.
\end{align*}
Expanding the covariant derivatives and absorbing the connection terms into the displayed $\nabla_3(\K,-\sic)$ equation gives
$|(\dual J_{343})^{(2)}|\les B|\Ssl_3^{(3)}|$.
Thus,
\begin{align*}
\nab_3\bigl(Z^i(\K,-\sic)\bigr)
= BZ^{\leq3}\bb
+B(\Ga_0\c\Ga_{-1})^{(3)}
+B\Ssl_3^{(3)}
+B\bigl((\Ga_0+\Ga_{-1})\c(\Ssl+\Ssl_3)\bigr)^{(2)}.
\end{align*}
Proposition \ref{estrb} and the Schouten estimates bound the source in $L^\infty_\ub L^2_uL^2(S)$ by $B^3$. Equation \eqref{ricci-incoming-transport} gives $1+\de^{1/2}B^3\les1$.
\end{proof}

\begin{prop}\label{chibLinfty}
We have the following estimate:
\begin{align*}
\left\|Z^{\leq3}\chib\right\|_{L^\infty_uL^\infty_\ub L^2(S)}
\les1.
\end{align*}
\end{prop}

\begin{proof}
By Proposition \ref{nulles},
$\nab_3\chib=\Ga_{-1}\c\Ga_{-1}-\aa-\S_{33}$.
Commuting with $Z^i$, $0\leq i\leq3$, gives
\begin{align*}
\nab_3(Z^i\chib)
=B(\Ga_{-1}\c\Ga_{-1})^{(3)}+B\aa^{(3)}+B\S_{33}^{(3)}.
\end{align*}
Apply \eqref{ricci-incoming-transport}. Sphere Sobolev puts the two product factors in $L^\infty_uL^\infty_\ub L^2(S)$ and $L^\infty_\ub L^2_uL^2(S)$, using $\Ga_{-1}^{(3)}$ for each. The curvature and Schouten terms have the latter norm, giving $1+\de^{1/2}B^3\les1$.
\end{proof}

\subsection{Top order flux estimates}

\begin{prop}\label{esttrchtrchb}
We have the following estimates:
\begin{align*}
\left\|h(\ub)Z^{\leq3}\nab\trch\right\|_{L^\infty_uL^2_\ub L^2(S)}
\les1,\qquad
\left\|Z^{\leq3}\nab\trchb\right\|_{L^\infty_\ub L^2_uL^2(S)}
\les1.
\end{align*}
\end{prop}

\begin{proof}
Commuting the trace equations in Proposition \ref{nulles} with $Z^i\nab$, $0\leq i\leq3$, gives
\begin{align*}
\nab_4(Z^i\nab\trch)&=B(\Ga_1\c\Ga_1)^{(4)}+B\S_{44}^{(4)},\\
\nab_3(Z^i\nab\trchb)&=B(\Ga_{-1}\c\Ga_{-1})^{(4)}+B\S_{33}^{(4)}.
\end{align*}
Equations \eqref{ricci-quadratic-primitive} and \eqref{curvature-product-split}, with factors of orders three and four, give
\begin{align*}
\left\|h(\ub)Z^{\leq3}\nab\trch\right\|_{L^\infty_uL^2_\ub L^2(S)}
\les{}&1+\ep B
\left\|h(\ub)\Ga_1^{(3)}\right\|_{L^2_\ub L^\infty_uL^2(S)}
\left\|h(\ub)\Ga_1^{(4)}\right\|_{L^\infty_uL^2_\ub L^2(S)}\\
&+\de B\left\|\S_{44}^{(4)}\right\|_{L^\infty_uL^2_\ub L^2(S)}
\les1+\ep B^3+\de B^2\les1.
\end{align*}
For $Z^i\nab\trchb$, use \eqref{ricci-incoming-transport} with both product factors in $L^2_u$:
\begin{align*}
\|Z^i\nab\trchb\|_{L^2(S_{u,\ub})}
\les1
+B\left\|\Ga_{-1}^{(3)}\right\|_{L^2_uL^2(S)}
\left\|\Ga_{-1}^{(4)}\right\|_{L^2_uL^2(S)}
+\de^{\frac12}B\left\|\S_{33}^{(4)}\right\|_{L^2_uL^2(S)}
\les1+B^3+\de^{\frac12}B^2.
\end{align*}
Hence
$\|Z^{\leq3}\nab\trchb\|_{L^\infty_\ub L^2_uL^2(S)}
\les\de^{1/2}(1+B^3)+\de B^2\les1$.
\end{proof}

\begin{prop}\label{estkakabmumub}
We have the following estimates:
\begin{align*}
\left\|Z^{\leq3}(\mu,\mub)\right\|_{L^\infty_uL^2_\ub L^2(S)}
&\les1,&
\left\|h(\ub)Z^{\leq3}(\mu,\mub)\right\|_{L^\infty_\ub L^2_uL^2(S)}
&\les1,\\
\left\|h(\ub)Z^{\leq3}\vkp\right\|_{L^\infty_uL^2_\ub L^2(S)}
&\les1.
\end{align*}
\end{prop}

\begin{proof}
By Proposition \ref{nullrenor} and \eqref{simplerGamma},
\begin{align*}
\nab_4\mu&=B\nab^{\leq1}(\Ga_1\c\Ga_0)+\nab^{\leq1}\Ssl_4+\Ga_0\c\Ssl_4,\\
\nab_4(Z^i\mu)&=B^2(\Ga_1\c\Ga_0)^{(4)}+B\Ssl_4^{(4)},\qquad0\leq i\leq3,
\end{align*}
where the last term absorbs the commuted $\Ga_0\c\Ssl_4$ terms.
By \eqref{ricci-outgoing-transport} and \eqref{curvature-product-split},
\begin{align}
\|Z^i\mu\|_{L^2(S_{u,\ub})}
\les{}&1+B^2\int_0^\ub\left(
\|\Ga_1^{(4)}\|_{L^2(S)}\|\Ga_0^{(3)}\|_{L^2(S)}
+\|\Ga_1^{(3)}\|_{L^2(S)}\|\Ga_0^{(4)}\|_{L^2(S)}
\right)d\ub'\\
&+B\int_0^\ub\|\Ssl_4^{(4)}\|_{L^2(S)}\,d\ub'\notag\\
\les{}&1+\ep B^3\|h(\ub)\Ga_1^{(4)}\|_{L^2_\ub L^2(S)}
+B^2h(\ub)^{-1}\|h(\ub)\Ga_1^{(3)}\|_{L^2_\ub L^2(S)}
\|\Ga_0^{(4)}\|_{L^2_\ub L^2(S)}
+\de^{\frac12}B^2\notag\\
\les{}&1+B^4h(\ub)^{-1}+\de^{\frac12}B^2.\label{ricci-mixed-primitive}
\end{align}
Using $\|h^{-1}\|_{L^2_\ub}\leq\ep$ and $h\leq1$, we obtain
\begin{align*}
\left\|Z^{\leq3}\mu\right\|_{L^\infty_uL^2_\ub L^2(S)}
&\les1+\ep B^4+\de B^2\les1,\qquad
\left\|h(\ub)Z^{\leq3}\mu\right\|_{L^\infty_\ub L^2_uL^2(S)}
\les1+\de^{\frac12}B^4+\de^{\frac12}B^2\les1.
\end{align*}
For $\mub$, Proposition \ref{nullrenor} gives
\begin{align*}
\nab_3(Z^i\mub)=B^2(\Ga_{-1}\c\Ga_0)^{(4)}+B\Ssl_3^{(4)}.
\end{align*}
The splitting in \eqref{ricci-mixed-primitive}, now integrated in $u$, gives
\begin{align*}
\|Z^i\mub\|_{L^2(S_{u,\ub})}
&\les1+\de^{\frac12}B^3\left(
\|\Ga_{-1}^{(4)}\|_{L^2_uL^2(S)}
+\|\Ga_0^{(4)}\|_{L^2_uL^2(S)}
\right)
+\de^{\frac12}B\|\Ssl_3^{(4)}\|_{L^2_uL^2(S)}\\
&\les1+\de^{\frac12}B^4h(\ub)^{-1}+\de^{\frac12}B^2.
\end{align*}
The same two flux norms give the bounds for $\mub$.

It remains to estimate $\vkp$. Proposition \ref{nullrenor} gives
\begin{align*}
\nab_3\vkp
&=\nab^{\leq1}(\Ga_1\c\Ga_{-1})
+\nab^{\leq1}(\Ssl,\S_{34})-J_{3\bu4},\\
\nab_3(Z^i\vkp)
&=B(\Ga_1\c\Ga_{-1})^{(4)}
+B(\Ssl,\S_{34})^{(4)}
+B(J_{3\bu4})^{(3)}.
\end{align*}
At top order, the product is estimated by
\begin{align*}
\left\|h(\ub)(\Ga_1\c\Ga_{-1})^{(4)}\right\|_{L^2_\ub L^2_uL^2(S)}
&\les B\left\|h(\ub)\Ga_1^{(4)}\right\|_{L^2_\ub L^2_uL^2(S)}
+B\left\|\Ga_{-1}^{(4)}\right\|_{L^\infty_\ub L^2_uL^2(S)}
\left\|h(\ub)\Ga_1^{(3)}\right\|_{L^2_\ub L^\infty_uL^2(S)}
\les B^3.
\end{align*}
Equation \eqref{ricci-incoming-weighted-flux}, with this fourth order product and the $h(\ub)L^2_\ub L^2_uL^2(S)$ bounds for the Schouten and Cotton sources, gives
$1+\de^{1/2}B^4+\de^{1/2}B^3\les1$.
\end{proof}
\begin{prop}\label{estetaetabomomb}
We have the following estimates:
\begin{align*}
\left\|h(\ub)Z^{\leq3}\nab(\eta,\etab)\right\|_{L^\infty_\ub L^2_uL^2(S)}
+\left\|Z^{\leq3}\nab(\eta,\etab)\right\|_{L^\infty_uL^2_\ub L^2(S)}
&\les1,\\
\left\|h(\ub)Z^{\leq3}\nab\om\right\|_{L^\infty_uL^2_\ub L^2(S)}
&\les1,\\
\left\|h(\ub)Z^{\leq3}\nab\hch\right\|_{L^\infty_uL^2_\ub L^2(S)}
+\left\|Z^{\leq3}\nab\hchb\right\|_{L^\infty_\ub L^2_uL^2(S)}
&\les1.
\end{align*}
\end{prop}
\begin{proof}
We first recall the Hodge systems
\begin{align*}
\sdiv\eta=-\mu+\K,\qquad
\curl\eta=\sic,\qquad
\sdiv\etab=-\mub+\K,\qquad
\curl\etab=-\sic.
\end{align*}
Commuting with $Z^{\leq3}$, applying
Propositions \ref{commutation} and \ref{ellipticLp}, and using the
lower order estimates to control the commutator terms, we obtain from
Propositions \ref{estkakabmumub}, \ref{estbr}, and \ref{estrb}
\begin{align*}
\left\|h(\ub)Z^{\leq3}\nab(\eta,\etab)\right\|_{L^\infty_\ub L^2_uL^2(S)}
+\left\|Z^{\leq3}\nab(\eta,\etab)\right\|_{L^\infty_uL^2_\ub L^2(S)}
\les1.
\end{align*}Next, from \eqref{renorvkp}, Propositions \ref{commutation} and
\ref{ellipticLp}, we have
\begin{align*}
\left\|h(\ub)Z^{\leq3}\nab\om\right\|_{L^\infty_uL^2_\ub L^2(S)}
\les
\left\|h(\ub)Z^{\leq3}(\b,\vkp)\right\|_{L^\infty_uL^2_\ub L^2(S)}
\les1,
\end{align*}
where we used Propositions \ref{estbr} and \ref{estkakabmumub}. Finally,
the Codazzi equations give
\begin{align*}
\sdiv\hch=
\frac12\nab\trch
+\Ga_0\c\Ga_1
-\b
+\frac12\Ssl_4,\qquad
\sdiv\hchb=
\frac12\nab\trchb
+\Ga_0\c\Ga_{-1}
+\bb
+\frac12\Ssl_3.
\end{align*}Commuting with $Z^{\leq3}$, applying
Propositions \ref{commutation} and \ref{ellipticLp}, and using the
lower order estimates for the commutator terms, we obtain
\begin{align*}
\left\|h(\ub)Z^{\leq3}\nab\hch\right\|_{L^\infty_uL^2_\ub L^2(S)}
&\les
\left\|h(\ub)Z^{\leq3}\nab\trch\right\|_{L^\infty_uL^2_\ub L^2(S)}
+\left\|h(\ub)Z^{\leq3}\b\right\|_{L^\infty_uL^2_\ub L^2(S)}\\
&\quad
+\left\|h(\ub)(\Ga_0\c\Ga_1)^{(3)}\right\|_{L^\infty_uL^2_\ub L^2(S)}
+\left\|h(\ub)\Ssl_4^{(3)}\right\|_{L^\infty_uL^2_\ub L^2(S)} \les1,
\end{align*}
where we used Propositions \ref{esttrchtrchb} and \ref{estbr}, together
with the bootstrap and Schouten estimates. Similarly,
\begin{align*}
\left\|Z^{\leq3}\nab\hchb\right\|_{L^\infty_\ub L^2_uL^2(S)}
&\les
\left\|Z^{\leq3}\nab\trchb\right\|_{L^\infty_\ub L^2_uL^2(S)}
+\left\|Z^{\leq3}\bb\right\|_{L^\infty_\ub L^2_uL^2(S)}\\
&\quad
+\left\|(\Ga_0\c\Ga_{-1})^{(3)}\right\|_{L^\infty_\ub L^2_uL^2(S)}
+\left\|\Ssl_3^{(3)}\right\|_{L^\infty_\ub L^2_uL^2(S)} \les1,
\end{align*}
where we used Propositions \ref{esttrchtrchb} and \ref{estrb}, together
with the bootstrap and Schouten estimates. This concludes the proof.
\end{proof}
\subsection{Estimates for metric components}
Recall that $\slg_{AB}$ and $\bbb^A$ are coordinate components in the transported angular coordinates $(\phi^1,\phi^2)$, and that $Z$ and $\nab$ act on them componentwise by ordinary frame differentiation; see Remark~\ref{rk:metric-components}.
\begin{prop}\label{estmetricangular}
We have the following estimate:
\begin{align*}
\left\|Z^{\leq3}\nab^{\leq1}\left(\log\Om,\slg_{AB},\bbb^A\right)\right\|_{L^\infty_uL^\infty_\ub L^2(S)}
\les1.
\end{align*}
\end{prop}
\begin{proof}
By \eqref{nullidentities} and Proposition \ref{metricequations},
\begin{align*}
e_4(\log\Om)=-\om,\qquad
e_4(\slg_{AB})=2\chi_{AB},\qquad
e_4(\bbb^A)=-2\Om^2(\eta^A-\etab^A),
\end{align*}
where the last two equations are understood in coordinate components.
Commuting componentwise with $Z^i\nab^{\leq1}$, for $0\leq i\leq3$,
and using Proposition \ref{commutation}, we obtain schematically
\begin{align*}
e_4\left(Z^i\nab^{\leq1}\left(\log\Om,\slg_{AB},\bbb^A\right)\right)
=
B^2\Ga_1^{(4)}+B^2\Ga_0^{(4)}.
\end{align*}
Applying \eqref{ricci-outgoing-transport} componentwise, we obtain
\begin{align*}
\left\|Z^i\nab^{\leq1}\left(\log\Om,\slg_{AB},\bbb^A\right)\right\|_{L^2(S_{u,\ub})}
\les{}&
1
+B^2\int_0^\ub
\left(
\|\Ga_1^{(4)}\|_{L^2(S)}
+\|\Ga_0^{(4)}\|_{L^2(S)}
\right)\,d\ub'
\les1+(\ep+\de^{\frac12})B^3\les1.
\end{align*}
This concludes the proof.
\end{proof}

\subsection{Estimates for Christoffel symbols}

\begin{prop}\label{estGamma}
We have the following estimates:
\begin{align*}
\left\|h(\ub)Z^{\leq3}\bGa_{4A}^B\right\|_{L^2_\ub L^\infty_uL^2(S)}
&\les1,&
\left\|h(\ub)Z^{\leq3}\nab^{\leq1}\bGa_{4A}^B\right\|_{L^\infty_uL^2_\ub L^2(S)}
&\les1,\\
\left\|Z^{\leq3}\bGa_{3A}^B\right\|_{L^\infty_uL^\infty_\ub L^2(S)}
&\les1,&
\left\|Z^{\leq3}\bGa_{AB}^C\right\|_{L^\infty_uL^\infty_\ub L^2(S)}
&\les1.
\end{align*}
\end{prop}
\begin{proof}
On each angular coordinate patch, we fix the positively oriented orthonormal frame  $e_A=a_{AC}\pr_{\phi^C}$, where the coefficients $a_{AC}$ depend smoothly only on the coordinate components of $\slg$. We first estimate $\bGa_{4A}^B$. Since $e_4=\pr_\ub$, we have $[e_4,\pr_{\phi^C}]=0$. Hence, by the torsion-free property of $\D$,
\begin{align*}
\g(\D_4\pr_{\phi^C},\pr_{\phi^D})
=
\g(\D_{\pr_{\phi^C}}e_4,\pr_{\phi^D})
=
\chi_{CD},
\end{align*}
where $\chi_{CD}$ denotes the coordinate components. Therefore,
\begin{align*}
\bGa_{4A}^B
&=
\g\left(
\D_4(a_{AC}\pr_{\phi^C}),
a_{BD}\pr_{\phi^D}
\right)=
a_{BD}\left(
e_4(a_{AC})\slg_{CD}
+a_{AC}\chi_{CD}
\right).
\end{align*}
Since $a_{AC}$ depends smoothly only on $\slg$ and $e_4(\slg_{CD})=2\chi_{CD}$, we obtain schematically
\begin{align*}
\bGa_{4A}^B=C_1(\slg)\c\chi.
\end{align*}
Consequently, by Propositions \ref{esttrchtrchb},
\ref{estetaetabomomb}, and \ref{estmetricangular},
\begin{align*}
\left\|h(\ub)Z^{\leq3}\bGa_{4A}^B\right\|_{L^2_\ub L^\infty_uL^2(S)}
\les1,\qquad
\left\|h(\ub)Z^{\leq3}\nab^{\leq1}\bGa_{4A}^B\right\|_{L^\infty_uL^2_\ub L^2(S)}
\les1.
\end{align*}
We next estimate $\bGa_{3A}^B$. Since
$e_3=\Om^{-2}(\pr_u+\bbb)$, we have
\begin{align*}
[e_3,\pr_{\phi^C}]
=
2\pr_{\phi^C}(\log\Om)e_3
-\Om^{-2}(\pr_{\phi^C}\bbb^E)\pr_{\phi^E}.
\end{align*}
The first term on the right hand side has no horizontal component.
Thus, by the torsion-free property,
\begin{align*}
\g(\D_3\pr_{\phi^C},\pr_{\phi^D})
=
\chib_{CD}
-\Om^{-2}(\pr_{\phi^C}\bbb^E)\slg_{ED}.
\end{align*}
It follows that
\begin{align*}
e_3(\slg_{CD})
&=
2\chib_{CD}
-\Om^{-2}(\pr_{\phi^C}\bbb^E)\slg_{ED}
-\Om^{-2}(\pr_{\phi^D}\bbb^E)\slg_{CE}.
\end{align*}
Since $a_{AC}$ depends smoothly only on $\slg$, we obtain
\begin{align*}
\bGa_{3A}^B
&=
\g\left(
\D_3(a_{AC}\pr_{\phi^C}),
a_{BD}\pr_{\phi^D}
\right)=
a_{BD}\left(
e_3(a_{AC})\slg_{CD}
+a_{AC}\chib_{CD}
-\Om^{-2}a_{AC}(\pr_{\phi^C}\bbb^E)\slg_{ED}
\right),
\end{align*}
and hence schematically
\begin{align*}
\bGa_{3A}^B
=
C_2(\log\Om,\slg)\c
\left(
\chib+\pr_\phi\bbb
\right).
\end{align*}
Since $\pr_\phi$ and the angular frame derivatives are uniformly
equivalent under the metric estimates, commuting with $Z^{\leq3}$
and using Propositions \ref{chibLinfty} and \ref{estmetricangular}, we obtain
\begin{align*}
\left\|Z^{\leq3}\bGa_{3A}^B\right\|_{L^\infty_uL^\infty_\ub L^2(S)}
&\les
\left\|Z^{\leq3}\chib\right\|_{L^\infty_uL^\infty_\ub L^2(S)}
+\left\|Z^{\leq3}\nab\bbb\right\|_{L^\infty_uL^\infty_\ub L^2(S)}
+1
\les1.
\end{align*}
Finally,
\begin{align*}
\bGa_{AB}^C
&=
\g(\D_Ae_B,e_C)=
a_{AD}a_{CE}
\g\left(
\D_{\pr_{\phi^D}}(a_{BF}\pr_{\phi^F}),
\pr_{\phi^E}
\right).
\end{align*}
Since both the coefficients $a_{BF}$ and the Christoffel symbols of
the coordinate metric depend smoothly on $\slg$ and its first
coordinate derivatives, we obtain schematically
\begin{align*}
\bGa_{AB}^C
=
C_3(\slg)\c\pr_\phi\slg.
\end{align*}
Therefore, Proposition~\ref{estmetricangular} gives
$\left\|Z^{\leq3}\bGa_{AB}^C\right\|_
{L^\infty_uL^\infty_\ub L^2(S)}\les1$.
This concludes the proof.
\end{proof}
Combining Propositions \ref{estetaetab}--\ref{estGamma}, this concludes the proof of Theorem \ref{M2}.
\section{Estimates for the Vlasov field}\label{secVlasov}
In this section, we prove Theorem~\ref{M3} by combining the characteristic formulation of the Vlasov equation with \eqref{eq:expressionTm-p4} and \eqref{eq:expressionTm-p3}. For $(x,p)\in P_m$ with $p^3,p^4>0$, let
\begin{equation}\label{dfcharacteristic}
\gamma_{x,p}(s):=(x(s),p(s)):=\exp_s(x,p),\qquad\quad \gamma_{x,p}(0)=(x,p),
\end{equation}
be the corresponding affinely parametrized characteristic of the geodesic flow. We denote by $s_*=s_*(x,p)>0$ the first backward hitting time of the initial hypersurfaces, so that
\begin{equation*}
x(-s_*)\in H_0\cup\Hb_0,
\end{equation*}
and set
\begin{equation*}
(x_*(x,p),p_*(x,p)):=\gamma_{x,p}(-s_*).
\end{equation*}
Let $S=S_{u,\ub}\subset V_*$ be a $2$--sphere.
\subsection{Geodesic equations and estimates for \texorpdfstring{$p^\mu$}{}}
\begin{prop}\label{geodesiceqs}
The momentum components along a characteristic satisfy
\begin{align*}
\frac{dp^3}{ds}+\frac12\chi_{AB}p^Ap^B+(\eta_A+\ze_A)p^3p^A+2\om p^3p^4&=0,\\
\frac{dp^4}{ds}+\frac12\chib_{AB}p^Ap^B+2\etab_Ap^4p^A-2\om(p^4)^2&=0,\\
\frac{dp^A}{ds}+\bGa_{BC}^Ap^Bp^C+(\bGa_{3B}^A+\chib_{BA})p^3p^B+(\bGa_{4B}^A+\chi_{BA})p^4p^B+2(\eta^A+\etab^A)p^3p^4&=0.
\end{align*}
Moreover,
\begin{equation}\label{eq:p3p4evolution}
\frac{d}{ds}(p^3p^4)+\frac12\left(p^4\chi_{AB}+p^3\chib_{AB}\right)p^Ap^B+(\eta_A+\etab_A)p^3p^4p^A=0.
\end{equation}
\end{prop}
\begin{proof}
The geodesic equations in the null frame are
\begin{equation}\label{geodesicp}
\dot{p}^\la+\bGa_{\mu\nu}^\la p^\mu p^\nu=0,\qquad \la=1,2,3,4,
\end{equation}
where
\begin{equation*}
\dot{x}=p^\mu e_\mu,\qquad \D_\mu e_\nu=\bGa_{\mu\nu}^\la e_\la.
\end{equation*}
Substituting the connection coefficients from Lemma~\ref{Gaexpression} gives the three displayed equations. Multiplying the first equation by $p^4$, the second by $p^3$, and adding them proves \eqref{eq:p3p4evolution}.
\end{proof}
\begin{lem}\label{estus}
Along every characteristic $\gamma_{x,p}(s)$, one has
\begin{equation*}
\frac{du}{ds}=\Om^{-2}p^3,\qquad\qquad \frac{d\ub}{ds}=p^4.
\end{equation*}
\end{lem}
\begin{proof}
The definitions of the optical functions and the null frame give
\begin{equation*}
e_3(u)=\Om^{-2},\qquad e_4(\ub)=1,\qquad e_A(u)=e_4(u)=e_A(\ub)=e_3(\ub)=0.
\end{equation*}
The conclusion follows from $\dot x=p^\mu e_\mu$.
\end{proof}

The following estimate will be used repeatedly throughout Section~\ref{secVlasov}. The sphere norm at parameter $s$ is understood to be taken on $S_{u(s),\ub(s)}$.

\begin{lem}\label{intC1}
Along every characteristic, one has
\begin{equation*}
\int_{-s_*}^0\left(p^3\|\Ga_{-1}^{(3)}\|_{L^2(S)}+q\|\Ga_0^{(3)}\|_{L^2(S)}+p^4\|\Ga_1^{(3)}\|_{L^2(S)}\right)ds\les\ep.
\end{equation*}
\end{lem}

\begin{proof}
By Lemma~\ref{estus} and the uniform lapse bounds, we have $p^3ds=\Om^2du\simeq du$, $p^4ds=d\ub$. Consequently, Cauchy--Schwarz gives
\begin{equation}\label{vlasov:null-characteristic}
\int_{-s_*}^0\left(p^3\|\Ga_{-1}^{(3)}\|_{L^2(S)}+p^4\|\Ga_1^{(3)}\|_{L^2(S)}\right)ds
\les\de^{\frac12}\|\Ga_{-1}^{(3)}\|_{L^2_uL^\infty_\ub L^2(S)}
+\|h(\ub)^{-1}\|_{L^2_\ub}\|h(\ub)\Ga_1^{(3)}\|_{L^2_\ub L^\infty_uL^2(S)}\les\ep.
\end{equation}
For the remaining term,
\begin{equation}\label{vlasov:mixed-characteristic}
\int_{-s_*}^0q\|\Ga_0^{(3)}\|_{L^2(S)}ds
\les\|\Ga_0^{(3)}\|_{L^\infty_uL^\infty_\ub L^2(S)}
\left(\int_{-s_*}^0p^3ds\right)^{\frac12}
\left(\int_{-s_*}^0p^4ds\right)^{\frac12} \les\de\|\Ga_0^{(3)}\|_{L^\infty_uL^\infty_\ub L^2(S)}\les\ep.
\end{equation}
Combining the two estimates proves the lemma.
\end{proof}

\begin{prop}\label{pmuest}
Along every characteristic intersecting $\supp f$, one has, for every $s\in[-s_*,0]$,
\begin{equation*}
\left|\log\frac{p^3(s)}{p^3(-s_*)}\right|+\left|\log\frac{p^4(s)}{p^4(-s_*)}\right|\les\ep.
\end{equation*}
In particular,
\begin{equation*}
p^3(s)\simeq p^3(-s_*),\qquad p^4(s)\simeq p^4(-s_*),
\end{equation*}
and, on $\supp f$,
\begin{equation*}
|p^\mu|\les1,\qquad \mu=1,2,3,4.
\end{equation*}
\end{prop}
\begin{proof}
Since $f$ is constant along the geodesic flow, the initial phase point of every characteristic intersecting $\supp f$ belongs to the initial support of $f$, so $0<p^3(-s_*),p^4(-s_*)\les1$. Next, Proposition~\ref{geodesiceqs} gives
\begin{align*}
\frac{d\log p^3}{ds}&=-\frac{\chi_{AB}p^Ap^B}{2p^3}-(\eta_A+\ze_A)p^A-2\om p^4,\\
\frac{d\log p^4}{ds}&=-\frac{\chib_{AB}p^Ap^B}{2p^4}-2\etab_Ap^A+2\om p^4.
\end{align*}
Using $|p^Ap^B|\les p^3p^4$, we obtain
\begin{equation*}
\left|\frac{d\log p^3}{ds}\right|+\left|\frac{d\log p^4}{ds}\right|\les p^3|\Ga_{-1}|+|p^A||\Ga_0|+p^4|\Ga_1|.
\end{equation*}
Therefore, Lemma~\ref{intC1} implies
\begin{align*}
\left|\log\frac{p^3(s)}{p^3(-s_*)}\right|+\left|\log\frac{p^4(s)}{p^4(-s_*)}\right|
\les\int_{-s_*}^{s}\left(p^3\|\Ga_{-1}^{(2)}\|_{L^2(S)}+|p^A|\|\Ga_0^{(2)}\|_{L^2(S)}+p^4\|\Ga_1^{(2)}\|_{L^2(S)}\right)ds' \les\ep.
\end{align*}
Thus $p^3(s)\simeq p^3(-s_*)$ and $p^4(s)\simeq p^4(-s_*)$. The initial momentum-support assumption gives $p^3+p^4\les1$ on $\supp f$, and the mass-shell relation yields $|p^A|^2\leq4p^3p^4\les1$, completing the proof.
\end{proof}
\subsection{Commutation identities}
We recall the index conventions used throughout this section:
\begin{equation*}
i,j,k\in\{1,2,3\},\qquad i',j',k'\in\{1,2,4\},\qquad \mu,\nu,\la\in\{1,2,3,4\},\qquad A,B,C\in\{1,2\}.
\end{equation*}
The horizontal lifts of the null-frame vector fields are denoted by
\begin{equation*}
E_\mu:=\hor(e_\mu)=e_\mu-\bGa_{\mu\nu}^{k'}p^\nu\pr_{p^{k'}},\qquad \mu\in\{1,2,3,4\},\qquad k'\in\{1,2,4\}.
\end{equation*}
\begin{lem}\label{Xecomm}
In the $p^4$--chart, set
\begin{equation*}
q:=\sqrt{p^3p^4},
\end{equation*}
and write
\begin{equation*}
X=p^\mu E_\mu=p^\mu e_\mu-\mathbf Q^B\pr_{p^B}-\mathbf Q^4\pr_{p^4},
\end{equation*}
where
\begin{align*}
\mathbf Q^B&=p^Cp^D\bGa_{CD}^B+p^Cp^3(\chib_{CB}+\bGa_{3C}^B)+p^Cp^4(\chi_{CB}+\bGa_{4C}^B)+2p^3p^4(\eta^B+\etab^B),\\
\mathbf Q^4&=\frac12\chib_{CD}p^Cp^D+2p^4p^C\etab_C-2\om(p^4)^2.
\end{align*}
Then
\begin{align*}
[X,e_A]&=\Bigl[p^C(\bGa_{CA}^B-\bGa_{AC}^B)+p^3(\bGa_{3A}^B-\chib_{AB})+p^4(\bGa_{4A}^B-\chi_{AB})\Bigr]e_B+p^3(\eta_A-\ze_A)e_3\\
&+e_A(\mathbf Q^B)\pr_{p^B}+e_A(\mathbf Q^4)\pr_{p^4},
\end{align*}
and
\begin{align*}
[X,e_3]=\Bigl[p^B(\chib_{BA}-\bGa_{3B}^A)+2p^4(\etab^A-\eta^A)\Bigr]e_A+\Bigl[p^B(\ze_B-\eta_B)+2\om p^4\Bigr]e_3+e_3(\mathbf Q^B)\pr_{p^B}+e_3(\mathbf Q^4)\pr_{p^4}.
\end{align*}
For the weighted momentum derivatives, one has
\begin{align*}
[X,p^4\pr_{p^4}]=p^3e_3-p^4e_4+\Bigl[p^4p^C(\chi_{CB}+\bGa_{4C}^B)-p^3p^C(\chib_{CB}+\bGa_{3C}^B)\Bigr]\pr_{p^B}-\Bigl[\frac12\chib_{CD}p^Cp^D+2\om(p^4)^2\Bigr]\pr_{p^4},
\end{align*}
and
\begin{align*}
[X,q\pr_{p^A}]&=-qe_A-\frac{p^A}{2q}p^3e_3+q\Bigl(\chib_{AC}p^C+2p^4\etab_A\Bigr)\pr_{p^4}\\
&+\Bigg\{q\Bigl[p^D(\bGa_{AD}^C+\bGa_{DA}^C)+p^3(\chib_{AC}+\bGa_{3A}^C)+p^4(\chi_{AC}+\bGa_{4A}^C)+\frac{p^Ap^D}{2p^4}(\chib_{DC}+\bGa_{3D}^C)+p^A(\eta^C+\etab^C)\Bigr]\\
&-\frac{\de_A^C}{4q}\Bigl[p^3\chib_{ED}p^Ep^D+p^4\chi_{ED}p^Ep^D+2q^2p^E(\eta_E+\etab_E)\Bigr]\Bigg\}\pr_{p^C}.
\end{align*}
\end{lem}
\begin{proof}
The expressions for $\mathbf Q^B$ and $\mathbf Q^4$ follow by contracting the connection coefficients in Lemma~\ref{Gaexpression}. Since the momentum coordinates are kept fixed under the action of the spacetime frame,
\begin{equation*}
[X,e_i]=p^\mu[e_\mu,e_i]+e_i(\mathbf Q^B)\pr_{p^B}+e_i(\mathbf Q^4)\pr_{p^4}.
\end{equation*}
Lemma~\ref{Gaexpression}, the symmetry of $\chi$ and $\chib$, and the gauge conditions $\omb=0$ and $\etab+\ze=0$ give
\begin{align*}
p^\mu[e_\mu,e_A]&=\Bigl[p^C(\bGa_{CA}^B-\bGa_{AC}^B)+p^3(\bGa_{3A}^B-\chib_{AB})+p^4(\bGa_{4A}^B-\chi_{AB})\Bigr]e_B+p^3(\eta_A-\ze_A)e_3,\\
p^\mu[e_\mu,e_3]&=\Bigl[p^B(\chib_{BA}-\bGa_{3B}^A)+2p^4(\etab^A-\eta^A)\Bigr]e_A+\Bigl[p^B(\ze_B-\eta_B)+2\om p^4\Bigr]e_3.
\end{align*}
This proves the first two formulas.

On the mass shell, $p^3=\frac{|p^C|^2+m^2}{4p^4}$, $q=\frac12\sqrt{|p^C|^2+m^2}$, and hence
\begin{equation*}
\pr_{p^A}p^3=\frac{p^A}{2p^4},\qquad \pr_{p^4}p^3=-\frac{p^3}{p^4},\qquad \pr_{p^A}q=\frac{p^A}{4q},\qquad \pr_{p^4}q=0.
\end{equation*}
Moreover,
\begin{equation*}
[X,\pr_{p^{k'}}]=-\pr_{p^{k'}}(p^\mu)e_\mu+\pr_{p^{k'}}(\mathbf Q^B)\pr_{p^B}+\pr_{p^{k'}}(\mathbf Q^4)\pr_{p^4}.
\end{equation*}
The required momentum derivatives are
\begin{align*}
\pr_{p^A}\mathbf Q^4&=\chib_{AC}p^C+2p^4\etab_A,\qquad\qquad\qquad \pr_{p^4}\mathbf Q^4=2p^C\etab_C-4\om p^4,\\
\pr_{p^A}\mathbf Q^C&=p^D(\bGa_{AD}^C+\bGa_{DA}^C)+p^3(\chib_{AC}+\bGa_{3A}^C)+p^4(\chi_{AC}+\bGa_{4A}^C)+\frac{p^Ap^D}{2p^4}(\chib_{DC}+\bGa_{3D}^C)+p^A(\eta^C+\etab^C),\\
\pr_{p^4}\mathbf Q^B&=-\frac{p^3}{p^4}p^C(\chib_{CB}+\bGa_{3C}^B)+p^C(\chi_{CB}+\bGa_{4C}^B).
\end{align*}
The formula for $[X,p^4\pr_{p^4}]$ follows from $[X,p^4\pr_{p^4}]=-\mathbf Q^4\pr_{p^4}+p^4[X,\pr_{p^4}]$, together with
\begin{align*}
p^4\pr_{p^4}\mathbf Q^4-\mathbf Q^4&=-\frac12\chib_{CD}p^Cp^D-2\om(p^4)^2,\\
p^4\pr_{p^4}\mathbf Q^B&=-p^3p^C(\chib_{CB}+\bGa_{3C}^B)+p^4p^C(\chi_{CB}+\bGa_{4C}^B).
\end{align*}
It remains to compute $X(q)$. Since $q$ depends only on the angular momentum variables in the $p^4$-chart,
\begin{equation*}
X(q)=-\mathbf Q^B\pr_{p^B}q=-\frac{p^B\mathbf Q^B}{4q}.
\end{equation*}
Metric compatibility implies that the contractions containing $\bGa_{CD}^B$, $\bGa_{3C}^B$, and $\bGa_{4C}^B$ vanish, and therefore
\begin{equation*}
p^B\mathbf Q^B=p^3\chib_{CD}p^Cp^D+p^4\chi_{CD}p^Cp^D+2q^2p^C(\eta_C+\etab_C).
\end{equation*}
This gives the stated expression for $X(q)$. Finally, $[X,q\pr_{p^A}]=X(q)\pr_{p^A}+q[X,\pr_{p^A}]$. Using  $-\frac{qp^A}{2p^4}e_3=-\frac{p^A}{2q}p^3e_3$ and substituting the momentum derivatives of $\mathbf Q^C$ and $\mathbf Q^4$ proves the final formula.
\end{proof}
\begin{lem}\label{lem:XE_commutator}
Restricted to the mass shell $P_m$, the weighted horizontal vector field $p^3E_3$ satisfies
\begin{align*}
[X,p^3E_3]&=p^3\bigl(p^C\chib_{CD}+2p^4\etab^D\bigr)E_D-\Bigl(\frac12\chi_{CD}p^Cp^D+p^3p^C\eta_C\Bigr)E_3\\
&+p^3\Bigl[p^Dp^E\R^C{}_{D3E}+p^Dp^3\R^C{}_{33D}+p^Dp^4\bigl(\R^C{}_{D34}+\R^C{}_{43D}\bigr)+p^3p^4\R^C{}_{334}+(p^4)^2\R^C{}_{434}\Bigr]\pr_{p^C}\\
&+p^3\Bigl[p^Dp^E\R^4{}_{D3E}+p^Dp^4\bigl(\R^4{}_{D34}+\R^4{}_{43D}\bigr)+(p^4)^2\R^4{}_{434}\Bigr]\pr_{p^4}.
\end{align*}
\end{lem}
\begin{proof}
We first compute on the ambient tangent bundle $T\M$, where $\pb^1,\pb^2,\pb^3,\pb^4$ are independent fiber coordinates. Recall that $E_\mu=\hor(e_\mu)=e_\mu-\pb^\nu\bGa_{\mu\nu}^\la\pr_{\pb^\la}$ and $X=\pb^\mu E_\mu$. Using
\begin{equation*}
E_\mu(\pb^\la)=-\bGa_{\mu\nu}^\la\pb^\nu,\qquad [E_\nu,E_\mu]=\hor([e_\nu,e_\mu])-\ver\bigl(\R(e_\nu,e_\mu)\pb\bigr),
\end{equation*}
we obtain
\begin{equation*}
[X,E_\mu]=\pb^\nu\bGa_{\nu\mu}^\la E_\la+\pb^\la\pb^\nu\R^\ka{}_{\la\mu\nu}\pr_{\pb^\ka}.
\end{equation*}
The vertical curvature vector is tangent to every mass shell, since
\begin{equation*}
p_\ka p^\la p^\nu\R^\ka{}_{\la\mu\nu}=p^\rho p^\la p^\nu\R_{\rho\la\mu\nu}=0.
\end{equation*}
After restricting to $P_m$ and using the $p^4$--chart, it follows that
\begin{equation*}
[X,E_\mu]=p^\nu\bGa_{\nu\mu}^\la E_\la+p^\la p^\nu\R^{k'}{}_{\la\mu\nu}\pr_{p^{k'}},\qquad k'\in\{1,2,4\}.
\end{equation*}
Consequently,
\begin{equation*}
[X,p^\alpha E_\mu]=-\mathbf Q^\alpha E_\mu+p^\alpha p^\nu\bGa_{\nu\mu}^\la E_\la+p^\alpha p^\la p^\nu\R^{k'}{}_{\la\mu\nu}\pr_{p^{k'}}, \qquad \mathbf Q^\alpha:=p^\mu p^\nu\bGa_{\mu\nu}^\alpha.
\end{equation*}
Taking $\alpha=\mu=3$ and using Lemma~\ref{Gaexpression} gives
\begin{equation*}
\mathbf Q^3=\frac12\chi_{CD}p^Cp^D+p^3p^C(\eta_C+\ze_C)+2\om p^3p^4.
\end{equation*}
The $E_D$-component is
\begin{equation*}
p^3p^\nu\bGa_{\nu3}^D=p^3\bigl(p^C\chib_{CD}+2p^4\etab^D\bigr),
\end{equation*}
whereas the $E_4$-component vanishes. For the $E_3$-component, we obtain
\begin{align*}
p^3p^\nu\bGa_{\nu3}^3-\mathbf Q^3&=p^3p^C\ze_C+2\om p^3p^4-\Bigl[\frac12\chi_{CD}p^Cp^D+p^3p^C(\eta_C+\ze_C)+2\om p^3p^4\Bigr]\\
&=-\frac12\chi_{CD}p^Cp^D-p^3p^C\eta_C.
\end{align*}
Finally, using $\R^{k'}{}_{\la33}=0$ and $\R^4{}_{3\mu\nu}=\g^{43}\R_{33\mu\nu}=0$, the curvature contractions reduce to
\begin{align*}
p^\la p^\nu\R^C{}_{\la3\nu}&=p^Dp^E\R^C{}_{D3E}+p^Dp^3\R^C{}_{33D}+p^Dp^4\bigl(\R^C{}_{D34}+\R^C{}_{43D}\bigr)+p^3p^4\R^C{}_{334}+(p^4)^2\R^C{}_{434},\\
p^\la p^\nu\R^4{}_{\la3\nu}&=p^Dp^E\R^4{}_{D3E}+p^Dp^4\bigl(\R^4{}_{D34}+\R^4{}_{43D}\bigr)+(p^4)^2\R^4{}_{434}.
\end{align*}
Substituting these expressions into the weighted commutator identity proves the formula.
\end{proof}
\begin{lem}\label{XpHcommutator}
Define
\begin{equation*}
H_A:=e_A-p^D\bGa_{AD}^C\pr_{p^C},\qquad A\in\{1,2\}.
\end{equation*}
Then, on the mass shell $P_m$,
\begin{align*}
[X,qH_A]
&=\Bigg\{q\Bigl[p^C\bGa_{CA}^D+p^3(\bGa_{3A}^D-\chib_{AD})+p^4(\bGa_{4A}^D-\chi_{AD})\Bigr]+\de_A^DX(q)\Bigg\}e_D+qp^3(\eta_A-\ze_A)e_3\\
&+\Bigg\{qp^Dp^3e_A(\chib_{DC})+qp^Dp^4e_A(\chi_{DC})+2qp^3p^4e_A(\eta^C+\etab^C)\\
&+qp^Dp^E\Bigl[\R^C{}_{DAE}-\bGa_{AF}^C\bGa_{ED}^F+\bGa_{EF}^C\bGa_{AD}^F+(\bGa_{AE}^F-\bGa_{EA}^F)\bGa_{FD}^C\\
&\hspace{34mm}-\frac12\chib_{AC}\chi_{ED}-\frac12\chi_{AC}\chib_{ED}+\frac12\chib_{EC}\chi_{AD}+\frac12\chi_{EC}\chib_{AD}\Bigr]\\
&+qp^Dp^3\Bigl[\R^C{}_{DA3}-\bGa_{AF}^C\bGa_{3D}^F-\chib_{AC}\eta_D+\bGa_{3F}^C\bGa_{AD}^F+\eta^C\chib_{AD}\\
&\hspace{34mm}+(\chib_{AF}-\bGa_{3A}^F)\bGa_{FD}^C+(\ze_A-\eta_A)\bGa_{3D}^C\Bigr]\\
&+qp^Dp^4\Bigl[\R^C{}_{DA4}-\bGa_{AF}^C\bGa_{4D}^F-\chi_{AC}\etab_D+\bGa_{4F}^C\bGa_{AD}^F+\etab^C\chi_{AD}+(\chi_{AF}-\bGa_{4A}^F)\bGa_{FD}^C\Bigr]\\
&+q\mathbf Q^D\bGa_{AD}^C-qp^D\bGa_{AD}^E\pr_{p^E}\mathbf Q^C-X(q)p^D\bGa_{AD}^C\Bigg\}\pr_{p^C}+q\Bigl[e_A(\mathbf Q^4)-p^D\bGa_{AD}^E\pr_{p^E}\mathbf Q^4\Bigr]\pr_{p^4},
\end{align*}
where
\begin{align*}
\mathbf Q^C&=p^Dp^E\bGa_{DE}^C+p^Dp^3(\chib_{DC}+\bGa_{3D}^C)+p^Dp^4(\chi_{DC}+\bGa_{4D}^C)+2p^3p^4(\eta^C+\etab^C),\\
\mathbf Q^4&=\frac12p^Dp^E\chib_{DE}+2p^4p^D\etab_D-2\om(p^4)^2,
\end{align*}
and
\begin{align*}
\pr_{p^E}\mathbf Q^C&=p^D(\bGa_{ED}^C+\bGa_{DE}^C)+p^3(\chib_{EC}+\bGa_{3E}^C)+p^4(\chi_{EC}+\bGa_{4E}^C)+\frac{p^Ep^D}{2p^4}(\chib_{DC}+\bGa_{3D}^C)+p^E(\eta^C+\etab^C),\\
\pr_{p^E}\mathbf Q^4&=p^D\chib_{ED}+2p^4\etab_E.
\end{align*}
Moreover,
\begin{equation*}
X(q)=-\frac{p^B\mathbf Q^B}{4q}=-\frac{1}{4q}\Bigl[p^3p^Cp^D\chib_{CD}+p^4p^Cp^D\chi_{CD}+2q^2p^C(\eta_C+\etab_C)\Bigr].
\end{equation*}
\end{lem}
\begin{proof}
Since
\begin{equation*}
X=p^\mu e_\mu-\mathbf Q^C\pr_{p^C}-\mathbf Q^4\pr_{p^4},
\end{equation*}
a direct expansion gives
\begin{align*}
[X,H_A]
&=p^\mu(\bGa_{\mu A}^\la-\bGa_{A\mu}^\la)e_\la+p^D\bGa_{AD}^C\pr_{p^C}(p^\mu)e_\mu\\
&+\Bigl[e_A(\mathbf Q^C)-p^\mu p^De_\mu(\bGa_{AD}^C)+\mathbf Q^D\bGa_{AD}^C-p^D\bGa_{AD}^E\pr_{p^E}\mathbf Q^C\Bigr]\pr_{p^C}\\
&+\Bigl[e_A(\mathbf Q^4)-p^D\bGa_{AD}^E\pr_{p^E}\mathbf Q^4\Bigr]\pr_{p^4}.
\end{align*}
With $\bGa_{ADC}:=\de_{CE}\bGa_{AD}^E$, metric compatibility gives
\begin{equation*}
\bGa_{ADC}+\bGa_{ACD}=e_A\g(e_D,e_C)=0,\qquad p^Cp^D\bGa_{AD}^C=0.
\end{equation*}
Consequently,
\begin{equation*}
p^D\bGa_{AD}^C\pr_{p^C}p^3=\frac{p^Cp^D}{2p^4}\bGa_{AD}^C=0,\qquad p^D\bGa_{AD}^C\pr_{p^C}p^4=0.
\end{equation*}
By Lemma~\ref{Gaexpression} and $\etab+\ze=0$, the base derivative part in the $p^4$--chart is
\begin{equation*}
([X,H_A])_{\mathrm{base}}=\Bigl[p^C\bGa_{CA}^D+p^3(\bGa_{3A}^D-\chib_{AD})+p^4(\bGa_{4A}^D-\chi_{AD})\Bigr]e_D+p^3(\eta_A-\ze_A)e_3.
\end{equation*}
For the vertical part, write
\begin{equation*}
e_A(\mathbf Q^C)=p^\mu p^De_A(\bGa_{\mu D}^C)+p^\mu p^3e_A(\bGa_{\mu3}^C)+p^\mu p^4e_A(\bGa_{\mu4}^C).
\end{equation*}
The last two terms satisfy
\begin{equation*}
p^\mu p^3e_A(\bGa_{\mu3}^C)+p^\mu p^4e_A(\bGa_{\mu4}^C)=p^Dp^3e_A(\chib_{DC})+p^Dp^4e_A(\chi_{DC})+2p^3p^4e_A(\eta^C+\etab^C).
\end{equation*}
Furthermore, the frame curvature identity gives
\begin{equation*}
e_A(\bGa_{\mu D}^C)-e_\mu(\bGa_{AD}^C)=\R^C{}_{DA\mu}-\bGa_{A\la}^C\bGa_{\mu D}^\la+\bGa_{\mu\la}^C\bGa_{AD}^\la+(\bGa_{A\mu}^\la-\bGa_{\mu A}^\la)\bGa_{\la D}^C.
\end{equation*}
Splitting $\mu$ into $E,3,4$, and substituting the coefficients from Lemma~\ref{Gaexpression}, yields the three curvature and connection-quadratic expressions displayed above.

It remains to differentiate the weight. In the $p^4$--chart, $q=\frac12\sqrt{|p^A|^2+m^2}$, $\pr_{p^A}q=\frac{p^A}{4q}$,  $\pr_{p^4}q=0$. Hence
\begin{equation*}
X(q)=-\frac{p^B\mathbf Q^B}{4q}.
\end{equation*}
Metric compatibility implies
\begin{equation*}
p^Bp^Cp^D\bGa_{CD}^B=p^Bp^C\bGa_{3C}^B=p^Bp^C\bGa_{4C}^B=0,
\end{equation*}
and therefore
\begin{equation*}
X(q)=-\frac{1}{4q}\Bigl[p^3p^Cp^D\chib_{CD}+p^4p^Cp^D\chi_{CD}+2q^2p^C(\eta_C+\etab_C)\Bigr].
\end{equation*}
Finally,
\begin{equation*}
[X,qH_A]=X(q)H_A+q[X,H_A],
\end{equation*}
and
\begin{equation*}
X(q)H_A=X(q)e_A-X(q)p^D\bGa_{AD}^C\pr_{p^C}.
\end{equation*}
Combining these identities with the preceding horizontal and vertical expressions proves the result.
\end{proof}
\begin{cor}\label{corVf}
Define the schematic families
\begin{equation*}
\VV_\pa:=\left\{qH_1,qH_2,p^3E_3\right\},\qquad
\VV_\perp:=\left\{q\pr_{p^1},q\pr_{p^2},p^4\pr_{p^4}\right\},
\end{equation*}
and set
\begin{align}
\begin{split}\label{dfCC1CC2}
\CC_1&:=p^4\c\Ga_1+(p^3+q)\c\GG^{(1)},\\
\CC_2&:=(p^3)^2\c\Ga_{-2}^{(1)}
+p^3q\c\Ga_{-1}^{(1)}
+q^2\c\Ga_0^{(1)}
+p^4q\c\Ga_1^{(1)}
+q^2\c\Ga_{-1}\c\Ga_1.
\end{split}
\end{align}
If $f$ satisfies the Vlasov equation $X(f)=0$, then, schematically,
\begin{align*}
X(\VV_\perp f)&=\VV_\pa f+\CC_1\c\VV_\perp f,\\
X(\VV_\pa f)&=\CC_1\c\VV_\pa f+\CC_2\c\VV_\perp f.
\end{align*}
\end{cor}
\begin{proof}
On the mass shell, $q^2=p^3p^4=\frac14\left(|p^A|^2+m^2\right)$, and hence
\begin{equation*}
|p^A|\les q,\qquad |p^Ap^B|\les q^2,\qquad
\left|\frac{p^A}{q}\right|\les1,\qquad
\left|\frac{p^Ap^B}{q^2}\right|\les1.
\end{equation*}
By the definitions of $H_A$ and $E_3$, together with
Lemma~\ref{Gaexpression}, we have schematically
\begin{align*}
qe_A&=qH_A+\CC_1\c\VV_\perp,&
p^Ae_A&=\VV_\pa+\CC_1\c\VV_\perp,\\
p^3e_3&=p^3E_3+\CC_1\c\VV_\perp,&
\mathbf Q^C\pr_{p^C}+\mathbf Q^4\pr_{p^4}
&=\CC_1\c\VV_\perp.
\end{align*}
Therefore, Lemma~\ref{Xecomm} gives
\begin{align*}
[X,q\pr_{p^A}]=\VV_\pa+\CC_1\c\VV_\perp,\qquad
[X,p^4\pr_{p^4}]=-X+\VV_\pa+\CC_1\c\VV_\perp.
\end{align*}
Since $X(f)=0$, this proves
\begin{equation*}
X(\VV_\perp f)=\VV_\pa f+\CC_1\c\VV_\perp f.
\end{equation*}
Next, Lemma~\ref{XpHcommutator},
Lemma~\ref{lem:XE_commutator}, and the preceding conversion formulas
give
\begin{align*}
[X,qH_A]=\CC_1\c\VV_\pa+\CC_2\c\VV_\perp,\qquad
[X,p^3E_3]=\CC_1\c\VV_\pa+\CC_2\c\VV_\perp.
\end{align*}
Consequently,
\begin{equation*}
X(\VV_\pa f)=\CC_1\c\VV_\pa f+\CC_2\c\VV_\perp f.
\end{equation*}
This completes the proof.
\end{proof}
\begin{cor}\label{corXDcomm}
Define
\begin{equation*}
\Dk:=\left\{e_1,e_2,e_3,q\pr_{p^1},q\pr_{p^2},
p^4\pr_{p^4}\right\}.
\end{equation*}
Then, for $0\leq i\leq3$, schematically,
\begin{align*}
X\left(\Dk^{\leq i}\VV_\perp f\right)
&=
\Dk^{\leq i}\left(
\VV_\pa f+\CC_1\c\VV_\perp f
\right)
+\Dk^{\leq i-1}\left(
\CC_1^{(1)}\c\Dk(\VV_\perp f)
\right),\\
X\left(\Dk^{\leq i}\VV_\pa f\right)
&=
\Dk^{\leq i}\left(
\CC_1\c\VV_\pa f+\CC_2\c\VV_\perp f
\right)
+\Dk^{\leq i-1}\left(
\CC_1^{(1)}\c\Dk(\VV_\pa f)
\right).
\end{align*}
\end{cor}
\begin{proof}
By $1\in\GG$, Lemma~\ref{Xecomm}, and the bounds
\begin{equation*}
|p^A|\les q,\qquad
\left|\frac{p^A}{q}\right|\les1,\qquad
\left|\frac{p^Ap^B}{q^2}\right|\les1,
\end{equation*}
we have schematically
\begin{equation*}
[X,e_A]=\CC_1^{(1)}\c\Dk,\qquad
[X,e_3]=\CC_1^{(1)}\c\Dk,\qquad
[X,q\pr_{p^A}]=\CC_1^{(1)}\c\Dk.
\end{equation*}
For the remaining momentum derivative, Lemma~\ref{Xecomm} gives
\begin{equation*}
[X,p^4\pr_{p^4}]=p^3e_3-p^4e_4+\CC_1^{(1)}\c\Dk.
\end{equation*}
Using
\begin{equation*}
p^4e_4
=
X-p^Ae_A-p^3e_3
+\mathbf Q^C\pr_{p^C}
+\mathbf Q^4\pr_{p^4},
\end{equation*}
together with
\begin{equation*}
p^Ae_A+p^3e_3
+\mathbf Q^C\pr_{p^C}
+\mathbf Q^4\pr_{p^4}
=
\CC_1^{(1)}\c\Dk,
\end{equation*}
we obtain
\begin{equation}\label{XDcommfirst}
[X,\Dk]=-X+\CC_1^{(1)}\c\Dk,
\end{equation}
where the $X$-term occurs only in the commutator with
$p^4\pr_{p^4}$, with coefficient $-1$.

Let $Z_1,\ldots,Z_n\in\Dk$. The identity
\begin{equation*}
[X,Z_1\cdots Z_n]
=
\sum_{j=1}^n
Z_1\cdots Z_{j-1}[X,Z_j]Z_{j+1}\cdots Z_n
\end{equation*}
and \eqref{XDcommfirst} give by induction
\begin{equation*}
X\left(\Dk^{\leq i}G\right)
=
\Dk^{\leq i}(XG)
+\Dk^{\leq i-1}\left(
\CC_1^{(1)}\c\Dk G
\right)
\end{equation*}
for every phase-space function $G$, where the last term is absent
when $i=0$. The terms generated by the $X$-component are included
in $\Dk^{\leq i}(XG)$. Applying this identity to
\begin{equation*}
G=\VV_\perp f,\qquad G=\VV_\pa f,
\end{equation*}
and using Corollary~\ref{corVf} proves the result.
\end{proof}
\subsection{Lower order estimates for the distribution function \texorpdfstring{$f(x,p)$}{}}
\begin{prop}\label{Lowerorderf}
We have
\begin{align*}
\|\VV_\perp f\|_{L^\infty_uL^\infty_\ub L^\infty(S;L^\infty(P_{x,m}))}
+\left\|\frac{\VV_\pa f}{p^3+q}\right\|_{L^\infty_uL^\infty_\ub L^\infty(S;L^\infty(P_{x,m}))}
\les1.
\end{align*}
\end{prop}
\begin{proof}
Along $\gamma_{x,p}$, write $p^\mu=p^\mu(0)$ and $q=q(0)$. Proposition~\ref{pmuest} and the initial bounds give
\begin{align*}
p^3(s)&\simeq p^3,\qquad p^4(s)\simeq p^4,\qquad q(s)\simeq q,\qquad |p^A(s)|\les q,\\
|\VV_\perp f(-s_*)|&\les1,\qquad |\VV_\pa f(-s_*)|\les p^3+q.
\end{align*}
By \eqref{vlasov:null-characteristic}--\eqref{vlasov:mixed-characteristic} and sphere Sobolev, the two parts of $\CC_1^{(1)}$ contribute at most $\ep$ and $\de\|\GG^{(4)}\|_{L^\infty_uL^\infty_\ub L^2(S)}\les\ep$. Hence
\begin{equation*}
\int_{-s_*}^0\bigl(p^3(s)+q(s)\bigr)\,ds\les\de,\qquad
\int_{-s_*}^0|\CC_1^{(1)}|\,ds\les\ep.
\end{equation*}
The definition of $\CC_2$, $\Ga_{-1}^{(1)},\Ga_0^{(1)}\in\GG^{(2)}$, and $q^2=p^3p^4$ imply
\begin{equation*}
|\CC_2|\les(p^3)^2|\Ga_{-2}^{(1)}|+q|\CC_1^{(1)}|+q^2|\Ga_{-1}||\Ga_1|.
\end{equation*}
Using $p^3ds\simeq du$, sphere Sobolev, and $\mo+\mr\les1$, we obtain
\begin{align*}
\int_{-s_*}^0|\CC_2|\,ds
&\les p^3\|\Ga_{-2}^{(1)}\|_{L^1_uL^\infty_\ub L^\infty(S)}
+q\int_{-s_*}^0|\CC_1^{(1)}|\,ds
+p^3\|\Ga_{-1}\|_{L^\infty_uL^\infty_\ub L^\infty(S)}
\int_{-s_*}^0p^4|\Ga_1|\,ds\\
&\les\de^{\frac12}p^3\|\Ga_{-2}^{(3)}\|_{L^2_uL^\infty_\ub L^2(S)}+\ep q+\ep p^3
\les(\de^{\frac12}+\ep)(p^3+q).
\end{align*}
Integrating the two equations of Corollary~\ref{corVf} along the characteristic and using \eqref{B1}, we obtain
\begin{align*}
|\VV_\perp f(x,p)|
&\les1+B\int_{-s_*}^0\bigl(p^3(s)+q(s)+|\CC_1|\bigr)\,ds
\les1+B(\de+\ep)\les1,\\
|\VV_\pa f(x,p)|
&\les p^3+q+B\int_{-s_*}^0\left[\bigl(p^3(s)+q(s)\bigr)|\CC_1|+|\CC_2|\right]ds\\
&\les\left[1+B(\de^{\frac12}+\ep)\right](p^3+q)\les p^3+q.
\end{align*}
Taking suprema proves the estimate.
\end{proof}
\subsection{Higher order estimates for the distribution function \texorpdfstring{$f(x,p)$}{}}\label{ssecL2}
\begin{lem}\label{VlasovL2}
Let $F,H:P_m\to\mathbb R$ be sufficiently regular and compactly supported in momentum, and let $w\geq0$ be sufficiently regular. Suppose that
\begin{equation*}
X(F)=H.
\end{equation*}
Then
\begin{align*}
&\int_0^\ub\int_{S_{u,\ub'}}\int_{P_m}p^3w|F|^2\dvol d\si d\ub'
+\int_0^u\int_{S_{u',\ub}}\int_{P_m}p^4w|F|^2\dvol d\si du'\\
&\les\int_0^\ub\int_{S_{0,\ub'}}\int_{P_m}p^3w|F|^2\dvol d\si d\ub'
+\int_0^u\int_{S_{u',0}}\int_{P_m}p^4w|F|^2\dvol d\si du'\\
&+\int_0^u\int_0^\ub\int_{S_{u',\ub'}}\int_{P_m}\left(|X(w)||F|^2+w|F||H|\right)\dvol d\si d\ub'du'.
\end{align*}
\end{lem}
\begin{proof}
Since the geodesic spray is divergence-free with respect to the natural measure on $P_m$,
\begin{equation*}
\bdiv_P\bigl(w|F|^2X\bigr)=X(w)|F|^2+2wFH.
\end{equation*}
Moreover,
\begin{equation*}
d\mu_{P_m}=2\Om^2dud\ub d\si\dvol,\qquad X(u)=\Om^{-2}p^3,\qquad X(\ub)=p^4.
\end{equation*}
Dividing by $2$, the boundary densities on $H_u$ and $\Hb_\ub$ are
\begin{equation*}
p^3w|F|^2d\ub d\si\dvol,\qquad \Om^2p^4w|F|^2du d\si\dvol.
\end{equation*}
The bulk density is $\Om^2(X(w)|F|^2+2wFH)dud\ub d\si\dvol$. Integrate over $P_m|_{V_{u,\ub}}$ and use $\Om\simeq1$.
\end{proof}
In the sequel, we use the shorthand notations
\begin{equation*}
\int_{H_u}:=\int_0^\ub\int_{S_{u,\ub'}}d\si\,d\ub',\qquad
\int_{\Hb_\ub}:=\int_0^u\int_{S_{u',\ub}}d\si\,du',\qquad
\int_V:=\int_0^u\int_0^\ub\int_{S_{u',\ub'}}d\si\,d\ub'\,du'.
\end{equation*}
Recall from $\mo+\mr\les1$ that
\begin{align*}
\|h(\ub)\Ga_1^{(4)}\|_{L^\infty_uL^2_\ub L^2(S)}&\les1,&
\|\Ga_0^{(4)}\|_{L^\infty_uL^2_\ub L^2(S)}&\les1,\\
\|h(\ub)\Ga_0^{(4)}\|_{L^\infty_\ub L^2_uL^2(S)}&\les1,&
\|\Ga_{-2}^{(4)}\|_{L^\infty_\ub L^2_uL^2(S)}
+\|\Ga_{-1}^{(4)}\|_{L^\infty_\ub L^2_uL^2(S)}&\les1.
\end{align*}
On $H_0\cup\Hb_0$, the initial assumptions, Lemma
\ref{initialtracesrepair}, Lemma \ref{Gaexpression}, and Proposition
\ref{sobolev} imply, for fixed momentum,
\begin{align*}
&\|\Dk^{\leq3}(\VV_\perp f)\|_{L^2(S)}
+(p^3+q)^{-1}\|\Dk^{\leq3}(qH_Af)\|_{L^2(S)}+(p^3+q)^{-1}\|\Dk^{\leq3}(p^3E_3f)\|_{L^2(S)}\\
&\les\TT_{(0)}\left(1+\|\GG^{(4)}\|_{L^2(S)}\right)
\les1.
\end{align*}
By Proposition~\ref{pmuest}, on $\supp f$ we have
$p^3,p^4\leq C=C(D)$. In the $p^4$--chart, with
$r=\sqrt{(p^1)^2+(p^2)^2}$ and $q=\frac12\sqrt{r^2+m^2}$,
\begin{align*}
\int_{\supp f}\frac{p^4}{q}\,\dvol\leq
4\pi\int_0^C\int_0^{2C}
\frac{r}{\sqrt{r^2+m^2}}\,dr\,dp^4\leq8\pi C^2.
\end{align*}
The same argument in the $p^3$--chart gives the corresponding bound for
$p^3/q$. Hence, also for $m=0$,
\begin{equation}\label{vlasov:moment-bound}
\int_{\supp f}\frac{p^3+p^4}{q}\,\dvol\les1.
\end{equation}
Therefore,
\begin{align*}
&\int_{H_0}\int_{\supp f}\frac{p^3}{q}
\left(
|\Dk^{\leq3}(\VV_\perp f)|^2
+(p^3+q)^{-2}|\Dk^{\leq3}(\VV_\pa f)|^2
\right)\dvol\\
&+\int_{\Hb_0}\int_{\supp f}\frac{p^4}{q}
\left(
|\Dk^{\leq3}(\VV_\perp f)|^2
+(p^3+q)^{-2}|\Dk^{\leq3}(\VV_\pa f)|^2
\right)\dvol
\les1.
\end{align*}
This gives the initial bounds required in Proposition \ref{fluxf}.
\begin{prop}\label{fluxf}
We have
\begin{align*}
&\int_{H_u}\int_{P_m}p^3q^{-1}\left(
|\Dk^{\leq3}(\VV_\perp f)|^2
+(p^3+q)^{-2}|\Dk^{\leq3}(\VV_\pa f)|^2
\right)\dvol\\
&+\int_{\Hb_\ub}\int_{P_m}p^4q^{-1}\left(
|\Dk^{\leq3}(\VV_\perp f)|^2
+(p^3+q)^{-2}|\Dk^{\leq3}(\VV_\pa f)|^2
\right)\dvol
\les1.
\end{align*}
\end{prop}
\begin{proof}
The proof is divided into six steps.
\paragraph{Step 1. Normalization and momentum weights.}
By the bootstrap assumption \eqref{B1}, the two fluxes
sum to at most $B^2$. Their initial values are bounded
above, and $|\VV_\perp f|\les1$ by
Proposition~\ref{Lowerorderf}. Set
\begin{equation*}
F_\perp:=\Dk^{\leq3}(\VV_\perp f),\qquad
F_\pa:=(p^3+q)^{-1}\Dk^{\leq3}(\VV_\pa f).
\end{equation*}
Proposition~\ref{geodesiceqs} and the mass-shell relation imply
\begin{equation*}
|X(q^{-1})|\les q^{-1}|\CC_1|,\qquad
\left|X\left(q^{-1}(p^3+q)^{-2}\right)\right|
\les q^{-1}(p^3+q)^{-2}|\CC_1|.
\end{equation*}
For $m=0$, insert
$\chi_\kappa=\chi(p^3/\kappa)\chi(p^4/\kappa)$,
where $\chi=0$ on $[0,1]$ and $\chi=1$ on $[2,\infty)$.
The logarithmic geodesic estimates give
\begin{equation*}
|X\chi_\kappa|
\les|\CC_1|
\mathbf1_{\{p^3\sim\kappa\}\cup\{p^4\sim\kappa\}}.
\end{equation*}
As $\kappa\to0$, the cutoff error tends to zero by the
integrable bound \eqref{vlasov:C1-bulk} established in
Step 3; the axes have zero fiber measure.
\paragraph{Step 2. Commuted energy inequalities.}
Weighted momentum differentiation preserves the momentum weights up to bounded factors. Corollary~\ref{corXDcomm} gives
\begin{align*}
X\left(\Dk^{\leq3}\VV_\perp f\right)
&=\Dk^{\leq3}(\VV_\pa f)
+\Dk^{\leq3}\left(\CC_1\c\VV_\perp f\right)
+\Dk^{\leq2}\left(
\CC_1^{(1)}\c\Dk(\VV_\perp f)\right),\\
X\left(\Dk^{\leq3}\VV_\pa f\right)
&=\Dk^{\leq3}\left(\CC_1\c\VV_\pa f\right)
+\Dk^{\leq3}\left(\CC_2\c\VV_\perp f\right)
+\Dk^{\leq2}\left(
\CC_1^{(1)}\c\Dk(\VV_\pa f)\right).
\end{align*}
The middle derivative splits contribute at most
$(\de^{1/2}+\ep)B^2\ll1$ by
\eqref{vlasov:C1-bulk}--\eqref{vlasov:C2-middle},
proved in Steps 3 and 4 below.
Lemma~\ref{VlasovL2}, with weights $q^{-1}$ and
$q^{-1}(p^3+q)^{-2}$, therefore gives
\begin{equation}\label{Vperpest}
\begin{aligned}
&\int_{H_u}\int_{P_m}\frac{p^3}{q}|F_\perp|^2\dvol
+\int_{\Hb_\ub}\int_{P_m}\frac{p^4}{q}|F_\perp|^2\dvol\\
&\les1+\int_V\int_{P_m}\frac{p^3+q}{q}
|F_\perp||F_\pa|\dvol+\int_V\int_{P_m}
\frac{
|F_\perp||\CC_1^{(3)}|
|\Dk^{\leq1}(\VV_\perp f)|
+|\CC_1^{(1)}||F_\perp|^2
}{q}\dvol,
\end{aligned}
\end{equation}
and
\begin{equation}\label{Vpaest}
\begin{aligned}
&\int_{H_u}\int_{P_m}\frac{p^3}{q}|F_\pa|^2\dvol
+\int_{\Hb_\ub}\int_{P_m}\frac{p^4}{q}|F_\pa|^2\dvol\\
&\les1+\int_V\int_{P_m}
\frac{
|F_\pa||\CC_1^{(3)}|(p^3+q)^{-1}
|\Dk^{\leq1}(\VV_\pa f)|
+|\CC_1^{(1)}||F_\pa|^2
}{q}\dvol\\
&\quad+\int_V\int_{P_m}\frac{|F_\pa|}{q(p^3+q)}
\left(
|\CC_2||F_\perp|
+|\CC_2^{(3)}||\VV_\perp f|
\right)\dvol.
\end{aligned}
\end{equation}
The coupling term in \eqref{Vperpest} is bounded by
\begin{equation*}
\begin{aligned}
\int_V\int_{P_m}\frac{p^3+q}{q}
|F_\perp||F_\pa|\dvol\les\int_V\int_{P_m}\frac{p^3+q}{q}
\bigl(|F_\perp|^2+|F_\pa|^2\bigr)\dvol\les\de B^2.
\end{aligned}
\end{equation*}
\paragraph{Step 3. Estimates for the $\CC_1$-terms.}
Use $L^2(S)\cdot L^\infty(S)$ at the endpoints and
$L^4(S)\cdot L^4(S)$ in the middle.
Sphere Sobolev requires at most three derivatives
of each family.
Lemma~\ref{Xecomm} involves only $\Ga_1$ and
$\{\Ga_0,\Ga_{-1},\bGa_{AB}^C,\bGa_{3A}^B\}$;
three derivatives of the latter are explicitly included
in $\GG^{(4)}$. Thus
\begin{equation*}
\CC_1^{(3)}
=p^4\c\Ga_1^{(3)}+(p^3+q)\c\GG^{(4)}.
\end{equation*}
For $F=F_\perp$ or $F=F_\pa$, the resulting bulk terms
have the common bound
\begin{equation}\label{vlasov:C1-bulk}
\begin{aligned}
&\int_0^u\int_0^\ub\int_{P_m}
\frac{
p^4\|\Ga_1^{(3)}\|_{L^2(S)}
+(p^3+q)\|\GG^{(4)}\|_{L^2(S)}
}{q}
\|F\|_{L^2(S)}^2\dvol\,d\ub' du'\\
&\les\left(
\|\Ga_1^{(3)}\|_{L^1_\ub L^\infty_uL^2(S)}
+\de\|\GG^{(4)}\|_{L^\infty_uL^\infty_\ub L^2(S)}
\right)B^2\les(\ep+\de)B^2.
\end{aligned}
\end{equation}
\paragraph{Step 4. Lower order and middle $\CC_2$-terms.}
Recall that
\begin{equation*}
\CC_2
=(p^3)^2\c\Ga_{-2}^{(1)}
+p^3q\c\Ga_{-1}^{(1)}
+q^2\c\Ga_0^{(1)}
+p^4q\c\Ga_1^{(1)}
+q^2\c\Ga_{-1}\c\Ga_1.
\end{equation*}
The boundedness of the lower order geometric quantities gives
\begin{equation*}
\frac{|\CC_2|}{p^3+q}
\les
p^3|\Ga_{-2}^{(1)}|
+(p^3+q)
+p^4|\Ga_1^{(1)}|.
\end{equation*}
Apply the argument of \eqref{vlasov:C1-bulk} with the
additional incoming coefficient
$\|\Ga_{-2}^{(3)}\|_{L^2(S)}$ and use
$2|F_\pa F_\perp|\leq|F_\pa|^2+|F_\perp|^2$.
The $p^3$--flux gives
\begin{equation}\label{vlasov:C2-low-bulk}
\begin{aligned}
\int_V\int_{P_m}
\frac{|F_\pa||\CC_2||F_\perp|}{q(p^3+q)}\dvol
&\les\left(
\de
+\|\Ga_{-2}^{(3)}\|_{L^1_uL^\infty_\ub L^2(S)}
+\|\Ga_1^{(3)}\|_{L^1_\ub L^\infty_uL^2(S)}
\right)B^2\\
&\les(\de^{\frac12}+\ep)B^2.
\end{aligned}
\end{equation}
For $G=\VV_\perp f$, the middle splits use
$L^4(S)\cdot L^4(S)$ and
$L^2(S)\cdot L^\infty(S)$, respectively:
\begin{equation}\label{vlasov:C2-middle}
\sum_{j=1}^2
\|\Dk^j\CC_2\c\Dk^{3-j}G\|_{L^2(S)}
\les
\|\CC_2^{(2)}\|_{L^2(S)}
\|\Dk^{\leq3}G\|_{L^2(S)}.
\end{equation}
The explicit formula for $\CC_2$ and the lower order
geometric bounds give
\begin{equation*}
\frac{\|\CC_2^{(2)}\|_{L^2(S)}}{p^3+q}
\les
p^3\|\Ga_{-2}^{(3)}\|_{L^2(S)}
+(p^3+q)
+p^4\|\Ga_1^{(3)}\|_{L^2(S)}.
\end{equation*}
Thus the bulk estimate \eqref{vlasov:C2-low-bulk}
also controls the middle splits in
\eqref{vlasov:C2-middle}; the additional Sobolev
derivatives are angular.
\paragraph{Step 5. Top order coefficient estimates.}
At the coefficient endpoint, sphere Sobolev and
the lower order bounds give
\begin{equation*}
\|(\Ga_{-1}\c\Ga_1)^{(3)}\|_{L^2(S)}
\les\|\Ga_1^{(3)}\|_{L^2(S)}.
\end{equation*}
Hence
\begin{equation*}
\begin{aligned}
\frac{\|\CC_2^{(3)}\|_{L^2(S)}}{p^3+q}
&\les
p^3\left(
\|\Ga_{-2}^{(4)}\|_{L^2(S)}
+\|\Ga_{-1}^{(4)}\|_{L^2(S)}
\right)
+q\|\Ga_0^{(4)}\|_{L^2(S)}
+p^4\|\Ga_1^{(4)}\|_{L^2(S)}.
\end{aligned}
\end{equation*}
Since $|\VV_\perp f|\les1$, Cauchy--Schwarz on $S$ gives
\begin{equation*}
\int_S\int_{P_m}
\frac{|F_\pa||\CC_2^{(3)}||\VV_\perp f|}
{q(p^3+q)}\dvol d\si
\les
\int_{P_m}
\frac{
\|F_\pa\|_{L^2(S)}\|\CC_2^{(3)}\|_{L^2(S)}
}{q(p^3+q)}\dvol.
\end{equation*}
The momentum pairing uses the following estimate:
for $\supp F\subset\supp f$ and weights $a\geq0$, $r>0$,
\begin{equation}\label{vlasov:moment-CS}
\int_{P_m}\frac{a}{q}|F|\,\dvol
\leq
\left(\int_{P_m}\frac{r}{q}|F|^2\dvol\right)^{\frac12}
\left(
\int_{P_m\cap\supp f}\frac{a^2}{rq}\dvol
\right)^{\frac12}.
\end{equation}
For a coefficient $G(x)$, Cauchy--Schwarz on $S$ gives
\begin{equation}\label{vlasov:sphere-moment-CS}
\begin{aligned}
\int_S\int_{P_m}\frac{a}{q}|F||G|\dvol\leq
\left(
\int_S\int_{P_m}\frac{r}{q}|F|^2\dvol
\right)^{\frac12}
\|G\|_{L^2(S)}
\left(
\sup_{x\in S}
\int_{P_m\cap\supp f}\frac{a^2}{rq}\dvol
\right)^{\frac12}.
\end{aligned}
\end{equation}
For $F=F_\pa$ and
$(a,r)=(p^3,p^3),(q,p^3),(p^4,p^4)$,
the dual weights in \eqref{vlasov:sphere-moment-CS}
are respectively
\begin{equation*}
\frac{a^2}{rq}=\frac{p^3}{q},\qquad\qquad
\frac{q}{p^3}=\frac{p^4}{q},\qquad\qquad
\frac{p^4}{q}.
\end{equation*}
By \eqref{vlasov:moment-bound}, the first two choices give
\begin{align*}
&\int_V\int_{P_m}\frac{|F_\pa|}{q}
\left[
p^3\left(|\Ga_{-2}^{(4)}|+|\Ga_{-1}^{(4)}|\right)
+q|\Ga_0^{(4)}|
\right]\dvol\\
&\les\de^{\frac12}\left(
\|\Ga_{-2}^{(4)}+\Ga_{-1}^{(4)}
\|_{L^\infty_\ub L^2_uL^2(S)}
+\|\Ga_0^{(4)}\|_{L^\infty_uL^2_\ub L^2(S)}
\right)
\left(
\int_V\int_{P_m}\frac{p^3}{q}|F_\pa|^2\dvol
\right)^{\frac12}\\
&\les\de B.
\end{align*}
For $\Ga_1^{(4)}$, use $(a,r)=(p^4,p^4)$
and place $h(\ub)^{-1}$ with $F_\pa$:
\begin{align*}
\int_V\int_{P_m}\frac{p^4}{q}
|F_\pa||\Ga_1^{(4)}|\dvol
&\les\de^{\frac12}
\|h(\ub)\Ga_1^{(4)}\|_{L^\infty_uL^2_\ub L^2(S)}
\left(
\int_V\int_{P_m}
\frac{p^4h(\ub)^{-2}}{q}|F_\pa|^2\dvol
\right)^{\frac12}\\
&\les\de^{\frac12}\|h(\ub)^{-1}\|_{L^2_\ub}B
\les\ep\de^{\frac12}B.
\end{align*}
Consequently,
\begin{equation*}
\int_V\int_{P_m}
\frac{|F_\pa||\CC_2^{(3)}||\VV_\perp f|}
{q(p^3+q)}\dvol
\les(\de+\ep\de^{\frac12})B.
\end{equation*}
\paragraph{Step 6. Closing the flux estimate.}
By Remark~\ref{Gawrk}, the top order Schouten terms
are bounded by $\TT_\S\leq B$.
Combining the preceding estimates with
\eqref{Vperpest}--\eqref{Vpaest}, we conclude
\begin{align*}
&\int_{H_u}\int_{P_m}\frac{p^3}{q}
\bigl(|F_\perp|^2+|F_\pa|^2\bigr)\dvol
+\int_{\Hb_\ub}\int_{P_m}\frac{p^4}{q}
\bigl(|F_\perp|^2+|F_\pa|^2\bigr)\dvol\\
&\les
1+(\de^{\frac12}+\ep)B^2
+(\de+\ep\de^{\frac12})(B+B^2)
\les1,
\end{align*}
where the last step uses \eqref{epep0}.
This proves Proposition~\ref{fluxf}.
\end{proof}
\subsection{Estimates for the Schouten tensor \texorpdfstring{$\S$}{}}
\begin{prop}\label{estS3i}
We have
\begin{equation*}
\left\|
Z^{\leq3}(\S_{33},\S_{34},\Ssl_3,\Ssl)
\right\|_{L^\infty_uL^\infty_\ub L^2(S)}
\les1.
\end{equation*}
\end{prop}
\begin{proof}
This is \eqref{curvature-schouten-trace}, obtained from incoming transport and $\TT_\S\leq B$.
\end{proof}
\begin{prop}\label{estTtop}
For $A,B\in\{1,2\}$ and $\nu\in\{1,2,3,4\}$, we have
\begin{equation*}
\left\|Z^{\leq4}(\S_{4\nu},\S_{AB})\right\|_{L^\infty_uL^2_\ub L^2(S)}
\les1,\qquad
\left\|Z^{\leq4}(\S_{3\nu},\S_{AB})\right\|_{L^\infty_\ub L^2_uL^2(S)}
\les1.
\end{equation*}
\end{prop}
\begin{proof}
Since $\S_{\mu\nu}=\T_{\mu\nu}-\frac13(\Tr\T)\g_{\mu\nu}$, it suffices to estimate $\T$ and its trace. The zeroth-order moments are bounded by $X(f)=0$, Proposition~\ref{pmuest}, and the initial bound for $f$. For the differentiated moments,
\begin{align*}
qe_A(f)&=qH_A(f)+qp^D\bGa_{AD}^C\pr_{p^C}(f)=qH_Af+q\GG^{(1)}\c\VV_\perp f,\\
p^3e_3(f)
&=p^3E_3(f)+p^3p^\nu\bGa_{3\nu}^{k'}\pr_{p^{k'}}(f)
=p^3E_3f+(p^3+q)\GG^{(1)}\c\VV_\perp f.
\end{align*}
Let $F_\perp,F_\pa$ be as in the proof of Proposition~\ref{fluxf}. The frame momentum coordinates and $\dvol=(p^4)^{-1}dp^1dp^2dp^4$ are independent of $x$. Differentiating \eqref{defT} and using the preceding identities gives the sphere estimates
\begin{align}
\|Z^{\leq3}\nab\T_{\mu\nu}\|_{L^2(S)}
&\les1+\|\GG^{(4)}\|_{L^2(S)}
+\int_{P_m}\frac{|p_\mu p_\nu|}{q}
\left((p^3+q)\|F_\pa\|_{L^2(S)}+q\|F_\perp\|_{L^2(S)}\right)\dvol,
\label{estTangular}\\
\|Z^{\leq3}e_3\T_{\mu\nu}\|_{L^2(S)}
&\les1+\|\GG^{(4)}\|_{L^2(S)}
+\int_{P_m}|p_\mu p_\nu|\frac{p^3+q}{p^3}
\left(\|F_\pa\|_{L^2(S)}+\|F_\perp\|_{L^2(S)}\right)\dvol.
\label{estT3}
\end{align}
Here the connection products use $L^\infty(S)\cdot L^2(S)$, $L^4(S)\cdot L^4(S)$, and $L^2(S)\cdot L^\infty(S)$. At most three derivatives fall on either the connection family or $\VV_\perp f$; Proposition~\ref{Lowerorderf} controls the undifferentiated factor, and $\|\GG^{(4)}\|_{L^\infty_uL^\infty_\ub L^2(S)}\les1$. 

Since $|p_3|=2p^4$, $|p_4|=2p^3$, and
$\frac{q(p^3+q)}{p^3}=p^4+q$,
Cauchy--Schwarz in momentum in
\eqref{estTangular}--\eqref{estT3} gives
\begin{align*}
\|Z^{\leq4}\T_{4\nu}\|_{L^2(S)}
&\les1+\left(\int_S\int_{P_m}\frac{p^3}{q}
\left(|F_\perp|^2+|F_\pa|^2\right)\dvol d\si\right)^{\frac12}\\
&\qquad\times\left(\int_{\{0<p^3,p^4\leq C(D)\}}\frac{p^3}{q}|p_\nu|^2
\left((p^3+q)^2+(p^4+q)^2\right)\dvol\right)^{\frac12},\\
\|Z^{\leq4}\T_{3\nu}\|_{L^2(S)}
&\les1+\left(\int_S\int_{P_m}\frac{p^4}{q}
\left(|F_\perp|^2+|F_\pa|^2\right)\dvol d\si\right)^{\frac12}\\
&\qquad\times\left(\int_{\{0<p^3,p^4\leq C(D)\}}\frac{p^4}{q}|p_\nu|^2
\left((p^3+q)^2+(p^4+q)^2\right)\dvol\right)^{\frac12}.
\end{align*}
The last factors are bounded by the calculation in
\eqref{vlasov:moment-bound}.
Proposition~\ref{fluxf}, using the $p^3$--flux on $H_u$
and the $p^4$--flux on $\Hb_\ub$, yields
\begin{equation*}
\left\|Z^{\leq4}\T_{4\nu}\right\|_{L^\infty_uL^2_\ub L^2(S)}\les1,\qquad
\left\|Z^{\leq4}\T_{3\nu}\right\|_{L^\infty_\ub L^2_uL^2(S)}\les1.
\end{equation*}
For the angular components and the trace, use
\begin{equation*}
|p^Ap^B|+m^2\les p^3p^4,\qquad
\Tr\T=-m^2\int_{P_m}f\,\dvol.
\end{equation*}
In the preceding moment estimates, replace $|p_\nu|$
by $p^4$ for the $p^3$--flux and by $p^3$ for the
$p^4$--flux. Then
\begin{align*}
&\left\|Z^{\leq4}\T_{AB}\right\|_{L^\infty_uL^2_\ub L^2(S)}
+\left\|Z^{\leq4}\T_{AB}\right\|_{L^\infty_\ub L^2_uL^2(S)}\\
&\quad+\left\|Z^{\leq4}\Tr\T\right\|_{L^\infty_uL^2_\ub L^2(S)}
+\left\|Z^{\leq4}\Tr\T\right\|_{L^\infty_\ub L^2_uL^2(S)}
\les1.
\end{align*}
The formula for $\S$ proves the proposition.

\end{proof}
\subsection{Estimates for the Cotton tensor \texorpdfstring{$J$}{}}
\begin{prop}\label{Jest}
We have
\begin{align*}
&\|h(\ub)J_{3\bu4}^{(3)}\|_{L^2_\ub L^2_uL^2(S)}
+\|h(\ub)J_{4\bu3}^{(3)}\|_{L^2_\ub L^2_uL^2(S)}
+\|h(\ub)\dual J_{434}^{(3)}\|_{L^2_\ub L^2_uL^2(S)}\\
&+\|\dual J_{343}^{(3)}\|_{L^2_\ub L^2_uL^2(S)}
+\|h(\ub)J_{3\bu3}^{(3)}\|_{L^2_\ub L^2_uL^2(S)}
+\|h(\ub)\Jh_3^{(3)}\|_{L^2_\ub L^2_uL^2(S)}\les1.
\end{align*}
\end{prop}
\begin{proof}
By the definition of the Cotton tensor and the symmetry of $\S$, we have
\begin{align*}
J_{3A4} &=\frac12\left(\D_A\S_{34}-\D_4\S_{3A}\right),&
J_{4A3} &=\frac12\left(\D_A\S_{43}-\D_3\S_{4A}\right),\\
J_{3A3} &=\frac12\left(\D_A\S_{33}-\D_3\S_{3A}\right),&
J_{AB3} &=\frac12\left(\D_B\S_{A3}-\D_3\S_{AB}\right),\\
\dual J_{434} &=\frac12\left(\D_B\S_{A4}-\D_A\S_{B4}\right)\slin^{AB},&
\dual J_{343} &=\frac12\left(\D_B\S_{A3}-\D_A\S_{B3}\right)\slin^{AB}.
\end{align*}
For the displayed components, Lemma~\ref{Gaexpression} gives
\begin{equation*}
\D_i\S_{\mu\nu}
=e_i(\S_{\mu\nu})
+\Ga_1\c\{\S_{3A},\S_{33},\S_{AB},\S_{34}\}
+\GG^{(1)}\c\S,\qquad i=1,2,3.
\end{equation*}
The $\Ga_1$-terms couple only to the four components in braces. Hence
\begin{equation}\label{vlasov:Schouten-derivative}
|(\D_i\S_{\mu\nu})^{(3)}|
\les
|Z^{\leq4}\S_{\mu\nu}|
+\left|
\left(\Ga_1\c\{\S_{3A},\S_{33},\S_{AB},\S_{34}\}\right)^{(3)}
\right|
+\left|(\GG^{(1)}\c\S)^{(3)}\right|.
\end{equation}
The $\GG$-term is controlled by Proposition~\ref{estTtop} and $\|\GG^{(4)}\|_{L^\infty_uL^\infty_\ub L^2(S)}\les1$. Proposition~\ref{estS3i} gives
\begin{equation}\label{vlasov:Cotton-product}
\begin{aligned}
&\left\|h(\ub)\left(\Ga_1\c\{\S_{3A},\S_{33},\S_{AB},\S_{34}\}\right)^{(3)}\right\|_{L^2_\ub L^2_uL^2(S)}\\
&\les\de^{\frac12}\|h(\ub)\Ga_1^{(3)}\|_{L^2_\ub L^\infty_uL^2(S)}
\|Z^{\leq3}(\S_{3A},\S_{33},\S_{AB},\S_{34})\|_{L^\infty_uL^\infty_\ub L^2(S)}\les1.
\end{aligned}
\end{equation}
The product uses $L^2(S)\cdot L^\infty(S)$ at the endpoints and $L^4(S)\cdot L^4(S)$ in the middle; three derivatives suffice. The $u$--integration gives $\de^{1/2}$. Thus
\begin{equation}\label{vlasov:weighted-Cotton}
\begin{aligned}
\|h(\ub)J_{4\bu3}^{(3)}\|_{L^2_\ub L^2_uL^2(S)}
+\|h(\ub)\dual J_{434}^{(3)}\|_{L^2_\ub L^2_uL^2(S)} +\|h(\ub)J_{3\bu3}^{(3)}\|_{L^2_\ub L^2_uL^2(S)}
+\|h(\ub)J_{AB3}^{(3)}\|_{L^2_\ub L^2_uL^2(S)}
\les1.
\end{aligned}
\end{equation}
For $\dual J_{343}$, we have the improved structure
\begin{align*}
\dual J_{343}
=(J_{AB3}-J_{BA3})\slin^{AB}=\frac12\left(\nab_B\S_{A3}-\nab_A\S_{B3}\right)\slin^{AB}
+\GG^{(1)}\c\S.
\end{align*}
In particular, no $\Ga_1$-term appears. Hence
\begin{equation*}
|\dual J_{343}^{(3)}|
\les
|Z^{\leq4}\S_{3A}|
+\left|(\GG^{(1)}\c\S)^{(3)}\right|.
\end{equation*}
Propositions~\ref{estTtop} and~\ref{estS3i} then give
\begin{align*}
\|\dual J_{343}^{(3)}\|_{L^2_\ub L^2_uL^2(S)}&\les\de^{\frac12}\left\|Z^{\leq4}\S_{3A}\right\|_{L^\infty_\ub L^2_uL^2(S)}
+\de^{\frac12}\|\GG^{(4)}\|_{L^\infty_uL^\infty_\ub L^2(S)}\les1.
\end{align*}
For $J_{3\bu4}$, the new derivative is $\D_4\S_{3A}$. Since $\T_{\mu\nu}=\S_{\mu\nu}-(\Tr\S)\g_{\mu\nu}$, the conservation law $\D^\mu\T_{\mu A}=0$ gives
\begin{equation*}
\D_4\S_{3A}
=-\D_3\S_{4A}+2\D^B\S_{BA}-2\D_A(\Tr\S).
\end{equation*}
Using $\Tr\S=\tr\S_{AB}-\S_{34}$ and applying \eqref{vlasov:Schouten-derivative} to the angular and incoming derivatives on the right gives
\begin{align*}
|(\D_4\S_{3A})^{(3)}|\les|Z^{\leq4}(\S_{4A},\S_{AB},\S_{34})|+\left|\left(\Ga_1\c\{\S_{3B},\S_{33},\S_{BC},\S_{34}\}+\GG^{(1)}\c\S\right)^{(3)}\right|.
\end{align*}
Equations~\eqref{vlasov:Schouten-derivative}--\eqref{vlasov:Cotton-product} give the $h(\ub)J_{3\bu4}^{(3)}$ bound. Finally, $J_{343}=2\de^{AB}J_{AB3}$ and the definition of $\Jh_3$ reduce its estimate to \eqref{vlasov:weighted-Cotton}.
\end{proof}
Combining the above propositions concludes the proof of Theorem~\ref{M3}.
\appendix
\section{Instantaneous blow-up for the Einstein--(real) scalar field system}
\label{Appendix}
In contrast to the boundedness conclusion of
Theorem~\ref{Luk3}, we establish the following
local instantaneous blow-up result for the Einstein--(real) scalar
field system.
We work in the characteristic setting of
Theorem~\ref{MainTheorem}, with freely prescribed geometric
data on $H_0$ as in \cite{Daluk} and regular scalar-field
data on $\Hb_0$.
The transverse derivatives are determined by the coupled
characteristic equations.

\begin{prop}
For the Einstein--(real) scalar field system
\begin{align*}
\Ric_{\mu\nu}-\frac12 \R\g_{\mu\nu}=\T_{\mu\nu}^{sc},
\qquad\quad \square_{\g}\phi=0,\qquad\quad
\T_{\mu\nu}^{sc}=\pr_\mu\phi\,\pr_\nu\phi
-\frac12\g_{\mu\nu}|\pr\phi|_{\g}^2,
\end{align*}
consider the characteristic initial value problem in the
geometric setting of Theorem~\ref{MainTheorem}, with regular
compatible scalar-field data replacing the Vlasov data and
$\phi|_{H_0}=0$.
Assume that there exist a connected open angular patch
$U\subset S_{0,0}$ and a constant $c_0>0$ such that
\begin{align*}
|\nab_3\phi(0,0,\vartheta)|
&\geq c_0,
\qquad\qquad\qquad\qquad\quad\vartheta\in U,\\
\int_0^\de|\hch(0,\ub,\vartheta)|^2\,d\ub
&=\infty
\qquad\text{for almost every }\vartheta\in U.
\end{align*}
Then there exists $u_1\in(0,\de)$ such that
\begin{equation*}
\lim_{\ub\to\de}
\|\T_{44}^{sc}\|_{L^\infty(S_{u,\ub})}=\infty,\qquad 0<u\leq u_1.
\end{equation*}
\end{prop}
\begin{proof}
The weighted wave and geometric estimates apply with
$\nab_3\phi$, $\nab\phi$, and $\nab_4\phi$
corresponding to $\Ga_{-1}$, $\Ga_0$, and $\Ga_1$,
respectively; see \cite[Footnote~8]{Luk18}.
In particular,
\begin{align*}
\mathcal E_3:=\left\|\left(h(\ub)Z^{\leq3}\nab_4\phi,Z^{\leq3}\nab\phi\right)\right\|_{L^\infty_uL^2_\ub L^2(S)}+
\left\|\left(Z^{\leq3}\nab_3\phi,
h(\ub)Z^{\leq3}\nab\phi\right)
\right\|_{L^\infty_\ub L^2_uL^2(S)}
\les1,
\end{align*}
where $Z=\{\nab_3,\nab\}$.
These estimates and sphere Sobolev give boundedness of
$\nab_3\phi$, $\nab\phi$, and the regular
geometric quantities.
Integrating $\phi$ from $H_0$ also gives
$\|\Delta\phi\|_{L^\infty_u L^\infty_\ub L^2(S)}\les1$.

Fix a nonempty relatively compact open patch $U'\Subset U$.
Work in coordinates transported by the flow $\Phi_u$ of
$\Om^2e_3$, suppressing the pullbacks below.
Since $\phi|_{H_0}=0$, the wave equation on $H_0$ reduces to
\begin{equation*}
\pr_\ub\nab_3\phi+\frac12\trch\,\nab_3\phi=0.
\end{equation*}
The uniform $L^1_\ub$ bound for $\trch$ therefore bounds
$|\nab_3\phi|$ away from zero on $H_0$ over $U'$.
Moreover, $\mathcal E_3\les1$ gives
\begin{equation*}
|\nab_3\phi(u,\ub,\vartheta)
-\nab_3\phi(0,\ub,\vartheta)|
\les u^{1/2}.
\end{equation*}
Thus, for sufficiently small $u_1>0$, $\nab_3\phi$
has a constant sign and satisfies
\begin{equation*}
|\nab_3\phi(u,\ub,\vartheta)|\geq c>0,
\qquad
0\leq u\leq u_1,\quad
0\leq\ub<\de,\quad \vartheta\in U'.
\end{equation*}
On $H_0$, Raychaudhuri's equation gives
\begin{equation*}
\pr_\ub\trch=-\frac12(\trch)^2-|\hch|^2.
\end{equation*}
Hence $\trch(0,\ub,\vartheta)\to-\infty$
almost everywhere on $U'$, and
$\|\trch(0,\ub)\|_{L^2(U')}\to\infty$.
The Gauss and null structure equations give
\begin{equation*}
\nab_3\trch+\trchb\,\trch
=-2\K+2\sdiv\eta+2|\eta|^2
+|\nab\phi|^2.
\end{equation*}
The right-hand side is bounded.
Consequently, with
\begin{equation*}
\lambda(u,\ub,\vartheta)
:=\exp\left(
\frac12\int_0^u\Om^2\trchb(s,\ub,\vartheta)\,ds
\right)\simeq1,
\end{equation*}
we have
\begin{equation*}
\lambda^2\trch(u,\ub,\vartheta)
=\trch(0,\ub,\vartheta)+O(1)
\end{equation*}
uniformly. Finally, the wave equation reads
\begin{equation*}
\nab_3\nab_4\phi+\frac12\trchb\,\nab_4\phi
=\Delta\phi
-\frac12\trch\,\nab_3\phi
+2\eta^A\nab_A\phi.
\end{equation*}
Since $\nab_4\phi|_{H_0}=0$, integration yields
\begin{equation*}
\lambda\nab_4\phi(u,\ub,\vartheta)
=-\frac12\trch(0,\ub,\vartheta)
\int_0^u\lambda^{-1}\Om^2
\nab_3\phi(s,\ub,\vartheta)\,ds
+\mathcal E(u,\ub,\vartheta),
\end{equation*}
where $\sup_{\ub<\de}\|\mathcal E(u,\ub)\|_{L^2(U')}\leq C_u$. The integral $\int_0^u\lambda^{-1}\Om^2\nab_3\phi(s,\ub,\vartheta)\,ds$ has constant sign and absolute value at least $cu$. Uniform equivalence of the pulled-back sphere measures therefore gives
\begin{equation*}
\|\nab_4\phi(u,\ub)\|_{L^2(S_{u,\ub})}
\geq cu\|\trch(0,\ub)\|_{L^2(U')}-C_u
\longrightarrow\infty
\end{equation*}
for every fixed $0<u\leq u_1$.
The bounded sphere areas and
$\T_{44}^{sc}=(\nab_4\phi)^2$ prove the claim.
\end{proof}

\end{document}